\documentclass[epj]{svjour}
\usepackage{graphics,epsfig}
\newcommand{\nn}{{\nonumber}\\}
\def\ket #1{|#1\rangle}
\def\tens{\otimes}

\def\Id{{\rm 1\kern-.3em I}}
\newcommand{\R}{{\rm I\kern-.2emR}}
\newcommand{\C}{{\ifmmode\mathchoice{{\rm C\kern-.4em\raisebox{.0ex}{\rule{.05ex}{.67em}}\kern.4em}}%
{{\rm C\kern-.4em\raisebox{.03ex}{\rule{.05ex}{.67em}}\kern.4em}}%
{{\rm C\kern-.3em\raisebox{.06ex}{\rule{.05ex}{.45em}}\kern.3em}}%
{{\rm C\kern-.3em\raisebox{.06ex}{\rule{.05ex}{.3em}}\kern.3em}}\else
{{\rm C\kern-.4em\raisebox{.03ex}{\rule{.05ex}{.67em}}\kern.4em}}\fi}}
\def\bra #1{\langle #1|}
\def\ket #1{|#1\rangle}
\def\SP #1 #2{\langle #1|#2\rangle}

\def\Expect #1{\langle #1\rangle}
\def\SpinorComp(#1,#2,#3){\Psi^{#1}_{#3}(#2_{#1})}
\def\AdSpinorComp(#1,#2,#3){\overline\Psi^{#1}_{#3}(#2_{#1})}
\def\spinorComp(#1,#2,#3){\Psi_{#3}(#2_{#1})}
\def\AdspinorComp(#1,#2,#3){\overline\Psi_{#3}(#2_{#1})}
\def\Spinor(#1,#2){\Psi^{#1}(#2_{#1})}
\def\AdSpinor(#1,#2){\bar\Psi^{#1}(#2_{#1})}
\def\SpinorI(#1,#2){\Psi_{Ip}^{#1}(#2_{#1})}
\def\AdSpinorI(#1,#2){\overline\Psi_{Ip}^{#1}(#2_{#1})}

\def\FeyProp(#1,#2,#3){S^{#1}_F(#2_{#1},#3_{#1})}
\def\FeyPropComp(#1,#2,#3,#4,#5){S^{#1}_{F\;#4 #5}(#2_{#1},#3_{#1})}
\def\metric(#1,#2){\langle #1, #2\rangle}
\def\bfgrk #1{\mbox{\boldmath$#1$}}
\def\MSD (#1,#2,#3,#4){[\Delta\frac{#1}{2}^#2]_{#3}(#4)}
\def\MSN (#1,#2,#3,#4){[N\frac{#1}{2}^#2]_{#3}(#4)}
\def\MSL (#1,#2,#3,#4){[\Lambda\frac{#1}{2}^#2]_{#3}(#4)}
\def\MSS (#1,#2,#3,#4){[\Sigma\frac{#1}{2}^#2]_{#3}(#4)}
\def\MSX (#1,#2,#3,#4){[\Xi\frac{#1}{2}^#2]_{#3}(#4)}
\def\MSO (#1,#2,#3,#4){[\Omega\frac{#1}{2}^#2]_{#3}(#4)}

\begin{document}
\onecolumn
\title{The light baryon spectrum in a relativistic quark model with
  instanton-induced quark forces}
\subtitle{The non-strange baryon spectrum and ground-states}
\author{Ulrich L\"oring\thanks{e-mail: {\tt loering@itkp.uni-bonn.de}}, Bernard Ch.~Metsch \and  Herbert R.~Petry
}                     
%
%
\institute{Institut f\"ur Theoretische Kernphysik, Universit\"at Bonn, Nu{\ss}allee 14--16, D--53115 Bonn, Germany}
%
\date{}
%
\authorrunning{U.~L\"oring {\it et al.}}  

\abstract{This is the second of a series of three papers treating light baryon
  resonances up to 3 GeV within a relativistically covariant quark model based
  on the three-fermion Bethe-Salpeter equation  with
  instantaneous two- and three-body forces. In this paper we apply the
  covariant Salpeter framework  (which we developed in the first
  paper \cite{Loe01a}) to specific quark model calculations.  Quark confinement is realized
  by a linearly rising three-body string potential with appropriate spinorial
  structures in Dirac-space.  To describe the hyperfine structure of the
  baryon spectrum we adopt 't~Hooft's residual interaction
  based on QCD-instanton effects and demonstrate that the
  alternative one-gluon-exchange is disfavored phenomenological
  grounds. Our fully relativistic framework allows to investigate the effects
  of the full Dirac structures of residual and confinement forces on the
  structure of the mass spectrum. In the present paper we present a detailed
  analysis of the complete non-strange baryon spectrum and show that several
  prominent features of the nucleon spectrum such as {\it e.g.} the Roper resonance
  and approximate ''parity doublets'' can be uniformly explained due to a
  specific interplay of relativistic effects, the confinement potential and
  't~Hooft's force. The results for the spectrum of strange baryons will be
  discussed in a subsequent paper \cite{Loe01c}.
} 
\maketitle
%
\section{Introduction}
\label{intro}
In the previous paper \cite{Loe01a} we analyzed the three-fermion Bethe-Salpeter
equation with instantaneous two- and three-body interaction
kernels. Without being too specific concerning the interaction
kernels, we derived the three-fermion Salpeter equation.  We now want
to apply this covariant formalism to a system of three light quarks
with flavors up, down and strange and thus use this fully relativistic
framework as basis for a quark model of light baryons.  In fact, the
reduced Salpeter equation provides a suitable, fully relativistic
framework, which nonetheless keeps as close as possible to the rather
successful non-relativistic potential models: On the one hand, we
found a one-to-one correspondence of the Salpeter amplitudes with the
ordinary states of the non-relativistic quark model meditated by the
embedding map of non-relativistic three-quark Pauli spinors to full
three-quark Dirac spinors.  On the other hand, this approach adopts
the concept of constituent quarks with an effective mass where the
underlying interactions are described by inter-quark potentials in the
rest-frame of the baryons.  Hence, we basically have the same input
describing the  quark dynamics as in non-relativistic quark
models, in particular the number of parameters remains exactly the
same as in a corresponding non-relativistic approach\footnote{It is
worth to mention here that this is quite in contrast to other attempts
like the so-called ''relativized'' quark models \cite{GoIs85,CaIs86}
which just parameterize relativistic effects and therefore
introduce additional parameters.}.  To be specific, we now have to
fill in the details of the underlying quark interactions, {\it i.e.}
we have to specify the three- and two-body potentials $V^{(3)}$
and $V^{(2)}$, respectively, which we use as instantaneous interaction
kernels. We then solve the resulting Salpeter equation
numerically. The fact that this equation can be cast in Hamiltonian
form allows for the use of a variational principle: We expand our wave
functions in terms of harmonic oscillator functions and diagonalize
the Hamiltonian with respect to a truncated wave function basis
checking carefully numerical stabilities.

One of our aims is to extend a covariant quark model for mesons
\cite{ReMu94,MuRe94,MuRe95,Mu96,MePe96,RiKo00,KoRi00} which is based on the
quark-antiquark Bethe-Salpeter equation with instantaneous two-body
interaction kernels. In this model the interaction between quark and antiquark
included a linearly rising (string-like) confinement potential provided with a
suitable spinorial form (Dirac structure), which was combined with the
effective residual interaction first computed by 't~Hooft from instanton
effects in QCD \cite{tHo76}. In fact it turned out that this relativistic
approach, which employs 't~Hooft's force as residual interaction, provides
significant improvements with respect to other, non-relativistic or
''relativized'' approaches which in general use parts of the residual
one-gluon-exchange in addition to a confining central potential. In particular,
this model allows for a consistent and complete description of the whole
mesonic mass spectrum but at the same time also for the description of
dynamical observables such as form factors, where a fully covariant treatment of the quark dynamics
becomes particularly crucial. The results are rather encouraging to extend
this approach to baryons.  Thus, our main choice for $V^{(2)}$ is 't~Hooft's
instanton-induced quark interaction.  However, in appendix \ref{sec:OGE} we
will also show the result of calculations with a one-gluon-exchange potential,
demonstrating that from a phenomenological point of view it should be discarded.

Similar to the Salpeter model for mesons, quarks in baryons shall be
confined by a linearly rising string potential. Generalizing the
linear quark-antiquark confinement for mesons our quark confinement
for baryons will be produced by a three-quark string potential
$V^{(3)}$ provided with an appropriate Dirac structure. As in the
meson model \cite{ReMu94,MuRe94,RiKo00,KoRi00} we use in addition
't~Hooft's two-quark interaction mentioned before as residual two-body
force $V^{(2)}$. A non-relativistic version with these dynamics has
been applied already by Blask {\it et al.}  \cite{BBHMP90,Bl90,Met93}
for the calculation of baryon (and meson) mass spectra.  We would like
to note that this model could satisfactorily account for the gross
features of the light baryon spectrum with only seven parameters. In
particular, it was able to explain the sign and the rough size of
hyperfine splittings of ground-state baryons as well as the right size
of splittings of negative parity excited baryons. However, a closer
look at the mass spectra reveals that special features, such as {\it
e.g.}  the conspicuously low-lying first scalar/isoscalar excitations
of the octet ground-states (Roper resonances) or the highest members
of Regge trajectories cannot be accounted for in the non-relativistic
framework. In particular, these issues improve in the present
relativistic approach. We will not always comment in detail on
calculations in alternative baryon models; for an excellent review see
ref. \cite{CaRo00}.

In this paper baryon resonances are treated as bound states and no
calculation of widths is performed. This is certainly questionable,
but so far is due to technical limitations. To improve this situation one has
to specify the decay channels and perform at least a perturbative
calculation of the decay widths \cite{Ma55}. Very often this is not
sufficient since final state interactions may also change
(nonperturbatively) the resonance positions appreciably
\cite{An97,An98a,An98b,AnSa97,AnSa96,AnProSa96}. As long as this shift
is approximately uniform, it can be absorbed in the potential
parameters of the quark model.  We are aware of the fact that this is
true in many, but not all cases \cite{An97,An98a,An98b,AnSa97,AnSa96,AnProSa96}.

Our paper is organized as follows: In section \ref{sec:3bodyConf} we
specify the explicit form of the instantaneous three-quark confinement kernel
$V^{(3)}$. We introduce two alternative confinement models (model
${\cal A}$ and model ${\cal B}$) which essentially differ in the choice
of the Dirac structures only. Both versions shall be tested in the subsequent
investigations in a comparison to the experimental mass spectrum.  In section
\ref{sec:eff_tHo_Lag} we introduce 't~Hooft's instanton-induced residual
two-quark interaction and discuss its specific structure. In the subsequent
sections a detailed discussion of our results in comparison to the
phenomenological non-strange baryon (and ground-state) spectrum is given.  We
start with some general comments concerning the parameter dependencies in
section \ref{sec:gen_comments}.  In section \ref{sec:delta} we investigate the
results of our calculations for the $\Delta$-spectrum where 't~Hooft's
interaction gives no contribution. These investigations constitute a first test
of the confinement models considered.  Section \ref{sec:ground_hyper} is
concerned with the hyperfine structure of the ground-state baryons and the
role of the instanton-induced interaction for generating this structure.
Section \ref{sec:Nuc} is devoted to an extensive discussion of the excited
nucleon spectrum. A principal objective of this discussion is to demonstrate
how instanton-induced effects along with the $\Delta-N$ hyperfine splitting
simultaneously generate several prominent structures seen in the experimental
nucleon spectrum such as {\it e.g.} the low position of the Roper resonance or the
occurrence of approximate ''parity doublets''.  Finally we give a summary and a
conclusion in section \ref{sec:concl}.  A detailed discussion of the
corresponding results for the strange baryons will be presented in a
subsequent paper \cite{Loe01c}.

\section{Three-body confinement}
\label{sec:3bodyConf}
It is well known that the global structure of the experimental baryon and
meson mass spectrum suggests a linearly rising, flavor-independent confinement
potential. The appearance of the experimentally observed baryon resonances as
nearly degenerate, alternating even- and odd-parity shells motivates the
picture that the quarks are moving in a local potential which roughly reflects
harmonic forces between the quarks. This picture led to the naive quark
oscillator shell model. Moreover, phenomenological analyses of the
experimental baryon (and also meson) spectra up to highest orbital excitations
show a remarkable empirical connection between the total spin $J$ and the
squared mass $M^2$ of the states, namely that certain states, within a
so-called Chew-Frautschi plot of $M^2$ {\it versus} $J$, lie on linear Regge
trajectories $M^2\sim J$. The interpretation of this empirical feature
motivates a string picture for the confinement mechanism, where the quarks are
connected by gluonic strings (flux tubes) such that the effective confining
potential rises linearly with the string length for large distances of the
quarks.  A further confirmation of this scenario stems from lattice QCD
calculations, which in fact indicate a string-like realization of the
confinement force in the static limit of heavy quarks; for a review of these
issues we refer to \cite{Bal00} and
references therein.\\
We therefore choose a three-body confinement force which in the rest-frame of
the baryon takes the form:
\begin{eqnarray}
V^{(3)}(x_1,x_2,x_3;\; x_1',x_2',x_3')
&=&
V^{(3)}_{\rm conf}({\bf x_1},{\bf x_2},{\bf x_3})\;
\delta^{(1)}(x_1^0 - x_2^0)\;
\delta^{(1)}(x_2^0 - x_3^0)\\
& &
\phantom{V^{(3)}_{\rm conf}({\bf x_1},{\bf x_2},{\bf x_3})\;}
\delta^{(4)}(x_1 - x_1')\;
\delta^{(4)}(x_2 - x_2')\;
\delta^{(4)}(x_3 - x_3').\nonumber
\end{eqnarray}
Here $V^{(3)}_{\rm conf}({\bf x_1},{\bf x_2},{\bf x_3}) = V^{(3)}_{\rm
  conf}(r_{3q}) \sim r_{3q}$ denotes the local three-quark potential whose
radial dependence is assumed to be linearly increasing with some 'collective'
radius $r_{3q} = r_{3q}({\bf x_1},{\bf x_2},{\bf x_3})$ (string picture). (As
will be discussed below such a three-quark distance can be realized in various
ways.)  In contrast to non-relativistic quark models, in our fully
relativistic framework the radial dependence has to be provided with an
appropriate spinorial three-quark Dirac structure ${\cal W}$.  Thus, the
most general ansatz for a flavor-independent confinement potential
$V^{(3)}_{\rm conf}$, which rises linearly with the inter-quark distance
$r_{3q}$, is given by
\begin{equation}
V^{(3)}_{\rm conf}({\bf x_1},{\bf x_2},{\bf x_3})
\;=\;
\left[
a \;{\cal W}_{\rm off}
\;\;+\;\; b\; r_{3q}({\bf x_1},{\bf x_2},{\bf x_3}) \;{\cal W}_{\rm str}\right],  
\end{equation}
where $a$ is an overall constant offset associated with a three-quark
Dirac structure ${\cal W}_{\rm off}$ and $b$ is the slope of
the linearly rising part which is associated with the Dirac structure
${\cal W}_{\rm str}$. In general the spin structures
${\cal W}_{\rm off}$ and ${\cal W}_{\rm str}$ of the constant
and linear part can be chosen differently (see discussion below).

\begin{figure}[!h]
  \begin{center}
    \epsfig{file={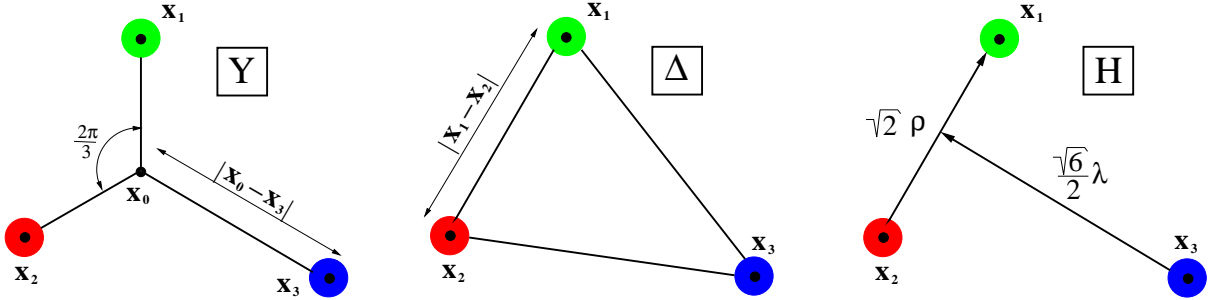},width=160mm}
  \end{center}
\caption{The Y-type (left),  $\Delta$-type (middle) and hyper-spherical (right) string-like
confinement interaction between all three quarks.}
\label{fig:3QConf}
\end{figure}

As illustrated in fig. \ref{fig:3QConf} there are several possibilities to
define the linear inter-quark distance, which we refer to as Y-type,
$\Delta$-type and hyperspherical (H) string. 
The first type was proposed by Carlson
{\it et al.} \cite{CaKoPa83a,CaKoPa83b} and is intimately related to the
color-$SU(3)$ group: In a color singlet $qqq$ system, each quark acts as a
source of one flux tube. For $SU(3)$ gauge fields the flux tubes can merge
at a single point $\bf x_0$, which in the adiabatic approximation is
chosen such that the energy and hence the length $r_{3q} = r_{Y}$ of the connecting
path is minimized to
\begin{equation}
\label{conf_Y}
r_{3q}({\bf x_1},{\bf x_2},{\bf x_3}) 
= r_{Y} 
:= \min_{\bf x_0}\;\sum_{i=1}^3 |{\bf x_i} - {\bf x_0}|.
\end{equation}
Then the straight lines emanating from the quarks will meet at an angle
$2\pi/3$ at the central junction point $\bf x_0$ (see fig. \ref{fig:3QConf}), unless
one of the angles within the baryonic triangle exceeds the value $2\pi/3$,
in which case a linear geometry is preferred. 
An alternative three-quark distance $r_{3q} =
r_{\Delta}$ is defined by the $\Delta$-string configuration, in which case the string
is formed by the sum of two-body strings between each quark pair:
\begin{equation}
\label{conf_Del}
r_{3q}({\bf x_1},{\bf x_2},{\bf x_3}) 
= 
r_{\Delta} := \sum_{i<j} |{\bf x_i} - {\bf x_j}|.
\end{equation}
Rescaling this quark distance with a factor $f \simeq 0.5493$
\cite{CaIs86,DoMu76,Bl90} its length $f\cdot r_{\Delta}$ constitutes a
fairly good approximation to the $Y$-type string potential 
\begin{equation}
\label{b_rescale}
r_Y \approx f\cdot r_\Delta, \qquad\textrm{where}\;\; f = 0.5493\qquad \left(\frac{1}{2} < f < \frac{1}{\sqrt{3}}\right).
\end{equation}
Obviously, whenever the three quarks are collinear, the factor is
exactly 1/2, while for the other extreme case of an equilateral
triangle the factor is $1/\sqrt{3}$.  The choice $f = 0.5493$ lies in
between these two extremes. This particular ratio is chosen such that
it minimizes the size of the expectation value of $|r_Y -f\;r_\Delta|$
in the harmonic-oscillator basis \cite{CaIs86,Bl90} and moreover is
also favored by some investigations of flux tubes on the lattice
\cite{DoMu76}. Only a few lattice results on static baryonic
potentials (see \cite{Bal00} and references therein) exist so far
with statistical errors too large to rule out either possibility. In a
more recent study (see \cite{Bal00}), however, clear evidence in
support of the $\Delta$-configuration has been found. A further
example is a hyperspherical ansatz \cite{FGPST95,SIG97}, which
increases linearly with the hyperradius
\begin{equation}
\label{conf_hyp}
r_{3q}({\bf x_1},{\bf x_2},{\bf x_3}) = r_{\rm hyp} := \sqrt{|{\bfgrk \rho}|^2+|{\bfgrk \lambda}|^2},\quad
\textrm{where}\quad
\begin{array}{rcl}
{\bfgrk \rho} &:=& \frac{1}{\sqrt{2}}({\bf x_1} - {\bf x_2}),\\[1mm]
{\bfgrk \lambda}&:=&\frac{1}{\sqrt{6}}({\bf x_1} + {\bf x_2}- 2 {\bf x_3}).
\end{array}
\end{equation}

We would like to note at this stage that we have tested the various radial
dependencies (\ref{conf_Y}), (\ref{conf_Del}) and (\ref{conf_hyp}) in our 
Salpeter model. Our investigations, however, clearly showed that the
structure of resulting spectra depends only slightly on the various radial
dependencies chosen. It turned out that the slope parameter $b$ can
always be appropriately rescaled (as {\it e.g.} in eq. (\ref{b_rescale}) with the factor
f) to obtain almost the same spectrum for all three choices. We
therefore prefer for our model the $\Delta$-shape string potential 
rising linearly with
$r_\Delta({\bf x_1},{\bf x_2},{\bf x_3}) = \sum_{i<j} |{\bf x_i} - {\bf x_j}|$
which, on the one hand, is favored by the most recent lattice studies anyway
and, on the other hand, is also much easier to handle numerically. We found,
however, that the structure of the resulting spectra depends much more
on the Dirac structure chosen, which we shall consider next.

The gross features of the baryon resonances seem to indicate that the
dominating confinement forces should be spin-independent, at least in the
non-relativistic limit. This property can be realized exactly if and only if
${\cal W}_{\rm off}$ and ${\cal W}_{\rm str}$ have the form
\begin{eqnarray}
{\cal W}_{\rm str} &=& \alpha_{\rm str}\Id\tens\Id\tens\Id + \beta_{\rm str}
\left( \gamma^0\tens\gamma^0\tens\Id + {\rm cycl. perm.}\right)\\
{\cal W}_{\rm off} &=& \alpha_{\rm off}\Id\tens\Id\tens\Id + \beta_{\rm off}
\left( \gamma^0\tens\gamma^0\tens\Id + {\rm cycl. perm.}\right)
\end{eqnarray}
(The appearance of $\gamma^0\tens\gamma^0$ is allowed
because the potential is defined in the rest-frame of the baryon!).
In this paper we show results only for the following model choices:
\begin{itemize}
\item
\textbf{\underline{Model ${\cal A}$}}
\begin{eqnarray}
V^{(3)}_{\rm conf}({\bf x_1}, {\bf x_2}, {\bf x_3})
&=& 
3\;a\;\;\frac{1}{4}\left[
\Id\tens\Id\tens\Id 
+
\gamma^0\tens\gamma^0\tens\Id
+ \textrm{cycl. perm.}
\right]\\
&&+\;\;
b\sum_{i<j} |{\bf x_i}-{\bf x_j}|\;\;
\frac{1}{2}\left[
-\Id\tens\Id\tens\Id 
+
\gamma^0\tens\gamma^0\tens\Id
+ \textrm{cycl. perm.}
\right],\nonumber
\end{eqnarray}
\item
\textbf{\underline{Model ${\cal B}$}}
\begin{eqnarray}
V^{(3)}_{\rm conf}({\bf x_1}, {\bf x_2}, {\bf x_3})
&=& 
\left[3a + b\sum_{i<j} |{\bf x_i}-{\bf x_j}| \right]\;\;
\frac{1}{4}
\left[
\Id\tens\Id\tens\Id 
+
\gamma^0\tens\gamma^0\tens\Id
+ \textrm{cycl. perm.}
\right].
\end{eqnarray}
\end{itemize}
which have in common that they give the same nonrelativistic limit
and that they produce equivalent results in flavor-symmetric states.
The offset constant $a$ and the slope $b$ enter as free parameters in
our model.
\section{'t~Hooft's instanton induced interaction}
\label{sec:eff_tHo_Lag}
As originally shown by 't~Hooft \cite{tHo76} instantons lead to an effective
contribution to the interaction of quarks. The corresponding effective
Lagrangian $\Delta{\cal L}_\textrm{\footnotesize eff}$ which has been
calculated by 't~Hooft for the case of the $SU(2)$ gauge group has been
generalized to the case of the $SU(3)$ gauge group by Shifman, Vainshtein and
Zakharov \cite{SVZ80}.  They have shown that the contribution of a single
instanton and antiinstanton configuration to the effective quark Lagrangian in
the case of the $SU(3)$ color group and three light quarks with flavors up
($u$), down ($d$) and strange ($s$) is given by \cite{SVZ80}
\begin{eqnarray}
\label{effL_SVZ}
\lefteqn{
\Delta{\cal L}_{\rm eff}(y) = \int_0^{\rho_c} \textrm{d}\rho\; \frac{d_0(\rho)}{\rho^5}
\Bigg\{}&&\nn
&&\phantom{+}
\left(m^0_u\rho -\frac{4}{3}\pi^2 \rho^3\;({\overline\Psi_u}_R{\Psi_u}_L)\right)
\left(m^0_d\rho -\frac{4}{3}\pi^2 \rho^3\;({\overline\Psi_d}_R{\Psi_d}_L)\right)
\left(m^0_s\rho -\frac{4}{3}\pi^2 \rho^3\;({\overline\Psi_s}_R{\Psi_s}_L)\right)\nn                                          
&&+ \frac{3}{32} \left(\frac{4}{3}\pi^2 \rho^3\right)^2
\Bigg[\left(({\overline\Psi_u}_R\lambda^a{\Psi_u}_L)({\overline\Psi_d}_R\lambda^a{\Psi_d}_L) 
-\frac{3}{4}({\overline\Psi_u}_R\sigma_{\mu\nu}\lambda^a{\Psi_u}_L)({\overline\Psi_d}_R\sigma^{\mu\nu}\lambda^a{\Psi_d}_L)\right)\nn
&&\phantom{+ \frac{3}{32} \left(\frac{4}{3}\pi^2\rho^3\right)^2a }
\times\left(m^0_s\rho - \frac{4}{3}\pi^2 \rho^3 \;({\overline\Psi_s}_R{\Psi_s}_L)\right)\nn
&&\phantom{+ \frac{3}{32} \left(\frac{4}{3}\pi^2\rho^3\right)^2a }
+ \frac{9}{40} \left(\frac{4}{3}\pi^2 \rho^3\right)
\;d^{abc}\;
({\overline\Psi_u}_R\sigma_{\mu\nu}\lambda^a{\Psi_u}_L)
({\overline\Psi_d}_R\sigma^{\mu\nu}\lambda^b{\Psi_d}_L)
({\overline\Psi_s}_R\lambda^c{\Psi_s}_L)\nn
&&\phantom{+ \frac{3}{32} \left(\frac{4}{3}\pi^2\rho^3\right)^2a }
+ \textrm{cycl.~perm.~of } (uds)\Bigg]\nn
&&+ \frac{9}{320} \left(\frac{4}{3}\pi^2 \rho^3\right)^3
\;d^{abc}\;
({\overline\Psi_u}_R\lambda^a{\Psi_u}_L)
({\overline\Psi_d}_R\lambda^b{\Psi_d}_L)
({\overline\Psi_s}_R\lambda^c{\Psi_s}_L)\nn
&&+ \frac{9}{256}\; \textrm{i} \left(\frac{4}{3}\pi^2 \rho^3\right)^3
\;f^{abc}\;
({\overline\Psi_u}_R\sigma_\mu^\nu \lambda^a{\Psi_u}_L)
({\overline\Psi_d}_R\sigma_\nu^\gamma \lambda^b{\Psi_d}_L)
({\overline\Psi_s}_R\sigma_\gamma^\mu \lambda^c{\Psi_s}_L) \;\;+\;\; (L \leftrightarrow R) \Bigg\}
\end{eqnarray}
Here $m^0_f$ are the current quark masses for the various light flavor degrees
of freedom $f = u,d$ and $s$. $\lambda^a$, ($a=1, \ldots, 8$) denote the
standard $SU(3)$ color Gell-Mann matrices and $f^{abc}$, $d^{abc}$ are
the $SU(3)$ structure constants defined by the commutators and
anticommutators
\begin{equation}
[\lambda^a,\lambda^b] = \textrm{i} 2 f^{abc} \lambda^c,\qquad
\{\lambda^a,\lambda^b\} = \frac{4}{3} \delta^{ab} \Id + 2 d^{abc} \lambda^c.
\end{equation}
${\Psi_f}_{L,R}\equiv{\Psi_f}_{L,R}(y)$  are the projections on left and right handed components of the quark field operators:
\begin{equation}
\label{LR_Psi}
{\Psi_f}_L := \frac{\Id + \gamma^5}{2}\;\Psi_f \qquad{\Psi_f}_R := \frac{\Id - \gamma^5}{2}\;\Psi_f.
\end{equation}
An important quantity entering eq. (\ref{effL_SVZ}) as a generic weight
is the reduced instanton density $D(\rho)=d_0(\rho)/\rho^5$ which describes
the vacuum-to-vacuum tunneling probability in the presence of an instanton
with size $\rho$. In the case of three colors and three flavors $d_0(\rho)$ is
given by \cite{Shu82}
\begin{equation}
\label{instdens}
d_0(\rho) = (3.63\; 10^{-3})\;\left(\frac{8\pi^2}{g^2(\rho)}\right)^{6}
\exp\left(-\frac{8\pi^2}{g^2(\rho)}\right) 
\end{equation}
The function $g(\rho)$ denotes the $\rho$-dependent running coupling constant,
which in two-loop accuracy reads
\begin{equation}
\label{thooftcoupling}
\frac{8\pi^2}{g^2(\rho)} 
=
9 \ln\left(\frac{1}{\rho\;\Lambda_{\rm QCD}}\right)
+ \frac{32}{9} \ln\left[\ln\left(\frac{1}{\rho\;\Lambda_{\rm QCD}} \right)\right],
\end{equation}
where $\Lambda_{\rm QCD}$ is the QCD scale parameter. Unfortunately, using the
semi-classical 't~Hooft formula (\ref{instdens}), large size instantons make
the integration diverge due to the power law behavior of the reduced instanton
size distribution $D(\rho)=d_0(\rho)/\rho^5$. This known as the so-called
'\textit{infrared problem}'.  To cure this problem, we cut the $\rho$ integral
for the moment by hand \cite{CDG78}, introducing a critical instanton size
$\rho_c$ for which the $\ln\ln$-term of (\ref{thooftcoupling}) is still
reasonably small compared to the $\ln$-term.\\

The interaction described before is chirally invariant if the current quark
masses are strictly zero. As is well known it is no longer $U_A(1)$-invariant
and thus exhibits the characteristic breaking of this symmetry in QCD. In
addition we must take the spontaneous breaking of chiral invariance into
account which is most easily done by normal ordering $\Delta{\cal L}_{\rm
  eff}$ with respect to a vacuum of massive quarks with constituent quark
masses. The result is
\begin{equation}
\Delta{\cal L}_{\rm eff} 
= 
\epsilon
+
\Delta{\cal L}_{\rm eff}^{(1)} 
+
\Delta{\cal L}_{\rm eff}^{(2)}
+
\Delta{\cal L}_{\rm eff}^{(3)}.
\end{equation}
\begin{enumerate}
\item
Here $\epsilon$ is a constant (a shift in the vacuum energy).
\item
$\Delta{\cal L}_{\rm eff}^{(1)}$  is a mass term: 
\begin{equation}
\Delta{\cal L}_{\rm eff}{(1)} 
= 
-\; \Delta m_n\; \left(\; :\overline\Psi_u \Psi_u: + :\overline\Psi_d \Psi_d:\; \right)
-\; \Delta m_s\; :\overline\Psi_s \Psi_s:,
\end{equation}
with
\begin{eqnarray}
\label{GapEqu_rep}
\Delta m_n &:=&  \int_0^{\rho_c} \textrm{d}\rho\; \frac{d_0(\rho)}{\rho^5}\;\;
\frac{4}{3}\pi^2 \rho^3 \;
\Big(m^0_n\rho - \frac{2}{3} \pi^2 \rho^3\; \Expect{\overline\Psi_n \Psi_n}\Big)\;
\Big(m^0_s\rho - \frac{2}{3} \pi^2 \rho^3\; \Expect{\overline\Psi_s \Psi_s} \Big),\\
\Delta m_s &:=& \int_0^{\rho_c} \textrm{d}\rho\; \frac{d_0(\rho)}{\rho^5}\;\;
\frac{4}{3}\pi^2 \rho^3 \;
\Big(m^0_n\rho - \frac{2}{3} \pi^2 \rho^3\; \Expect{\overline\Psi_n \Psi_n} \Big)^2,
\end{eqnarray}
where $m_n:=m_n^0+\Delta m_n$ and $m_s:=m_s^0+\Delta m_s$ are
naturally identified with the non-strange and strange constituent
quark masses (if possible contributions from the confining forces are
excluded). In the following we will treat the effective constituent quark
masses $m_n$ and $m_s$ as free parameters which we fit to the experimental
baryon spectrum.
\item
$\Delta{\cal L}_{\rm eff}^{(2)}$ is a two-body
interaction which can be cast in the more convenient form
\begin{eqnarray}
\label{L2tHooftFierzSymProj}
\Delta{\cal L}_{\rm eff}^{(2)} &=&
- 2
:\overline\Psi \tens  \overline\Psi
\left[
\left(\;\Id \tens \Id + \gamma^5 \tens \gamma^5\;\right)
{\cal P}_{S=0}^{\cal D}
\tens
\left(g_{nn}\;{\cal P}_{\cal A}^{\cal F}(nn)
+
g_{ns}\;{\cal P}_{\cal A}^{\cal F}(ns)\right)
\tens
{\cal P}_{\bf \bar 3}^{\cal C}
\right]
\Psi\tens \Psi:\nn[1mm]
&&
-\phantom{2}
:\overline\Psi \tens  \overline\Psi
\left[
\left(\;\Id \tens \Id + \gamma^5 \tens \gamma^5\;\right)
{\cal P}_{S=1}^{\cal D}
\tens
\left(g_{nn}\;{\cal P}_{\cal A}^{\cal F}(nn)
+
g_{ns}\;{\cal P}_{\cal A}^{\cal F}(ns)\right)
\tens
{\cal P}_{\bf 6}^{\cal C}
\right]
\Psi\tens \Psi:\;.
\end{eqnarray}
where the effective coupling constants $g_{nn}:=3/8\;g_{\rm eff}(s)$
and $g_{ns}:=3/8\;g_{\rm eff}(n)$ are given by
\begin{eqnarray}
\label{geff_rep}
g_{\rm eff} (f) &:=& \int_0^{\rho_c} \textrm{d}\rho\;
\frac{d_0(\rho)}{\rho^5}\;\; \left( \frac{4}{3}\pi^2 \rho^3 \right)^2
\; \Big(m^0_f\rho - \frac{2}{3} \pi^2 \rho^3\; \Expect{\overline\Psi_f
\Psi_f} \Big).
\end{eqnarray}
This expression in terms of two-quark projection operators in Dirac, flavor
and color spaces allows an immediate identification of the diquark channels
that in fact are affected by the interaction: In color space ${\cal P}_{\bf
  \bar 3}^{\cal C}$ and ${\cal P}_{\bf 6}^{\cal C}$ denote the
projectors onto color antitriplet and color sextet pairs.  Since the three
quarks within the baryon have to constitute a color singlet two quarks always
have to be in a color antitriplet state. For this reason the second term of
the Lagrangian (\ref{L2tHooftFierzSymProj}), which affects color-sextet quark
pairs only, does not contribute in lowest order to the baryon dynamics.  In
Dirac space, ${\cal P}_{S=0}^{\cal D}$ and
${\cal P}_{S=1}^{\cal D}$ are the projectors onto antisymmetric
spin-singlet and symmetric spin-triplet states, respectively, defined by
\begin{equation}
{\cal P}_{S=0}^{\cal D}
\;:=\;
\frac{1}{4}\; \Id\tens\Id - \frac{1}{4}\;
{\bfgrk\Sigma}\cdot\tens\;{\bfgrk\Sigma}
\qquad{\rm and}\qquad
{\cal P}_{S=1}^{\cal D}
\;:=\;
\frac{3}{4}\; \Id\tens\Id + \frac{1}{4}\; {\bfgrk\Sigma}\cdot\tens\;{\bfgrk\Sigma},
\end{equation}
with
${\bfgrk\Sigma}\cdot\tens\;{\bfgrk\Sigma}:=\sum_{i=1}^3\Sigma^i\tens\Sigma^i$
and $\Sigma^i= {\rm diag}(\sigma^i,\sigma^i)$, where $\sigma^i$ are the usual Pauli
matrices.
In flavor space the operators ${\cal P}_{\cal A}^{\cal F}(nn)$ and
  ${\cal P}_{\cal A}^{\cal F}(ns)$ denote the projectors onto
flavor-antisymmetric quarks which either are non-strange ($nn$), {\it i.e.} with isospin
zero, or non-strange-strange ($ns$):
\begin{equation}
\begin{array}{rcl}
{\cal P}_{\cal A}^{\cal F}(nn) 
&:=& 
\left(
{\cal P}_n^{\cal F} \tens {\cal P}_n^{\cal F}
\right)\;{\cal P}_{\cal A}^{\cal F}
\\[2mm]
{\cal P}_{\cal A}^{\cal F}(ns) 
&:=& 
\left(
{\cal P}_n^{\cal F} \tens {\cal P}_s ^{\cal F}
+
{\cal P}_s^{\cal F} \tens {\cal P}_n ^{\cal F}
\right)\;{\cal P}_{\cal A}^{\cal F}
\end{array}
\end{equation}
Here ${\cal P}_n^{\cal F} = \ket{u}\bra{u} + \ket{d}\bra{d}$ and
${\cal P}_s^{\cal F} = \ket{s}\bra{s}$ are the projectors for single
quark flavors and
${\cal P}_{\cal A}^{\cal F}=\frac{1}{2}(\Id^{\cal F}-{\cal P}^{\cal F}_{12})$
is the antisymmetrizer in the two-particle flavor space. Accordingly,
't~Hooft's force acts exclusively on quark pairs which are antisymmetric in
flavor and it distinguishes between diquarks of different flavor content ($nn$)
and ($n$s) by means of the different effective couplings $g_{nn}$ and $g_{ns}$.
\item
$\Delta{\cal L}_{\rm eff}^{(3)}$ is just the original interaction
but now with constituent quark field operators. It yields interaction vertices
whose contribution within the Bethe-Salpeter equation vanishes for
color-free three-quark states, and hence needs no further discussion.
\end{enumerate}
The most interesting contribution stems from the two-body
interaction Lagrangian. It is usually used in connection with the solution
of the $U_A(1)$-problem.  In the framework of calculations for mesons \cite{MuRe94,KoRi00}
we have indeed shown that it yields the correct splitting of the
lowest meson nonet and in general of all low-lying meson states. Hence
we have good reasons to use it also for the calculation of baryon
masses. As in the case of mesons we have to regularize the spatial
$\delta$-function dependence of this two-body Lagrangian by a form
factor, which we give a Gaussian form:
\begin{equation}
\delta^{(4)}(x) \longrightarrow 
\frac{1}{\lambda^3\pi^\frac{3}{2}}\;e^{-\frac{|{\bf x}|^2}{\lambda^2}}\;\;\delta^{(1)}(x^0).
\end{equation}
For the moment we make no effort to derive it and treat the effective range
$\lambda$ as well as the couplings $g_{nn}$ and $g_{ns}$ as phenomenological
parameters which we fit to the experimental baryon spectrum. Our candidate for
the instantaneous two-quark interaction kernel $V^{(2)}$ has then the
following form:
\begin{equation}
V^{(2)}(x_1,x_2;\;x_1',x_2')
=
V^{(2)}_{\rm 't~Hooft}({\bf x})\;
\delta^{(1)}(x^0)\; \delta^{(4)}(x_1-x_1')\;\delta^{(4)}(x_2-x_2'),
\end{equation}
with $x:=x_1-x_2$ and 
\begin{equation}
\label{tHooftPot}
V^{(2)}_{\rm 't~Hooft}({\bf x})
=
\frac{-4}{\lambda^3\pi^\frac{3}{2}}\;e^{-\frac{|{\bf x}|^2}{\lambda^2}}
 \left[\Id\tens\Id+\gamma^5\tens\gamma^5\right]{\cal P}^{\cal D}_{S_{12}=0}
\tens 
\left(
g_{nn}{\cal P}^{\cal F}_{\cal A}(nn)
+
g_{ns}{\cal P}^{\cal F}_{\cal A}(ns)
\right)\tens
{\cal P}_{\bar3}^{\cal C}.
\end{equation}
\section{Parameters and general comments}
\label{sec:gen_comments}
In this (and a consecutive \cite{Loe01c}) paper we will present a detailed discussion of the resulting mass
spectra of light baryons (with the light quark flavors 'up', 'down' and
'strange') calculated within our covariant Salpeter framework using the
confinement models ${\cal A}$ and ${\cal B}$ and employing 't~Hooft's
instanton-induced force as residual interaction.

The seven free parameters entering in the present calculations are listed in table
\ref{tab:ModelParam}. These parameters involve the effective non-strange  and strange constituent quark masses
$m_n$ and $m_s$, and the confinement parameters $a$ and $b$ describing the off-set
and the string tension of the three-body confinement potential. 't~Hooft's
instanton-induced interaction is determined by the couplings $g_{nn}$ and
$g_{ns}$ and the effective range $\lambda$.
\begin{table}[h]
\center
\begin{tabular}{ccccc}
\hline
             &             & Parameter & Model ${\cal A}$ & Model ${\cal B}$\\
\hline
Constituent  & non-strange     & $m_n$     & $330$ MeV          & $300$ MeV\\
quark masses & strange         & $m_s$     & $670$ MeV          & $620$ MeV\\
\\
Confinement  & offset          & $a$       & $-744$ MeV         & $-1086$ MeV\\
parameters   & slope           & $b$       & $470$ MeV fm$^{-1}$&  $1193$ MeV fm$^{-1}$\\
\\
't~Hooft     & effective range & $\lambda$ & $0.4$ fm           & $0.4$ fm\\
interaction  & $nn$-coupling     & $g_{nn}$  & $136.0$ MeV fm$^3$ & $89.6$ MeV fm$^3$\\                  
             & $ns$-coupling     & $g_{ns}$  &  $94.0$ MeV fm$^3$ & $61.7$ MeV fm$^3$\\
\hline
\end{tabular}
\caption{The parameters of the confinement force, the 't~Hooft interaction and the constituent
quark masses fitted in the models ${\cal A}$ and ${\cal B}$.}
\label{tab:ModelParam}
\end{table}

Our detailed discussion of the results for the light flavor baryons will be
organized according to the specific parameter dependence  dictated by the
simple characteristic and selective action of the instanton-induced
interaction. Due to the simplicity of 't~Hooft's force and the
induced strong selection rules, a clear cut identification of the various
effects of the long range confining potential and the short range residual
interaction can be made by comparison with the experimentally observed
baryon spectra. Accordingly, the discussion of our results in the subsequent
sections will be organized as follows:\\

As 't~Hooft's force acts only on flavor-antisymmetric quark pairs, this
residual interaction does not contribute to the non-strange spectrum of
$\Delta$-resonances due to their common totally symmetric flavor wave function
with isospin $T=3/2$.  Therefore, in our approach (without any
additional residual interactions apart from 't~Hooft's force) the whole
$\Delta$-resonance\footnote{Of course the same also applies to the whole
  $\Omega$-spectrum. Unfortunately the only well-established $\Omega$-state 
  still is just the decuplet ground-state $\Omega\frac{3}{2}^+(1672)$} spectrum
is determined by the confining three-body force and the relativistic dynamics
alone!  This feature thus provides a first test of our
phenomenological ansatz for the confinement potential.  Furthermore, we can
fix the confinement parameters, {\it i.e.}  the off-set parameter $a$ and the
slope $b$, as well as the non-strange quark mass $m_n$ by the positions of the
experimentally best established $\Delta$-resonances without any influence of
the residual interaction. In this respect, the confining three-body
interaction kernel, which is a string-like, {\it i.e.} linearly rising three-body
potential with the two distinct Dirac structures of the models
${\cal A}$ and ${\cal B}$, has to account for the correct description of
the positive parity $\Delta$-Regge trajectory $M^2 \sim J$ up to the
highest spin known $J=15/2^+$. At the same time, it
should not induce too large spin-orbit effects, for which there is hardly any
evidence in the experimental baryon spectrum. The results for the complete
$\Delta$-spectrum in both model versions will be discussed in sect. \ref{sec:delta}.\\
 
Like the $\Delta$-resonances with the corresponding ground-state
$\Delta(1232)$, also the remaining strange spin-$3/2^+$ decuplet
ground-states, {\it i.e.} the hyperons $\Sigma^*(1385)$, $\Xi^*(1583)$ and
$\Omega(1672)$, are unaffected by 't~Hooft's interaction. Hence, the position
of these states then determines the strange constituent quark mass $m_s$.  In this
way, all parameters except for those of the 't~Hooft interaction are fixed,
and in the second step we then can analyze the effect of the instanton-induced
interaction on the structure of the remaining baryon spectra.

As already known from non-relativistic potential models the long range
confinement potential cannot describe the splittings of the spin-$1/2$ octet
and spin-$3/2$ decuplet ground-states. Thus, the first feature the 't~Hooft
interaction has to account for is the hyperfine structure of the ground
states, {\it i.e.}  the mass splittings $\Delta(1232)-N(939)$ of the
non-strange ground-states and the hyperon splittings
$\Sigma^*(1385)-\Sigma(1193)$, $\Xi^*(1583)-\Xi(1318)$ and
$\Sigma(1193)-\Lambda(1116)$.  In this respect we will show that the
instanton-induced interaction indeed leads to an at least as satisfactory
description as the short-range spin-spin part of the Fermi-Breit interaction,
which is generated by the one-gluon exchange (OGE) and is commonly used in
non-relativistic (or 'relativized') quark models \cite{IsKa78,IsKa79,CaIs86}.
In the non-strange sector the $\Delta-N$ mass splitting is explained by
't~Hooft's force being attractive for the nucleon, where two (non-strange)
quarks can be in a state with trivial spin and isospin.  Thus the $\Delta-N$
splitting fixes the coupling $g_{nn}$.  In a similar manner 't~Hooft's force
is attractive for $\Sigma$ and $\Xi$ with the strength $g_{ns}$ where a
non-strange-strange ($ns$) quark pair can be in a state with trivial spin and
antisymmetric flavor. In this way the splittings $\Sigma^*-\Sigma$ and
$\Xi^*-\Xi$ determine the coupling $g_{ns}$.  At the same time, the position
of $\Lambda(1116)$, which contains both types ($nn$ and $ns$) of quark pairs and
thus is influenced by both couplings $g_{nn}$ and $g_{ns}$, should then be
properly described in order to get the right experimentally observed
$\Sigma-\Lambda$ mass difference.  Generating the hyperfine structure of
ground-state baryons by 't~Hooft's
force will be investigated in detail in sect. \ref{sec:ground_hyper}.\\

All the parameters being fixed, the calculation of all other baryon masses is
then parameter-free, {\it i.e.} the majority of excited mass spectra of $N$-resonances,
$\Lambda$-resonances, $\Sigma$-resonances, $\Xi$-resonances and the $\Omega$
baryons are thus predictions of our models ${\cal A}$ and ${\cal B}$. In this
paper we restrict our extensive discussion to the predictions for the nucleon
sector (sect. \ref{sec:Nuc}).  The parameter-free predictions for the
complete spectrum of excited strange baryons will be discussed in a subsequent
paper \cite{Loe01c}.

A principal objective of our discussion is to demonstrate that the
instanton-induced interaction along with a fully relativistic treatment of the
quark dynamics within our covariant Salpeter approach plays an important role
in the description of the complete spectrum of light baryons, {\it i.e.}  not
only for the hyperfine structure of the ground-state baryons as {\it e.g.}
the $\Delta-N$-splitting, but especially for distinctive features of the
excited spectra. In both confinement models we will explore in detail how and
to what extent the striking features of the excited spectra can really be
understood by the effect of 't~Hooft's force with its parameters $g_{nn}$,
$g_{ns}$ and $\lambda$ fixed from the hyperfine structure of the octet and
decuplet ground state baryons.  Prominent features of the excited baryon
spectra that shall be discussed in this context are for instance the
conspicuously low positions of the first isoscalar/scalar excitations of octet
ground-states, {\it i.e.} the Roper-resonance $N\frac{1}{2}^+(1440)$ and its
strange counter parts $\Lambda\frac{1}{2}^+(1600)$ and
$\Sigma\frac{1}{2}^+(1660)$, or the appearance of approximately degenerate
states of the same spin but opposite parity (approximate ''parity doublets'').\\

Now let us turn to the discussion of the light baryon spectra. We
begin with the discussion of the $\Delta$-spectrum, which
phenomenology reflects the role of the three-body confinement kernel
in the Salpeter equation.
\section{The $\Delta$-spectrum -- A first test of the confinement kernel}
\label{sec:delta}
\subsection{Introductory remarks}
In this subsection we investigate and discuss the results of both
confinement models ${\cal A}$ and ${\cal B}$ in the sector of
$\Delta$-baryons (with isospin $T=\frac{3}{2}$ and strangeness
$S^*=0$). We compare our results with the main features of the
complete, hitherto measured $\Delta$-resonance spectrum up to the
highest orbital excitations with total spin $J=\frac{15}{2}$.  As
already mentioned, there is no effect of 't~Hooft's force in this
sector, because the flavor part of the amplitudes is given by the
totally symmetric isospin-$3/2$ flavor function. Therefore, the
internal quark dynamics of the $\Delta$-baryons is described by the
relativistic kinetic energy and the three-body confinement potential
only.  This offers the possibility to study the role of the
confinement force only, {\it i.e.}  without the influence of the
residual instanton-induced interaction.  Therefore, the calculation of
the $\Delta$-spectrum constitutes a first important test of the
different three-body confinement kernels ${\cal A}$ and
${\cal B}$. As mentioned before the phenomenology of Regge
trajectories motivates the string picture for confinement, {\it i.e.}
in our case a three-body potential, which rises linearly with the
distance between the quarks.  In contrast to the radial (string-like)
dependence of the confinement mechanism originating {\it e.g.} from
the flux tube model \cite{CaKoPa83a,CaKoPa83b}, but also motivated by
recent lattice QCD calculations \cite{Bal00}, the spinorial (Dirac)
structure of three-quark confinement lacks any clue from first
principles or lattice QCD.  Thus, an appropriate ansatz for the Dirac
structure of the confinement kernel is motivated by phenomenological
arguments only. Here the aim is to reproduce the global structure of
the $\Delta$-spectrum, especially the well-established positive parity
$\Delta$-Regge trajectory, with the empirical characteristic
$M^2\propto J$. In particular, the slope of the trajectory should be
quantitatively correctly described in order to reproduce the resonance
positions even of the highest orbital excitations. At the same time
the confinement force has to account for the correct energy gaps
between the approximately degenerate bands, which are formed by the
radially excited states with different total angular momenta $J$
(corresponding to the even-parity $2\hbar\omega$, $4\hbar\omega$,
$6\hbar\omega$ and odd-parity $1\hbar\omega$, $3\hbar\omega$,
$5\hbar\omega$ 'oscillator shells' of the naive quark oscillator shell
model\footnote{Due to the one-to-one correspondence of the Salpeter
amplitudes with the states of the non-relativistic quark model via the
embedding mapping of ordinary non-relativistic Pauli wave functions
(see sect. \ref{NconfigPauli}), we will use throughout this work the
notation that we assign a model state to that $N\hbar\omega$
oscillator shell at which this state would first appear by counting
the states in the non-relativistic oscillator spectrum.  Note
however that this is really just a convenient notation, {\it i.e.} it
does not imply that the wave function of this state is restricted to
the configuration subspace of this assigned $N\hbar\omega$ shell.
Expanding this state in a (finite $N < N_{\rm max}$) basis of
oscillator states with given flavor, spin and parity, the
states of all shells in general mix. On the other hand
it is very likely that the dominant part of the corresponding wave
function is made up of $N\hbar\omega$ states.}).  Moreover, the
observed approximate degeneracy in these band structures indicates
that spin-orbit effects are rather small.  This is an additional
experimental feature that we aim to describe in our
models. Consequently, the Dirac structures chosen should be such that
spin-orbit forces are not too large and compatible with the
experimentally observed small intra-band splittings of the different
shells.  In this respect, the spin structures of model ${\cal A}$
and ${\cal B}$, which in both cases consist of a special
combination of a scalar ($\Id\tens\Id\tens\Id$) and a time-like vector
part ($\gamma^0\tens\gamma^0\tens\Id+\ldots$), have already been
selected from a wider class of possible Dirac structures that are all
compatible with Lorentz covariance, time-reversal-, parity- and
${\cal CPT}$-invariance, as well as the hermiticity requirements of
the Salpeter equation. Here we want to verify, that both confinement
versions can indeed account for the phenomenological features of the
$\Delta$-spectrum.  Then, (due to the assumed flavor independence of
the confinement force) these should be also appropriate for the other
baryon sectors, where in addition instanton effects become
substantial.  The phenomenology of the $\Delta$-spectrum already fixes
the offset parameter $a$ and the slope $b$ of the confinement
potential, as well as the non-strange quark mass $m_n$.  This is
mainly done by fitting the positive-parity $\Delta$-Regge trajectory,
{\it i.e.} the ground-state $\Delta{\frac{3}{2}^+}(1232,\mbox{****})$,
and its orbital excitations $\Delta{\frac{7}{2}^+}(1950,\mbox{****})$
and $\Delta{\frac{11}{2}^+}(2420,\mbox{****})$, which all are
well-established four-star resonances with moderate uncertainties in
the observed resonance positions. There is evidence of a fourth member
of this trajectory, namely the next spin-$15/2$ resonance
$\Delta{\frac{15}{2}^+}(2950,\mbox{**})$, which, however, is only a
two-star resonance with quite large experimental
uncertainties. Therefore, this member is not used in the fit and thus
will be a prediction. In order to achieve the correct gap between the
positive-parity ground state $\Delta{\frac{3}{2}^+}(1232,\mbox{****})$
and the first negative parity (1 $\hbar\omega$) band, we additionally
consider also the lowest two, well-established four-star resonances of
negative parity, {\it i.e.}  $\Delta{\frac{1}{2}^-}(1620,\mbox{****})$
and $\Delta{\frac{3}{2}^-}(1700,\mbox{****})$, for fixing the three
parameters $a$, $b$ and $m_n$.
\subsection{The positive-parity $\Delta$-Regge trajectory}
\begin{figure}[h]
  \begin{center}
    \input{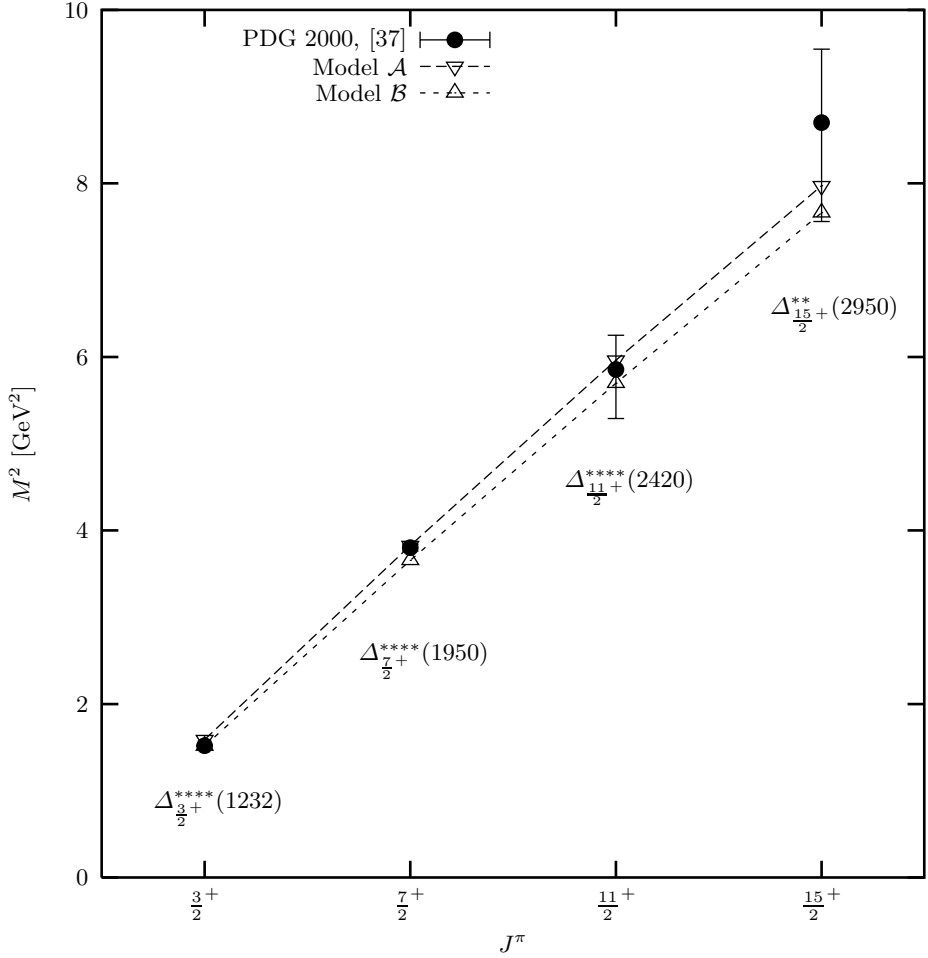}
  \end{center}
\caption{Chew-Frautschi plot ($M^2$ {\it vs.} $J$) of the
positive-parity $\Delta$-Regge trajectory
($\Delta{\frac{3}{2}^+}$, $\Delta{\frac{7}{2}^+}$,
$\Delta{\frac{11}{2}^+}$, $\Delta{\frac{15}{2}^+}$), with the
parameters of model ${\cal A}$ and model ${\cal B}$, compared to
experimental masses from the Particle Data Group (see
\cite{PDG00}). Both models yield the correct linear Regge
characteristic $M^2 \sim J$ in good agreement with experiment. See
table \ref{tab:del_regge} for numerical values.}
\label{fig:del_regge}
\end{figure}

Figure \ref{fig:del_regge} shows the Chew-Frautschi plot (M$^2$
{\it versus} $J$) of the positive parity $\Delta$-Regge trajectory as obtained
in models ${\cal A}$ and ${\cal B}$ with the
parameters $a$, $b$ and $m_n$ given in table \ref{tab:ModelParam}.  Indeed,
both models yield excellent Regge trajectories, which show the
qualitatively correct linear characteristics.

Due to the correctly reproduced trajectory slope, the experimentally observed resonance positions of the three
four-star states $\Delta{\frac{3}{2}^+}(1232,\mbox{****})$,
$\Delta{\frac{7}{2}^+}(1950,\mbox{****})$, and $\Delta{\frac{11}{2}^+}(2420,\mbox{****})$ are fairly
well described and even the highest observed orbital excitation
$\Delta{\frac{15}{2}^+}(2950,\mbox{**})$ with total spin $J^\pi=\frac{15}{2}^+$, which
is the highest  orbital excitation of the whole light baryon
spectrum measured at all, is predicted within the errors of the measured
resonance position.

In table \ref{tab:del_regge} the masses calculated in model ${\cal A}$ and
${\cal B}$ for the $\Delta$-Regge states are given
explicitly in comparison with the corresponding experimental resonance
positions taken from the Particle Data Group \cite{PDG00}.
\begin{table}[h]
\center
\begin{tabular}{cccccc}
\hline
State          & Rating & $J^\pi$           & exp. Mass [MeV] & Mass [MeV]         & Mass [MeV]\\
               &        &                   &                 &  Model ${\cal A}$& Model ${\cal B}$\\
\hline
$\Delta(1232)$ & ****   & $\frac{3}{2}^+$   & 1230 - 1234     & 1261              & 1231 \\ 
$\Delta(1950)$ & ****   & $\frac{7}{2}^+$   & 1940 - 1960     & 1956              & 1912 \\
$\Delta(2420)$ & ****   & $\frac{11}{2}^+$  & 2300 - 2500     & 2442              & 2387 \\
$\Delta(2950)$ & **     & $\frac{15}{2}^+$  & 2750 - 3090     & 2824              & 2768 \\
\hline
\end{tabular}
\caption{Position of states belonging to the positive-parity $\Delta$-Regge trajectory calculated in the models ${\cal A}$ and
${\cal B}$ in comparison to the experimental resonance positions \cite{PDG00}.
For a graphical presentation see fig. \ref{fig:del_regge}.}
\label{tab:del_regge}
\end{table}

Comparing both models, we find model ${\cal A}$ yielding a slightly larger
slope of the trajectory (see fig. \ref{fig:del_regge})  in 
somewhat better agreement with experiment than model ${\cal B}$
for the high-spin states. On the other hand model ${\cal A}$
predicts the position of the $\Delta$-ground-state
$\Delta{\frac{3}{2}^+}(1232,\mbox{****})$ slightly too high in comparison to the
experimentally determined position, whereas model ${\cal B}$ fits this
resonance position exactly. Altogether, we thus find both models to be of the
same good quality. Note that the states $\Delta{\frac{3}{2}^+}$,
$\Delta{\frac{7}{2}^+}$, $\Delta{\frac{11}{2}^+}$, $\Delta{\frac{15}{2}^+}$,
$\ldots$, belonging to the Regge trajectory, correspond to a sequence of
baryons with increasingly large separations of the quarks. Accordingly, this result
strongly supports our ansatz of the string-like confinement mechanism
in both models. In the following a more detailed discussion of these models
concerning the structure of the complete $\Delta$-spectrum is presented.

\subsection{Discussion of the complete $\Delta$-spectrum}
\label{subsec:disc_Del}
\begin{figure}[!h]
  \begin{center}
    \epsfig{file={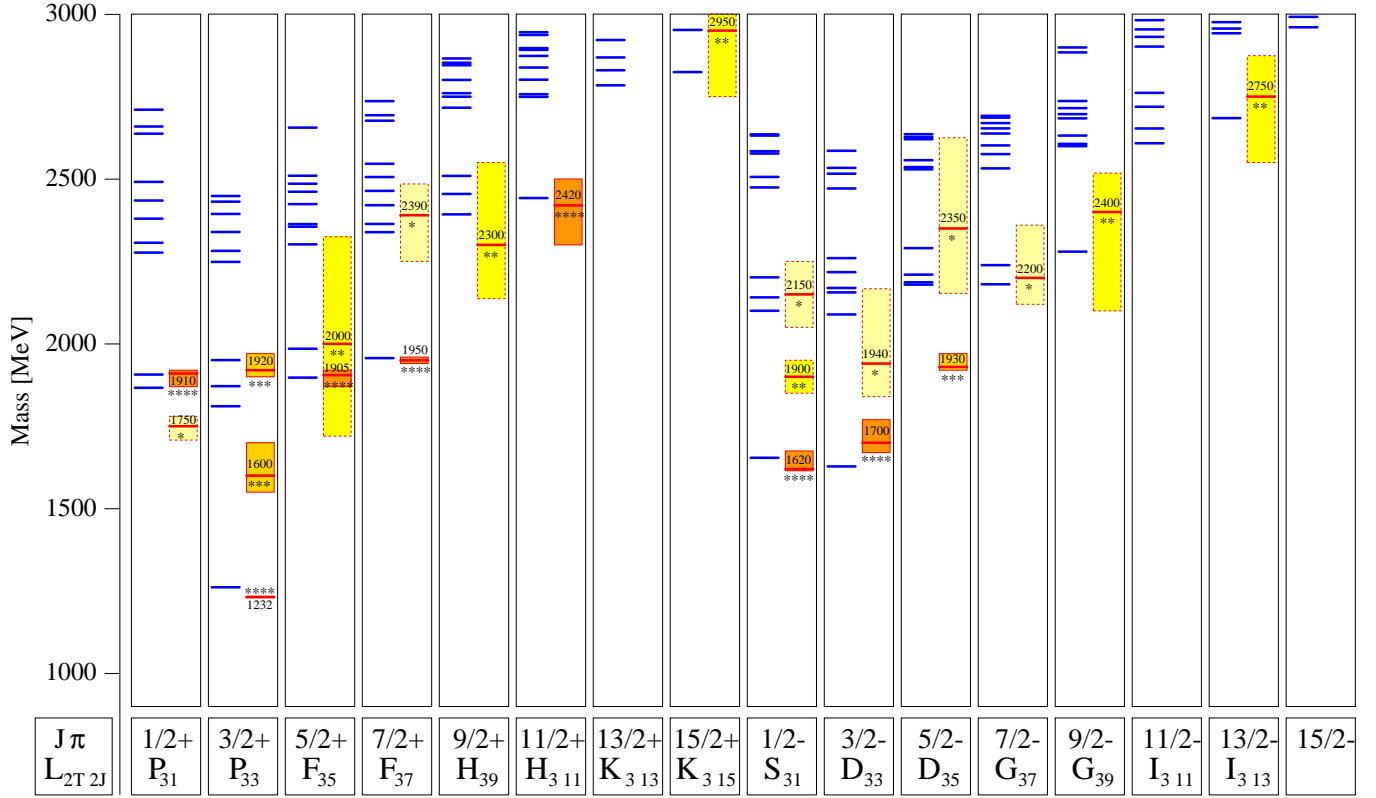},width=180mm}
  \end{center}
\caption{The calculated positive and negative parity
\textbf{$\Delta$-resonance spectrum} (isospin $T=\frac{3}{2}$ and
strangeness $S^* = 0$) in \textbf{model ${\cal A}$} (left part of each column)
in comparison to the experimental spectrum taken from Particle Data Group \cite{PDG00} (right part of each
column).  The resonances are classified by the total angular momentum
$J$ and parity $\pi$. The experimental resonance position is indicated
by a bar, the corresponding uncertainty by the shaded box which is
darker for better established resonances; the status of each
resonance is additionally indicated by stars.}
\label{fig:DeltaM2}
\end{figure}

\begin{figure}[!h]
  \begin{center}
    \epsfig{file={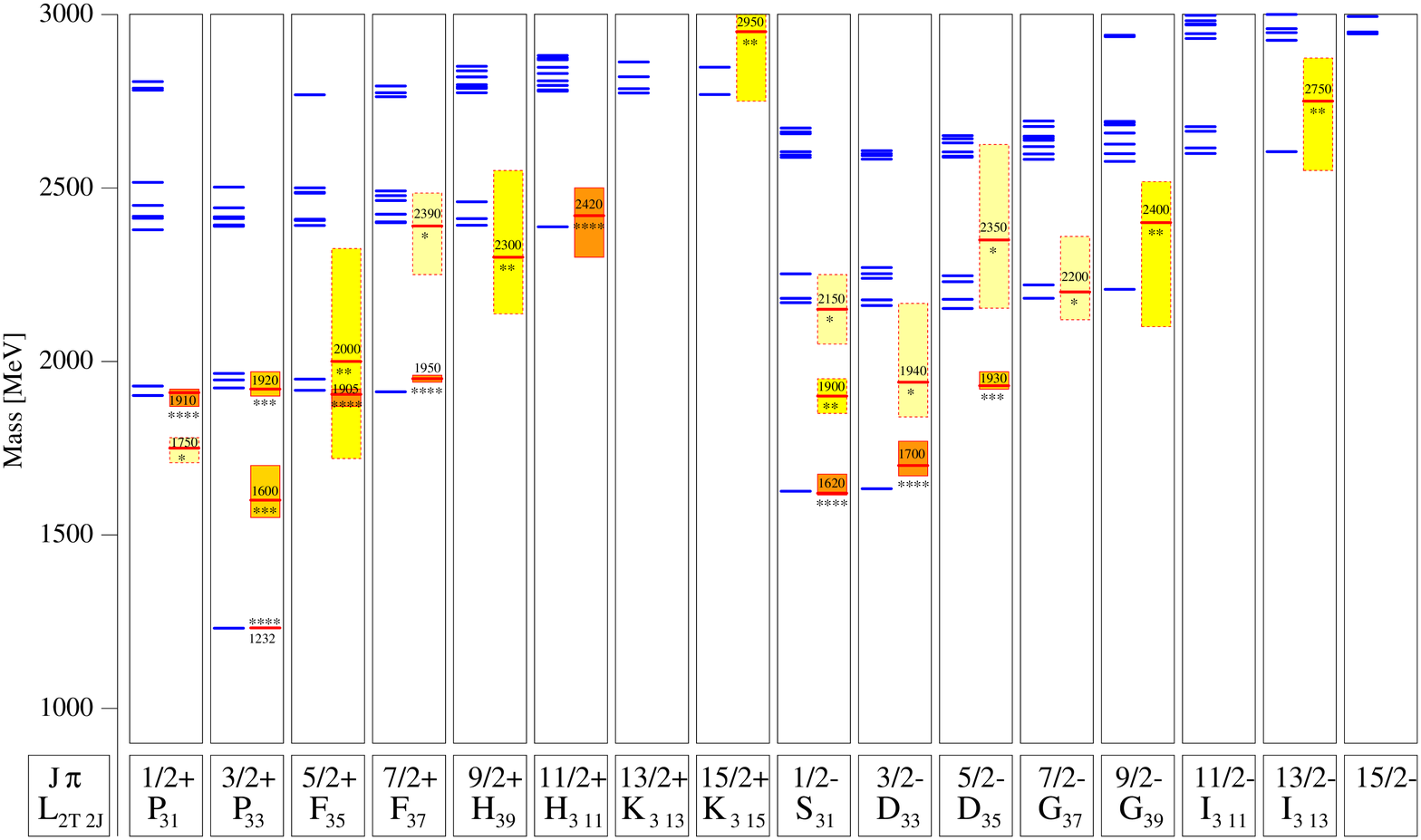},width=180mm}
  \end{center}
\caption{The calculated positive and negative parity
\textbf{$\Delta$-resonance spectrum} (isospin $T=\frac{3}{2}$ and
strangeness $S^* = 0$) in \textbf{model ${\cal B}$} (left part of each column)
in comparison to the experimental spectrum taken from Particle Data Group \cite{PDG00} (right part of each
column). See also caption to fig. \ref{fig:DeltaM2}}
\label{fig:DeltaM1}
\end{figure}

Figure \ref{fig:DeltaM2} shows the resulting positions of positive and
negative parity $\Delta$-baryons (isospin $T=\frac{3}{2}$ and strangeness
$S^*=0$) with total angular momenta up to $J=\frac{15}{2}$ obtained in model
${\cal A}$. These are compared to all presently known resonances
quoted by the Particle Data Group \cite{PDG00}.  Likewise, figure
\ref{fig:DeltaM1} displays the corresponding results of model ${\cal B}$.
The resonances in each column are classified by their total spin $J$ and
parity $\pi$. On the left of each column at most the first ten calculated
states are shown and can be compared with the experimentally observed positions
on the right hand side of each column. The corresponding uncertainties
in the measured resonance positions are indicated by the shaded areas. The
status of each resonance is denoted by the corresponding number of stars
following the notation of the Particle Data Group \cite{PDG00} and
moreover, by the shading of the error box which is darker for better
established resonances.\\ In addition to figs. \ref{fig:DeltaM2}
and \ref{fig:DeltaM1}, the calculated masses of the excited
states are summarized according to their assignment to a particular shell
in one of the tables \ref{tab:1hwband} -- \ref{tab:Del6hwband}.


\subsubsection{Global structure of the $\Delta$-spectrum}
Before discussing single resonance positions in each shell in detail
let us first focus on the gross structure of the global
$\Delta$-spectrum. Figures \ref{fig:DeltaM2} and \ref{fig:DeltaM1}
show that with both variants ${\cal A}$ and ${\cal B}$, indeed
a satisfactory  overall description of the global features of the
experimentally known resonances can be obtained.\\ In both models the
positions of the shell structures are well reproduced together with the
correct position of states that belong to the
positive-parity $\Delta$-Regge trajectory. The centroids of the various even- and odd-parity
bands are in good agreement with the centroids of the corresponding
experimentally observed band structures with the exception of the
negative-parity states around 1900 MeV. Concerning the structures of
the shells themselves it should be noted that model ${\cal A}$ exhibits
larger intra-band splittings than model ${\cal B}$ due to slightly
larger spin-orbit effects in model ${\cal A}$.  In
this respect, the Dirac structure (consisting of the scalar and
time-like vector part in both models) is combined in model
${\cal B}$ such that the spin-orbit forces of both parts
cancel and thus these relativistic
effects, which arise in connection with the embedding map in the Salpeter
amplitudes, are really minimized.  Nevertheless, also the intra-band splittings of model
${\cal A}$ are still moderate enough to be compatible with the central values for 
the experimentally observed resonance masses. They
even generate the desired feature that the average mass values in each
shell rise with total angular momentum $J$, in contrast
to model ${\cal B}$, where the states with different spin $J$ are
almost degenerate. This produces a slight tilt in
the shells of model ${\cal A}$ (see {\it e.g.}  the positive parity
$2\hbar\omega$ shell), which generally yields higher resonance
positions for the states with maximum angular momentum, thus producing
positive- and negative-parity Regge trajectories with a slightly
bigger slope in model ${\cal A}$ in better agreement with the
experiment. 

Now let us discuss the single resonance positions and the pattern of
splittings in the positive- and negative-parity bands of the excited
$\Delta$-spectrum in some more detail.
\subsubsection{States of the 1${\bfgrk\hbar}{\bfgrk\omega}$ band}
We start with the negative-parity $1\hbar\omega$ band, where our models
predict (as usual in constituent quark models) two states which can be
uniquely assigned to the two well-established four-star resonances
$\Delta{\frac{1}{2}^-}(1620,\mbox{****})$ and
$\Delta{\frac{3}{2}^-}(1700,\mbox{****})$. Our results for these states are
given explicitly in table \ref{tab:1hwband}.

\begin{table}[h]
\center
\begin{tabular}{ccccccc}
\hline
Exp. state   &PW&${J^\pi}$        & Rating     & Mass range [MeV]& Model state &  Model state \\
\cite{PDG00} &  &                      &            & \cite{PDG00}    & in model ${\cal A}$& in model ${\cal B}$  \\
\hline
$\Delta(1620)$&$S_{31}$&${\frac{1}{2}^-}$& ****     &1615-1675  & $\MSD(1,-,1,1654)$  & $\MSD(1,-,1,1625)$\\
\hline
$\Delta(1700)$&$D_{33}$&${\frac{3}{2}^-}$& ****     &1670-1770  & $\MSD(3,-,1,1628)$  & $\MSD(3,-,1,1633)$ \\
\hline
\end{tabular}
\caption{Calculated positions of $\Delta$-states assigned to the negative parity $1\hbar\omega$ shell
         in comparison to the corresponding experimental mass values taken
         from \cite{PDG00}. PW denotes the partial wave and the rating is
         given according to the PDG classification \cite{PDG00}. Here and throughout this work we use the notation
         $[B\;J^\pi]_n({M})$ for the predicted model states in model ${\cal A}$ and
         ${\cal B}$, respectively, where $B$ denotes the baryon ({\it i.e.} the
         flavor), $J^\pi$ are spin and parity, and $M$ is the predicted mass
         given in MeV. 
         $n = 1, 2, 3, \ldots$ is the principal quantum
         number counting the states in each sector $J^\pi$ beginning with the
         lowest state.}
\label{tab:1hwband}
\end{table}
Both models yield the correct center-of-gravity of this $1 \hbar\omega$ band. In particular, the position of
$\Delta{\frac{1}{2}^-}(1620,\mbox{****})$ is fairly well reproduced
within the experimental range: model ${\cal A}$ predicts 1654 MeV
and model ${\cal B}$ gives 1625 MeV.  Note that in quark models,
which use the one-gluon exchange as a residual interaction, the
$\Delta$ ground-state is shifted upwards relative to the negative
parity excited states $\Delta{\frac{1}{2}^-}$ and
$\Delta{\frac{3}{2}^-}$, thus in general predicting the
centroid of the $1 \hbar\omega$ band too low.  Unfortunately,
in both models the calculated position of the
$\Delta{\frac{3}{2}^-}$ state turns out to be roughly $70$ MeV too low
compared to the observed position of the
$\Delta{\frac{3}{2}^-}(1700,\mbox{****})$.  The splitting of
$\Delta{\frac{1}{2}^-}(1620,\mbox{****})$ and
$\Delta{\frac{3}{2}^-}(1700,\mbox{****})$ is often interpreted as one
of a few possible hints for the relevance of spin-orbit forces in
baryon spectroscopy. In this respect, both models cannot account for
this splitting: In model ${\cal B}$ both states are nearly
degenerate and in model ${\cal A}$ the splitting
$\Delta{\frac{3}{2}^-}-\Delta{\frac{1}{2}^-}$ even has the wrong
sign.  Note however that the experimental indications for the true
size of this splitting itself are less clear, since the ranges of
possible values for both resonances are even overlapping.

\subsubsection{States of the 2${\bfgrk\hbar}{\bfgrk\omega}$ band}
We now focus on the structure of the positive parity
  $2\hbar\omega$ shell.  
Our results for the states of this shell are
given explicitly in table \ref{tab:2hwband}.
Disregarding for the moment the
puzzling low position of the three-star $\Delta{\frac{3}{2}^+}(1600)$
resonance, figs. \ref{fig:DeltaM2} and \ref{fig:DeltaM1} show that both
confinement versions predict the centroid of the
$2\hbar\omega$ band at around 1900 MeV in  very good agreement with
experiment. Due to the minimization of spin-orbit effects by the
Dirac structure of confinement ${\cal B}$, the
$2\hbar\omega$ states in model ${\cal B}$ hardly show any splitting and
all states are clustered within a narrow region between 1900 and 1970 MeV.
We thus find for each well-established three- and four-star resonance in this
region, {\it i.e.} the $\Delta{\frac{1}{2}^+}(1910,\mbox{****})$, the
$\Delta{\frac{3}{2}^+}(1920,\mbox{***})$, the $\Delta{\frac{5}{2}^+}(1905,\mbox{****})$
and the $\Delta{\frac{7}{2}^+}(1950,\mbox{****})$, a predicted state for a possible
assignment. Due to the strong clustering of the predicted states, the
experimentally indicated pattern of splittings, {\it e.g.}
$\Delta{\frac{1}{2}^+}(1910,\mbox{****})-\Delta{\frac{1}{2}^+}(1750,\mbox{*})$ and
$\Delta{\frac{5}{2}^+}(1905,\mbox{***})-\Delta{\frac{5}{2}^+}(2000,\mbox{**})$ or even the low
position $\Delta{\frac{3}{2}^+}(1600,\mbox{***})$ cannot be described.  Concerning the
$\Delta{\frac{1}{2}^+}(1750,\mbox{*})$ and the $\Delta{\frac{5}{2}^+}(2000,\mbox{**})$ note,
however, that their rating is only one- and two-star, respectively. On the
other hand, the intra-band splittings in model ${\cal A}$, which are
induced by moderate spin-orbit effects, reproduce the experimentally
indicated pattern of splittings in this shell quite well.
Consequently, a tentative assignment of the states on the basis of their positions
compared to the experimental values is less ambiguous.  
Let us discuss the
intra-band pattern in some more detail (see also table \ref{tab:2hwband}):

\begin{table}[h]
\center
\begin{tabular}{ccccccc}
\hline
Exp. state   &PW&${J^\pi}$        & Rating     & Mass range [MeV]& Model state &  Model state \\
\cite{PDG00} &  &                      &            & \cite{PDG00}    & in model ${\cal A}$& in model ${\cal B}$  \\
\hline
$\Delta(1750)$ &$P_{31}$& ${\frac{1}{2}^+}$& *     & 1708-1780 & $\MSD(1,+,1,1866)$ & \\
$\Delta(1910)$ &$P_{31}$& ${\frac{1}{2}^+}$& ****  & 1870-1920 & $\MSD(1,+,2,1906)$& $\begin{array}{c}
                                                                                     \MSD(1,+,1,1901)\\
                                                                                     \MSD(1,+,2,1928)
                                                                                     \end{array}$\\
\hline
$\Delta(1600)$ &$P_{33}$& ${\frac{3}{2}^+}$& ***   & 1550-1700 & $\MSD(3,+,2,1810)$  & \\
$\Delta(1920)$ &$P_{33}$& ${\frac{3}{2}^+}$& ***   & 1900-1970 &
$\begin{array}{c}\MSD(3,+,3,1871)\\\MSD(3,+,4,1950)\end{array}$& 
$\begin{array}{c}\MSD(3,+,2,1923)\\\MSD(3,+,3,1946)\\\MSD(3,+,4,1965)\end{array}$ \\
\hline
$\Delta(1905)$ &$F_{35}$& ${\frac{5}{2}^+}$& ****  & 1870-1920 & $\MSD(5,+,1,1897)$  & $\MSD(5,+,1,1916)$\\
$\Delta(2000)$ &$F_{35}$& ${\frac{5}{2}^+}$& **    & 1720-2325 & $\MSD(5,+,2,1985)$  & $\MSD(5,+,2,1948)$\\
\hline
$\Delta(1950)$ &$F_{37}$& ${\frac{7}{2}^+}$& ****  & 1940-1960  & $\MSD(7,+,1,1956)$  & $\MSD(7,+,1,1912)$\\
\hline
\end{tabular}
\caption{Calculated positions of $\Delta$-states assigned to the positive parity $2\hbar\omega$ shell
  and their tentative assignment to observed resonances due to comparison with
  experimental mass values taken from \cite{PDG00}. Notation as in table \ref{tab:1hwband}.}
\label{tab:2hwband}
\end{table}
In the spin-$1/2^+$ sector models ${\cal A}$ and ${\cal B}$
describe the situation very similarly.  In model ${\cal A}$ we find
two close states at 1866 MeV and 1906 MeV, where the latter fits
nicely the well-established four-star resonance
$\Delta{\frac{1}{2}^+}(1910,\mbox{****})$ within the assigned errors,
while the other one overestimates the position of the poorly
determined one-star resonance
$\Delta{\frac{1}{2}^+}(1750,\mbox{*})$. In the sector with
$J^\pi=3/2^+$ model ${\cal A}$ generates the largest splittings
within the $2\hbar\omega$ band in at least qualitative agreement with
the low position of the $\Delta{\frac{3}{2}^+}(1600,\mbox{***})$
resonance.  Nevertheless, despite this improvement relative to model
${\cal B}$, the role of this state remains still unclear also in
model ${\cal A}$, since the lowest calculated state in this sector
at 1810 MeV is still far away from the PDG reported mass value
centered around $1600$ MeV. We would like to comment here that this
state is often viewed as the analogue of the low-lying Roper
resonance. Both states have in common that they are the first scalar
excitation of the corresponding ground-state, positioned even below
the lowest excitations of negative parity.  However, in the spirit of
our model, which in addition to a proper confinement force uses
exclusively 't~Hooft's instanton-induced force as residual
interaction, the low positions of
$\Delta{\frac{3}{2}^+}(1600,\mbox{***})$ and the Roper resonance
cannot originate from the same dynamics. In this respect, we should
anticipate already here that in the case of the Roper resonance and
its strange counter parts (which all appear in the sectors of the
spin-${\frac{1}{2}^+}$ octet ground-state baryons) the low positions
can be nicely explained as an instanton effect due to the attractive
action of 't~Hooft's force in these sectors. Thus, quite in contrast
to the $N$-sector we do not find a similar selective lowering for the
$\Delta{\frac{3}{2}^+}(1600,\mbox{***})$. Models which use the OGE
interaction \cite{CaIs86}, have similar problems to account for this
resonance \cite{CaRo00}. In other phenomenological approaches
\cite{GlRi96,Gl00} the low position of the
$\Delta{\frac{3}{2}^+}(1600,\mbox{***})$ is explained due to a
flavor-dependent Goldstone-boson exchange interaction.  It thus seems
worthwhile to comment on the experimental situation of this somewhat
puzzling resonance: Its position is extracted from tedious analyses of
mostly older $\pi N\rightarrow \pi N$ scattering data.  The different
analyses do often not agree, but even exhibit very large discrepancies
with a big range of possible values. The mass range quoted by the
Particle Data Group \cite{PDG00} for this state is 1550 to 1700
MeV. In our opinion it is astonishing that the Particle Data Group
\cite{PDG00} has given this state a three-star rating although the
various analyses are not in good agreement. Thus, despite a three-star
rating, the current experimental evidence concerning the position of
$\Delta{\frac{3}{2}^+}(1600,\mbox{***})$ is in our opinion not
compulsory but rather unclear. Hence it becomes questionable if our
calculated mass value of 1810 MeV for this resonance really
constitutes a serious discrepancy, especially in view of more recent
analysis \cite{MaSa92,VrDyLe00} that predict the resonance position at
the upper end of this range at about 1700 MeV. In this respect we hope
that with the new generation of experimental facilities and the
corresponding new efforts in baryon spectroscopy, the situation for
this state will soon be clarified. In agreement with other
(non-relativistic or 'relativized') constituent quark models our
framework predicts two further states with $J^\pi=\frac{3}{2}^+$ in
the $2\hbar\omega$ shell. The state predicted at 1950 MeV in model
${\cal A}$ matches the observed three-star state
$\Delta{\frac{3}{2}^+}(1920,\mbox{***})$ within its uncertainties. The
other state, whose mass value is predicted at 1871 MeV lies near this
mass range.  In model ${\cal B}$ all three states are clustered in
the mass range of the $\Delta{\frac{3}{2}^+}(1920,\mbox{***})$.  In
the spin-$5/2^+$ sector two states are predicted within the
$2\hbar\omega$ band, as required by the experimental findings. The
positions of the well-established four-star resonance
$\Delta{\frac{5}{2}^+}(1905,\mbox{****})$ and the two-star state
$\Delta{\frac{5}{2}^+}(2000,\mbox{**})$ and accordingly also their
mass difference are fairly well described by the predictions at 1897
MeV and 1985 MeV in model ${\cal A}$. Again the situation is better
than in model ${\cal B}$, which cannot reproduce this splitting due
its minimal relativistic spin-orbit effects. But we should not attach
too much importance to the $\Delta{\frac{5}{2}^+}(2000,\mbox{**})$,
which has only a two-star rating and moreover reveals big
uncertainties in its determined mass position.  In the sector
$J^\pi=\frac{7}{2}^+$ the well-established four-star resonance
$\Delta{\frac{7}{2}^+}(1950,\mbox{****})$ is the only state seen in
this mass range of the $F_{37}$ partial wave, compatible with our
prediction. The position of this state, which is a member of the
positive-parity $\Delta$-Regge trajectory discussed previously, is
exactly reproduced by the prediction at 1956 MeV in model
${\cal A}$, whereas model ${\cal B}$, which yields the mass
value 1912 MeV, slightly underestimates this position.

\subsubsection{States of the 3${\bfgrk\hbar}{\bfgrk\omega}$ band}
Apart from the  puzzling, low-lying
$\Delta\frac{3}{2}^+(1600,\mbox{***})$ resonance discussed previously, the present experimental
$\Delta$-spectrum also shows a curious structure in the negative parity
sector, namely the three resonances $\Delta\frac{1}{2}^-(1900,\mbox{**})$,
$\Delta\frac{3}{2}^-(1940, *)$ and $\Delta\frac{5}{2}^-(1930,\mbox{***})$
around 1900 MeV, which are nearly degenerate with the states of the
positive-parity $2\hbar\omega$ shell. Taking these states seriously within a
constituent quark model, they have to be naturally assigned to the
$3\hbar\omega$ band, since the two states predicted in the $1\hbar\omega$
shell could already be uniquely assigned to the well-established resonances
$\Delta\frac{1}{2}^-(1620,\mbox{****})$ and
$\Delta\frac{1}{2}^-(1700,\mbox{****})$. As can be seen in figs.
\ref{fig:DeltaM2} and \ref{fig:DeltaM1} neither model ${\cal A}$ nor model
${\cal B}$ can account for these rather low-lying states. In
both models the rich structure of states assigned to the $3\hbar\omega$ shell
is spread around 2200 to 2300 MeV, which agrees with the center-of-gravity of
the other one- and two-star resonances observed in the
$3\hbar\omega$ band; see also table \ref{tab:Del3hwband}, where the predicted
masses for all $\Delta$-states in the $3\hbar\omega$ band are 
given explicitly.  The only state of the $3\hbar\omega$ band that can be uniquely
assigned to an experimentally observed resonance is the single state in the
$\Delta\frac{9}{2}^-$ sector at 2280 MeV in model ${\cal A}$ and at 2207
MeV in ${\cal B}$: the predicted positions agree within the very big range
of possible values of the single two-star resonance
$\Delta\frac{9}{2}^-(2400,\mbox{**})$ seen in the $G_{39}$ partial wave.
All other negative-parity resonances observed in this mass region have only a
one-star rating. Their quite big ranges of possible values agree with the
positions of several $3\hbar\omega$ states predicted by both models.

\begin{table}[!h]
\center
\begin{tabular}{ccccccc}
\hline
Exp. state   &PW&${J^\pi}$        & Rating     & Mass range [MeV]& Model state &  Model state \\
\cite{PDG00} &  &                      &            & \cite{PDG00}    & in model ${\cal A}$& in model ${\cal B}$  \\
\hline
$\Delta(1900)$&$S_{31}$&${\frac{1}{2}^-}$& **     &1850-1950& &\\
$\Delta(2150)$&$S_{31}$&${\frac{1}{2}^-}$& *      &2050-2250&$\begin{array}{c}\MSD(1,-,2,2100)\\
                                                                              \MSD(1,-,3,2141)\\
                                                                              \MSD(1,-,4,2202)
                                                              \end{array}$
                                                            &$\begin{array}{c}
                                                                              \MSD(1,-,2,2169)\\
                                                                              \MSD(1,-,3,2182)\\              
                                                                              \MSD(1,-,4,2252)
                                                              \end{array}$\\
\hline
$\Delta(1940)$&$D_{33}$&${\frac{3}{2}^-}$& *      &1840-2167&$\begin{array}{c}
                                                              \MSD(3,-,2,2089)\\
                                                              \MSD(3,-,3,2156)\\
                                                              \end{array}$
                                                             &$\begin{array}{c}
                                                               \MSD(3,-,2,2161)
                                                               \end{array}$\\
              &        &                 &        &         &$\begin{array}{c}
                                                              \MSD(3,-,4,2170)\\ 
                                                              \MSD(3,-,5,2218)\\   
                                                              \MSD(3,-,6,2260)
                                                              \end{array}$  
                                                            &$\begin{array}{c}
                                                               \MSD(3,-,3,2177)\\       
                                                               \MSD(3,-,4,2239)\\
                                                               \MSD(3,-,5,2253)\\
                                                               \MSD(3,-,6,2270)
                                                              \end{array}$\\
\hline
$\Delta(1930)$&$D_{35}$&${\frac{5}{2}^-}$& ***    &1920-1970& &\\
$\Delta(2350)$&$D_{35}$&${\frac{5}{2}^-}$& *      &2153-2625&$\begin{array}{c}
                                                               \MSD(5,-,1,2170)\\
                                                               \MSD(5,-,2,2187)\\    
                                                               \MSD(5,-,3,2210)\\ 
                                                               \MSD(5,-,4,2290)
                                                              \end{array}$ 
                                                            &$\begin{array}{c} 
                                                                \MSD(5,-,1,2152)\\
                                                                \MSD(5,-,2,2179)\\
                                                                \MSD(5,-,3,2230)\\
                                                                \MSD(5,-,4,2247)
                                                              \end{array}$\\
\hline
$\Delta(2200)$&$G_{37}$&${\frac{7}{2}^-}$& *      &2120-2360&$\begin{array}{c}
                                                               \MSD(7,-,1,2181)\\
                                                               \MSD(7,-,2,2239)
                                                              \end{array}$ 
                                                            &$\begin{array}{c}
                                                               \MSD(7,-,1,2182)\\
                                                               \MSD(7,-,2,2220)
                                                              \end{array}$\\
\hline
$\Delta(2400)$&$G_{39}$&${\frac{9}{2}^-}$& **      &2100-2518&$\MSD(9,-,1,2280)$ &$\MSD(9,-,1,2207)$\\
\hline
\end{tabular}
\caption{Calculated positions of negative-parity
$\Delta$ states in the $3\hbar\omega$ shell in comparison to the
corresponding experimental mass values taken from \cite{PDG00}. Notation as in
table \ref{tab:1hwband}.}
\label{tab:Del3hwband}
\end{table}
The lowest $3\hbar\omega$ states calculated in our models lie at about 2100
MeV, rather far above the masses of the three conspicuously low-lying
states. Several other constituent quark models also cannot account for this
puzzling structure ({\it e.g.} models with OGE forces \ref{CaIs86,CaRo00}): in
general quark model predictions for the masses of these states are
consistently too high by about 150-250 MeV.  As in the case of the
$\Delta\frac{3}{2}^+(1600,\mbox{***})$, a comment concerning the experimental status
of these mysterious resonances is necessary at this stage: Also here, the
experimental situation is rather unclear and unsatisfactory.
The existence of the one-star $\Delta\frac{3}{2}^-(1940,\mbox{*})$ resonance anyway is questionable and
concerning the two-star $\Delta\frac{1}{2}^-(1900,\mbox{**})$ resonance it is worth
emphasizing that in the 1998 edition of the {\it 'Review of Particle Physics'}
\cite{PDG98} this resonance has already been downgraded  from three stars to
two due to its weak signal in speed plots.  Moreover, it should be mentioned that
$\Delta\frac{3}{2}^-(1940,\mbox{*})$ and $\Delta\frac{1}{2}^-(1900,\mbox{**})$ have not been seen
in various partial wave analyses of $\pi N\rightarrow \pi N$ scattering
data \cite{Kan99}. Furthermore, additional corrections to the existing partial wave solutions due
to additional new data from recently measured spin rotation parameters in
$\pi^+ p$ elastic scattering at (ITEP)-PNPI \cite{Al95} indicate that the signal
of the $\Delta\frac{3}{2}^-(1940,\mbox{*})$ resonance even completely disappears (see
\cite{Kan99,Al97} and references therein). Concerning the three-star resonance
$\Delta\frac{5}{2}^-(1930,\mbox{***})$ the Particle Data Group \cite{PDG00}
states that various analyses are not in good agreement.  Due to this quite
unclear experimental situation, we again do not pay too much attention to
this puzzling structure when evaluating the quality of our
confinement mechanisms. But we hope that also in this case the situation will
soon be clarified by the new experimental investigations in baryon
spectroscopy. Let us emphasize that the confirmation of these resonances would
strongly disfavor our model as well as several other models \cite{CaRo00}.

\subsubsection{Beyond the 3${\bfgrk\hbar}{\bfgrk\omega}$ band}
The high mass part of the experimental $\Delta$-spectrum is still
hardly explored.  Explicit mass values for some of the states in model
${\cal A}$ and ${\cal B}$ assigned to the
$4\hbar\omega$, $5\hbar\omega$ and $6\hbar\omega$ shell, are
summarized in tables \ref{tab:Del4hwband}, \ref{tab:Del5hwband} and
\ref{tab:Del6hwband}, respectively.\\ In the positive-parity
$4\hbar\omega$ band three resonances with spins $J^\pi=\frac{7}{2}^+$,
$\frac{9}{2}^+$ and $\frac{11}{2}^+$ are found experimentally.  In the
$\Delta\frac{11}{2}^+$ sector this is the well-established four-star
resonance $\Delta\frac{11}{2}^+(2420,\mbox{****})$ which belongs to
the positive-parity Regge trajectory discussed before.  Both
models predict a single $\Delta\frac{11}{2}^+$ state in the
$4\hbar\omega$ shell which can be uniquely assigned to
$\Delta\frac{11}{2}^+(2420,\mbox{****})$.  The position at 2442 MeV in
model ${\cal A}$ and at 2388 MeV in model ${\cal B}$ nicely
agrees within the errors of the empirically determined position.  The
big ranges of possible values of the observed resonances in
$\Delta\frac{7}{2}^+$ and $\Delta\frac{9}{2}^+$ agree with several
states predicted in both models (see table \ref{tab:Del4hwband}).\\ In
the energy region of the negative parity $5\hbar\omega$ band only the
resonance $\Delta\frac{13}{2}^-(2750,\mbox{**})$ with a two-star
rating has been observed in the $I_{3\;13}$ partial wave. Both models
predict a single $\Delta\frac{13}{2}^-$ state  (belonging to the
$5\hbar\omega$ shell) at 2685 MeV in model ${\cal A}$ and at 2604
MeV in model ${\cal B}$, compatible with this single observed
state. In the $\Delta\frac{11}{2}^-$ sector no resonance has been
seen so far; the predicted states of model ${\cal A}$ and
model ${\cal B}$ in this sector are given in table
\ref{tab:Del5hwband}.\\ Finally, the highest observed excitation of
the light baryon spectrum, {\it i.e} the two-star resonance
$\Delta\frac{15}{2}^+(2950,\mbox{**})$, which belongs to the
positive-parity Regge trajectory, is the only observed state of
the $6\hbar\omega$ shell.  Both models predict in the
$\Delta\frac{15}{2}^+$ sector two states whose positions at
2824 and 2952 MeV in model ${\cal A}$ and at 2769 and 2848 MeV in
model ${\cal B}$ are within the uncertainty range of the
$\Delta\frac{15}{2}^+(2950,\mbox{**})$ (see table
\ref{tab:Del6hwband}).

\begin{table}[!h]
\center
\begin{tabular}{ccccccc}
\hline
Exp. state   &PW&${J^\pi}$        & Rating     & Mass range [MeV]& Model state &  Model state \\
\cite{PDG00} &  &                      &            & \cite{PDG00}    & in model ${\cal A}$& in model ${\cal B}$  \\
\hline
$\Delta(2390)$&$F_{37}$&${\frac{7}{2}^+}$& *     &2250-2485&$\begin{array}{c}
                                                              \MSD(7,+,2,2339)\\
                                                              \MSD(7,+,3,2364)\\
                                                              \MSD(7,+,4,2421)\\
                                                              \MSD(7,+,5,2464)\\
                                                              \MSD(7,+,6,2506)\\
                                                              \MSD(7,+,7,2546)
                                                             \end{array}$ 
                                                           &$\begin{array}{c}
                                                              \MSD(7,+,2,2400)\\
                                                              \MSD(7,+,3,2402)\\
                                                              \MSD(7,+,4,2424)\\
                                                              \MSD(7,+,5,2463)\\
                                                              \MSD(7,+,6,2477)\\
                                                              \MSD(7,+,7,2491)
                                                             \end{array}$\\

\hline
$\Delta(2300)$&$H_{39}$&${\frac{9}{2}^+}$& **    &2137-2550&$\MSD(9,+,1,2393)$&$\MSD(9,+,1,2392)$\\
              &        &                 &       &         &$\MSD(9,+,2,2455)$&$\MSD(9,+,2,2411)$\\
              &        &                 &       &         &$\MSD(9,+,3,2509)$&$\MSD(9,+,3,2460)$\\
\hline
$\Delta(2420)$&$H_{3\;11}$&${\frac{11}{2}^+}$& ****     &2300-2500&$\MSD(11,+,1,2442)$ &$\MSD(11,+,1,2388)$\\
\hline
\end{tabular}
\caption{Calculated positions of the  positive-parity
$\Delta$--states in the $4\hbar\omega$ shell with $J\geq\frac{7}{2}$ in comparison to the
corresponding experimental mass values taken from \cite{PDG00}. Notation as in
table \ref{tab:1hwband}.}
\label{tab:Del4hwband}
\end{table}

\begin{table}[!h]
\center
\begin{tabular}{ccccccc}
\hline
Exp. state   &PW&${J^\pi}$        & Rating     & Mass range [MeV]& Model state &  Model state \\
\cite{PDG00} &  &                      &            & \cite{PDG00}    & in model ${\cal A}$& in model ${\cal B}$  \\
\hline
&$I_{3\;11}$&${\frac{11}{2}^-}$& & & $\MSD(11,-,1,2608)$ & $\MSD(11,-,1,2599)$\\
&           &                  & & & $\MSD(11,-,2,2653)$ & $\MSD(11,-,2,2615)$\\
&           &                  & & & $\MSD(11,-,3,2719)$ & $\MSD(11,-,3,2663)$\\
&           &                  & & & $\MSD(11,-,4,2761)$ & $\MSD(11,-,4,2676)$\\
\hline
$\Delta(2750)$&$I_{3\;13}$&${\frac{13}{2}^-}$& **   &2550-2874 & $\MSD(13,-,1,2685)$ & $\MSD(13,-,1,2604)$ \\
\hline
\end{tabular}
\caption{Calculated positions of the negative-parity
$\Delta$--states in the $5\hbar\omega$ shell with $J\geq\frac{11}{2}$ in comparison to the
corresponding experimental mass values taken from \cite{PDG00}. Notation as in
table \ref{tab:1hwband}.}
\label{tab:Del5hwband}
\end{table}

\begin{table}[!h]
\center
\begin{tabular}{ccccccc}
\hline
Exp. state   &PW&${J^\pi}$        & Rating     & Mass range [MeV]& Model state &  Model state \\
\cite{PDG00} &  &                      &            & \cite{PDG00}    & in model ${\cal A}$& in model ${\cal B}$  \\
\hline
&$K_{3\;13}$&${\frac{13}{2}^+}$& & & $\MSD(13,+,1,2784)$ & $\MSD(13,+,1,2773)$ \\
&           &                  & & & $\MSD(13,+,2,2830)$ & $\MSD(13,+,2,2786)$\\
&           &                  & & & $\MSD(13,+,3,2869)$ & $\MSD(13,+,3,2820)$\\
&           &                  & & & $\MSD(13,+,4,2922)$ & $\MSD(13,+,4,2863)$\\
\hline
$\Delta(2950)$&$K_{3\;15}$&${\frac{15}{2}^+}$& **   &2750-3090 & $\MSD(15,+,1,2824)$  & $\MSD(15,+,1,2769)$ \\
&           &                  & & & $\MSD(15,+,2,2952)$ & $\MSD(15,+,2,2848)$\\
\hline
\end{tabular}
\caption{Calculated positions of the positive-parity
$\Delta$ states in the $6\hbar\omega$ shell with $J\geq\frac{13}{2}$ in comparison to the
corresponding experimental mass values taken from \cite{PDG00}. Notation as in
table \ref{tab:1hwband}.}
\label{tab:Del6hwband}
\end{table} 

\subsection{Summary for the $\Delta$-spectrum}
To summarize our discussion of the $\Delta$-spectrum, we have presented the
results for two different confinement models, which essentially differ only in
the Dirac structure of their linearly rising part. 
In both models this part consists of a combination of a
scalar and a time-like vector part. In model ${\cal B}$ this combination is
such that relativistic spin-orbit forces of both parts almost cancel, whereas the combination in model
${\cal A}$ produces spin-orbit forces which are small enough to be still
compatible with experimental findings.\\
Once the Dirac structures are fixed, we obtain in both models a quite good description of the complete
$\Delta$-spectrum up to highest orbital excitations $J\leq\frac{15}{2}$.
This is achieved by adjusting only three (!) parameters: the non-strange quark mass
$m_n$, the confinement offset-parameter $a$ and the slope $b$, see table \ref{tab:ModelParam}.
\begin{itemize}
\item
Both models yield excellent Regge trajectories with the correct
phenomenological characteristic $M^2\propto J$ and the right slope, indicating that baryons with
increasingly large separations of the quarks are well described. This feature
strongly supports our string-like ansatz for confinement.
\item
In both variants ${\cal A}$ and ${\cal B}$, the resulting positions of
the even and odd parity shell structures agree fairly well with the centroids
of the experimentally observed band structures.
\item
Note that all of the seven well-established resonances of the
$\Delta$-spectrum which have a four-star rating \cite{PDG00} are very
well described in both confinement models.
\item
The results of the confinement models ${\cal A}$ and ${\cal B}$ mainly
differ in the intra-band splittings. Due to the minimization of spin-orbit
effects, model ${\cal B}$ shows hardly any splitting and the different states
in each shell are nearly degenerate. Model ${\cal A}$, however, shows moderate
intra-band splittings due to rather small relativistic spin-orbit
effects. These splittings, {\it e.g.} in the $2\hbar\omega$
band, even tend to agree with experimentally observed structures as far as
they actually can be disentangled from the rather large experimental uncertainties.  
\end{itemize}  
Thus we have shown that the major, well-established structures of the
complete $\Delta$-resonance spectrum in fact can satisfactorily be
reproduced by proper choices of the confinement potential alone, {\it
i.e.} without any need of an additional residual interaction in this
flavor sector. This is quite in the spirit of our model which uses the
instanton-induced interaction as residual force.  Moreover, it is
worth to emphasize once more the economical simplicity of our model:
with the spinorial Dirac structure of the confinement kernel
fixed, the bulk of all structures is parameterized by three parameters
only!\\ However, some puzzling structures of the $\Delta$-spectrum,
such as the low-lying resonance $\Delta\frac{3}{2}^+(1600,\mbox{***})$
in the positive $2\hbar\omega$ band, as well as the states
$\Delta\frac{1}{2}^-(1900,\mbox{**})$, $\Delta\frac{3}{2}^-(1940, *)$
and $\Delta\frac{5}{2}^-(1930,\mbox{***})$ assigned to the negative
parity $3\hbar\omega$ band do not fit in. Despite the three-star
rating of $\Delta\frac{3}{2}^+(1600,\mbox{***})$ and
$\Delta\frac{5}{2}^-(1930,\mbox{***})$, we have the impression that
the currently available experimental evidence for these structures
from $\pi N$-phase shift analysis is not very convincing. Therefore a
verification of the positions of these resonances in reactions
complementary to $\pi-N$ scattering such as the electro-production of
mesons off nucleons is highly desirable to decide if the lack of these
structures in our models (and more generally in all other quark models)
really reflects a serious discrepancy.\\ Unfortunately, the present
data basis of the other states, which also is extracted almost
entirely from partial-wave analyses of older $\pi N\rightarrow \pi N$
total, elastic and charge exchange scattering data, is partly of
rather limited quality and hence does not allow to favor one of the
models at this stage. Hence, both models ${\cal A}$ and
${\cal B}$ are of almost the same quality in describing the present
experimental situation of the $\Delta$-spectrum.

\section{The hyperfine structure of the light ground-state baryons}
\label{sec:ground_hyper}
In the foregoing section, we focused on the role of the confinement
force alone.  We considered the $\Delta$-spectrum which is not
influenced by 't~Hooft's force and thus we could fix in each model the
three parameters $a$, $b$ and $m_n$ to get a satisfactory description
of the whole $\Delta$-spectrum.\\ In order to fix models
${\cal A}$ and ${\cal B}$ for those strange baryons that are
likewise not influenced by 't~Hooft's force either, it remains to
determine the strange constituent quark mass $m_s$. This is done by
the positions of the other (strangeness $S^* \not = 0$) spin-$3/2$
decuplet ground-states, {\it i.e.} the hyperons $\Sigma^*(1385)$,
$\Xi^*(1583)$ and $\Omega(1672)$.  In this respect, the heavier
strange quark mass $m_s>m_n$ then leads to the pattern of
approximately equally spaced distances between decuplet ground-states
differing by $\Delta S^* = 1$. Then all parameters in table
\ref{tab:ModelParam} apart from those of the 't~Hooft interaction are
fixed for both models.  This is a proper starting point for analyzing
the effect of 't~Hooft's instanton-induced interaction on the
remaining states, when turning on the 't~Hooft couplings $g_{nn}>0$
and $g_{ns}>0$.\\

We will start this discussion with the spin-$1/2$ octet ground-states
$N(939)$, $\Lambda(1116)$, $\Sigma(1193)$ and $\Xi(1318)$: In this respect,
the first prominent feature of the baryon spectrum the instanton-induced
interaction has to account for is generating the hyperfine structure of the
ground-states. In the present approach this means the lowering of the octet
ground-states relative to the unaffected decuplet ground-states, in order to
realize the mass differences $\Delta-N$, $\Sigma^*-\Sigma$,
$\Xi^*-\Xi$ and $\Sigma-\Lambda$. As in non-relativistic quark models, the
confinement potentials of models ${\cal A}$ and ${\cal B}$ cannot
describe this splitting and, without a residual interaction, the octet and
decuplet states with the same strangeness content are nearly degenerate.\\ In
earlier (mostly non-relativistic) attempts, as initially suggested by De
Rujula, Georgy and Glashow \cite{DGG75} and subsequently applied by Isgur {\it
  et al.} and others \cite{IsKa78,IsKa79,CaIs86}, this structure was explained
quantitatively by the short-range spin-spin hyperfine part (Fermi contact
term) of the Fermi-Breit interaction due to perturbative one-gluon-exchange.
Despite the success in describing the ground-state hyperfine structure by this
spin-dependent contact interaction, there remain a lot of objections, for
instance the spin-orbit problem connected with additional strong spin-orbit
forces of the Fermi-Breit interaction or inconsistently large values of the
strong coupling constant $\alpha_s$. As mentioned before this calls into
question the justification for applying perturbative one-gluon exchange at
least for light quarks. The most convincing argument against one-gluon
exchange and in favor of the non-perturbative and explicitly flavor dependent
instanton-induced interaction is that the latter provides a natural solution
of the $\pi-\eta-\eta'$ puzzle, as well as an explanation of the $\eta-\eta'$
mixing in the mesonic sector, see for instance \cite{MuRe94}. In leading order the OGE,
however, is flavor independent and thus yields degenerate $\pi$ and $\eta$
mesons in clear contradiction to experiment.  Moreover, the instanton force
does not have a (phenomenologically unwanted) large spin-orbit part.\\ 
Ground-state baryon mass splittings with the alternative instanton-induced
interactions alone have been first explored in a simple model by Shuryak and
Rosner \cite{ShRo89}.  A first, more extensive study of the baryon (and meson)
spectra in the framework of a non-relativistic quark model\footnote{We should
  mention here that this model is regained just in the non-relativistic limit
  of our fully covariant models ${\cal A}$ and ${\cal B}$.} using a
string-based confining interaction and instanton-induced interactions was made
by Blask {\it et al.} \cite{BBHMP90,Met92}. Both attempts have shown that
't~Hooft's interaction (which explains the $\rho-\pi$, $K^*-K$ and
$\pi-\eta-\eta'$ mass splittings in the mesonic sector), also provides an
appropriate description of the spin splittings in the octet and decuplet which
is at least as good as that from the short-range spin-spin hyperfine part of
the Fermi-Breit interaction.

\subsection{Naive considerations}
\begin{figure}[!h]
  \begin{center}
    \epsfig{file={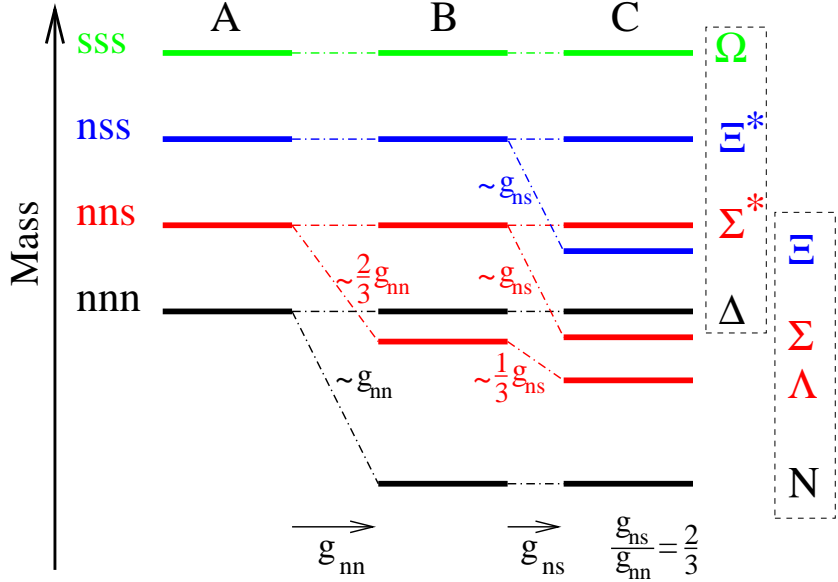},width=110mm}
  \end{center}
\caption{Schematic mass spectra of the baryon ground-states (due to a naive
  non-relativistic first order perturbation theory, see text)
  with (A) no residual interaction, (B) the pairing force between non-strange
  quarks and (C) in addition the pairing force between a non-strange and a
  strange quark.}
\label{fig:scemeGoundSplit}
\end{figure}

For the further discussion it is instructive to illustrate first in a
simple (naive) non-relativistic first order perturbative calculation,
how the octet-decuplet mass splittings is generated in
principle. Therefore, consider the non-relativistic limit of the
(unregularized) 't~Hooft interaction (in lowest order
$(\frac{p}{m})^0$ of a $(\frac{p}{m})$-expansion) leading for color
anti-triplet quark pairs to a spin-flavor dependent pairing force
\begin{equation}
V^{\rm NR}_{\rm 't~Hooft} ({\bf x_1 -x_2})
:=
-4\; 
{\cal P}^{\cal S}_{S_{12}=0}\tens
\left( g_{nn}\;{\cal P}^{\cal F}_{\cal A}(nn) 
     + g_{ns}\;{\cal P}^{\cal F}_{\cal A}(ns)
\right)
\delta^{(3)}({\bf x_1 -x_2}),
\end{equation}
where ${\cal P}^{\cal S}_{S_{12}=0} := \frac{1}{4}(\Id\tens\Id -
{\bfgrk\sigma}\!\cdot\!\tens\; {\bfgrk\sigma})$ is the projector on
quark pairs with trivial spin $S_{12}=0$, and
${\cal P}^{\cal F}_{\cal A}(nn)$ and
${\cal P}^{\cal F}_{\cal A}(ns)$ denote the flavor projectors on
antisymmetric non-strange ($nn$) and non-strange-strange ($ns$) quark pairs,
respectively. For simplicity we assume the normalized wave functions $\ket
{B_{\bf 8}}$ of all octet ground-states $B_{\bf 8} = N$, $\Lambda$, $\Sigma$
and $\Xi$ with total spin and parity $J^\pi=\frac{1}{2}^+$ to be given
approximately by the $S_3$-invariant $\bf 56$-plet states
\begin{equation}
\ket {B_{\bf 8}} 
\;\simeq\;
\ket{\psi^{L=0\;+}_{\cal S}}\tens
 \frac{1}{\sqrt 2}\bigg[
\ket {\chi^\frac{1}{2}_{{\cal M}_{\cal S}}}
\tens
\ket {\phi^{B_{\bf 8}}_{{\cal M}_{\cal S}}}
+
\ket {\chi^\frac{1}{2}_{{\cal M}_{\cal A}}}
\tens
\ket {\phi^{B_{\bf 8}}_{{\cal M}_{\cal A}}}
\bigg],
\end{equation} 
where the totally symmetric S-wave ground-state function
$\ket{\psi^{L=0\;+}_{\cal S}}$ in coordinate space is assumed to be the
same for all states, {\it i.e.}  distortions of the wave functions due to the
heavier strange quark are ignored. $\ket
{\chi^{1/2}_{\cal M_{{\cal S}/{\cal A}}}}$ denotes the mixed
symmetric/antisymmetric spin-$1/2$ wave function and $\ket
{\phi^{B_{\bf 8}}_{\cal M_{{\cal S}/{\cal A}}}}$ the mixed
symmetric/antisymmetric flavor octet wave functions. 
Accordingly, we assume the wave functions $\ket {B_{\bf
    10}}$ of all decuplet ground-states $B_{\bf 10} = \Delta$, $\Sigma^*$,
$\Xi^*$ and $\Omega$ with total spin and parity $J^\pi=\frac{3}{2}^+$
to be given approximately by the $S_3$-invariant $\bf 56$-plet states
\begin{equation}
\ket {B_{\bf 10}} 
\;\simeq\;
\ket{\psi^{L=0\;+}_{\cal S}}\tens
\ket {\chi^\frac{3}{2}_{\cal S}}
\tens
\ket {\phi^{B_{\bf 10}}_{\cal S}},
\end{equation} 
where $\chi^{3/2}_{\cal S}$ denotes the totally symmetric
spin-$3/2$ wave function and $\ket {\phi^{B_{\bf
10}}_{\cal S}}$ the totally symmetric flavor decuplet wave
functions. Without 't~Hooft's force, the octet and decuplet states
with the same strangeness content are degenerate.  Due to its
spin-flavor projector structure, the pairing force causes no shifts of
the masses of decuplet baryons:
\begin{equation}
\bra{\Delta} \;V^{\rm NR}_{\rm 't~Hooft}\tens\Id\;\ket{\Delta} 
=\bra{\Sigma^*} \;V^{\rm NR}_{\rm 't~Hooft}\tens\Id\;\ket{\Sigma^*} 
=\bra{\Xi^*} \;V^{\rm NR}_{\rm 't~Hooft}\tens\Id\;\ket{\Xi^*} 
=\bra{\Omega} \;V^{\rm NR}_{\rm 't~Hooft}\tens\Id\;\ket{\Omega} 
=
0,  
\end{equation}
but it shifts the octet states downward as they contain spin-singlet, S-wave, flavor
antisymmetric quark pairs ('scalar diquarks'). In this simplified
picture a first order perturbative calculation yields the following
decuplet-octet mass splittings:
\begin{equation}
\label{DecOc_split}
\begin{array}{rcccl}
M_\Delta-M_N\; (\simeq 300 \textrm{ MeV})
&\simeq&  
-3\; \bra {N} \;V^{\rm NR}_{\rm 't~Hooft}\tens\Id\;\ket {N} 
&=& 
C\;g_{nn},\\[1mm]
M_{\Sigma^*}-M_\Lambda\; (\simeq 270 \textrm{ MeV}) 
&\simeq& 
-3\;\bra {\Lambda}\; V^{\rm NR}_{\rm 't~Hooft}\tens\Id\;\ket {\Lambda} 
&=& 
C\;\left(\frac{2}{3}\; g_{nn} +
\frac{1}{3}\; g_{ns}\right),\\[1mm]
M_{\Sigma^*}-M_\Sigma\; (\simeq 200 \textrm{ MeV})
&\simeq&
-3\; \bra{\Sigma}\;  V^{\rm NR}_{\rm 't~Hooft}\tens\Id\;\ket {\Sigma} 
&=& C\;g_{ns},\\[1mm]
M_{\Xi^*}-M_\Xi\; (\simeq 200 \textrm{ MeV})
&\simeq&
- 3\; \bra {\Xi} \;V^{\rm NR}_{\rm 't~Hooft}\tens\Id\;\ket {\Xi} 
&=& C\;g_{ns},
\end{array}
\end{equation}
where $C := 6\;\bra{\psi^{L=0\;+}_{\cal S}} \delta^{(3)}({\bf x_1
-x_2}) \ket{\psi^{L=0\;+}_{\cal S}} > 0$ is a constant factor which
is the same for all mass splittings.  Here we have indicated roughly
the experimental values in parentheses. Let us discuss in this naive model the
implications of the particular flavor dependence for the baryon
ground-state spectrum as sketched schematically
in fig. \ref{fig:scemeGoundSplit}:
\begin{itemize}
\item
The equally spaced mass differences of about 150 MeV between the
decuplet states are due to their different strange quark content and
the explicit $SU(3)$ breaking caused by the heavier strange quark;
without residual interaction octet and decuplet states of the same
strangeness are degenerate (see A in fig.  \ref{fig:scemeGoundSplit}).
\item
The nucleon contains only a non-strange scalar diquark and hence the
downward mass shift relative to the $\Delta$ is proportional to the coupling
$g_{nn}>0$ (see B in fig.  \ref{fig:scemeGoundSplit}).
\item In a similar way the equally big downward mass shifts of $\Sigma$ and $\Xi$
  relative to $\Sigma^*$ and $\Xi^*$, respectively, are proportional to the
  strength $g_{ns}>0$, since both these octet states contain a scalar
  non-strange-strange diquark only (see C in fig. \ref{fig:scemeGoundSplit}).
\item
The $\Lambda$-hyperon, however, contains both types of diquarks, where
the scalar non-strange diquark content is twice that of the scalar
non-strange-strange diquark content. This implies:  concerning the non-strange
coupling $g_{nn}$ the mass shift of $\Lambda$ is a factor $2/3$
weaker than the nucleon mass shift and the lowering of the $\Lambda$
mass due to the non-strange-strange coupling is only $1/3$ of the
corresponding lowering of $\Sigma$ and $\Xi$, respectively (see B and C
in fig. \ref{fig:scemeGoundSplit}). In particular, we find
the $\Sigma-\Lambda$ mass splitting to be given by the difference of
the couplings $g_{nn}$ and $g_{ns}$:
\begin{equation}
\label{SiLa_split}
\begin{array}{rcl}
M_\Sigma-M_\Lambda\; (\simeq 70 \textrm{ MeV})
&\simeq& 
C\;\frac{2}{3}\left(g_{nn} - g_{ns}\right).
\end{array}
\end{equation}
\end{itemize}
Note that in the limit $g_{nn} = g_{ns}$ the 't~Hooft interaction
becomes flavor $SU(3)$ symmetric, such that the decuplet-octet mass
splittings are then all the same. In particular, the
$\Sigma-\Lambda$ mass difference then vanishes.  Consequently, the correct
description of the phenomenological value and sign of the
$\Sigma-\Lambda$ mass difference within this simplified model requires
the non-strange coupling $g_{nn}$ to be bigger than the
non-strange-strange coupling $g_{ns}$, thus implying a ratio
$g_{ns}/g_{nn}< 1$. Using eqs. (\ref{DecOc_split}) and
(\ref{SiLa_split}) the explicit flavor dependence of the pairing force
leads to the simple relations
\begin{equation}
\frac{M_\Sigma-M_\Lambda}{M_\Delta-M_N}=\frac{2}{3}\left(1-\frac{g_{ns}}{g_{nn}}\right)
\qquad
\textrm{and}
\qquad
\frac{M_{\Sigma^*}-M_\Sigma}{M_\Delta-M_N}
= 
\frac{M_{\Xi^*}-M_\Xi}{M_\Delta-M_N}
=
\frac{g_{ns}}{g_{nn}}
\end{equation}
between the $\Delta-N$ mass splitting and the splittings $\Sigma-\Lambda$,
$\Sigma^*-\Sigma$ and $\Xi^*-\Xi$, respectively. This implies that all three
relative mass differences have to be described by just one parameter, namely
the ratio $g_{ns}/g_{nn}<1$.  Relating the experimental estimates of the
$\Sigma-\Lambda$ splitting ($\simeq 70$ MeV) to the $\Delta-N$ mass difference
($\simeq 300$ MeV) due to the first equation yields a ratio of about
$g_{ns}/g_{nn} \simeq 2/3$. This is consistent with the same ratio
$g_{ns}/g_{nn} = 2/3$ that we get from the second relation by comparing the
unequal mass shifts $\Sigma^*-\Sigma$, $\Xi^*-\Xi$ ($\simeq 200$ MeV) and
$\Delta-N$ ($\simeq 300$ MeV). Consequently the instanton-induced, flavor-dependent pairing force indeed  provides a consistent explanation of the
ground-state pattern even in this crude naive model. As illustrated
schematically in the last column C of  fig. \ref{fig:scemeGoundSplit} the
choice of the ratio $g_{ns}/g_{nn} = 2/3$  can indeed account for the correct
level ordering of octet and decuplet ground-states in accordance with the
experimental findings: 
$M_N < M_\Lambda < M_\Sigma < M_\Delta < M_\Xi < M_{\Sigma^*} < M_{\Xi^*} < M_\Omega$.\\ 
Finally we want to add a
remark here concerning the ratio $g_{ns}/g_{ns} < 1$. The fact that the
instanton-induced attraction is weaker between a strange and a non-strange
quark than between two non-strange quarks, is exactly what one anticipates from the
normal ordering of the original three-flavor 't~Hooft vertex (see sect.
\ref{sec:eff_tHo_Lag}): The effective coupling $g_{nn}$ results from a Wick contraction
of the strange quark fields with the heavier effective strange quark mass,
whereas $g_{ns}$ is obtained by a contraction of a non-strange quark field with
the lighter effective non-strange quark mass, thus in fact
yielding $g_{ns} < g_{nn}$ as required by the phenomenology of the hyperfine
structure (see eq. (\ref{geff_rep})).
\subsection{Generating the hyperfine splittings in models ${\cal A}$ and ${\cal B}$}
\begin{figure}[!h]
  \begin{center}
    \epsfig{file={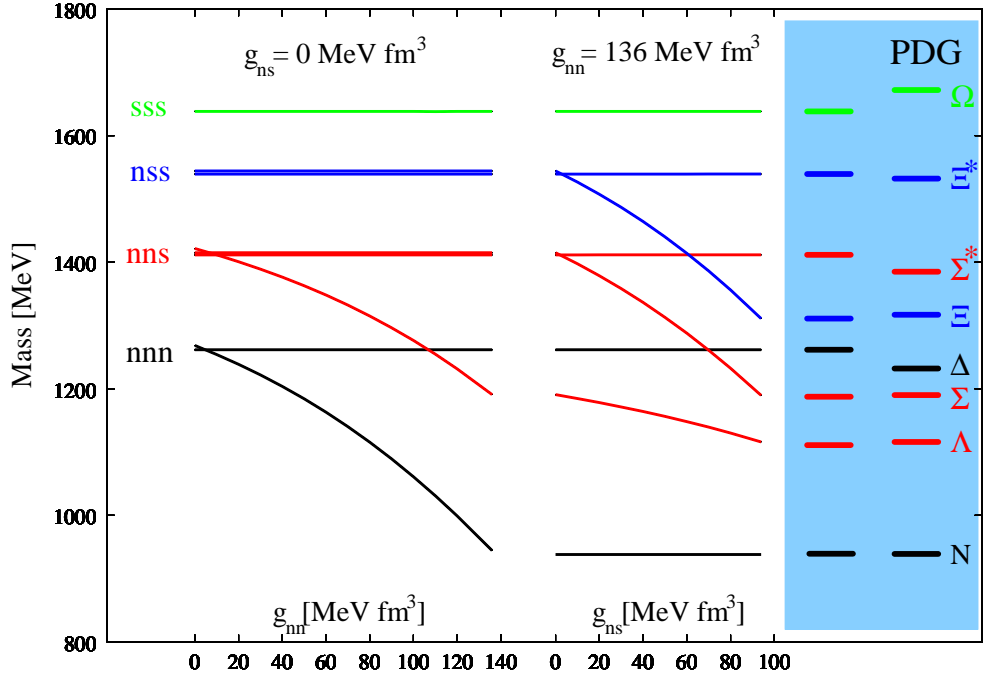},width=130mm}
  \end{center}
\caption{Generating the hyperfine structure of the octet and decuplet
  ground-state baryons by the instanton force in model ${\cal A}$. The last
  column headed with 'PDG' shows for comparison the experimental ground-state
  positions \cite{PDG00}. For a detailed explanation see text.}
\label{fig:groundVarM2}
\end{figure}

\begin{figure}[!h]
  \begin{center}
    \epsfig{file={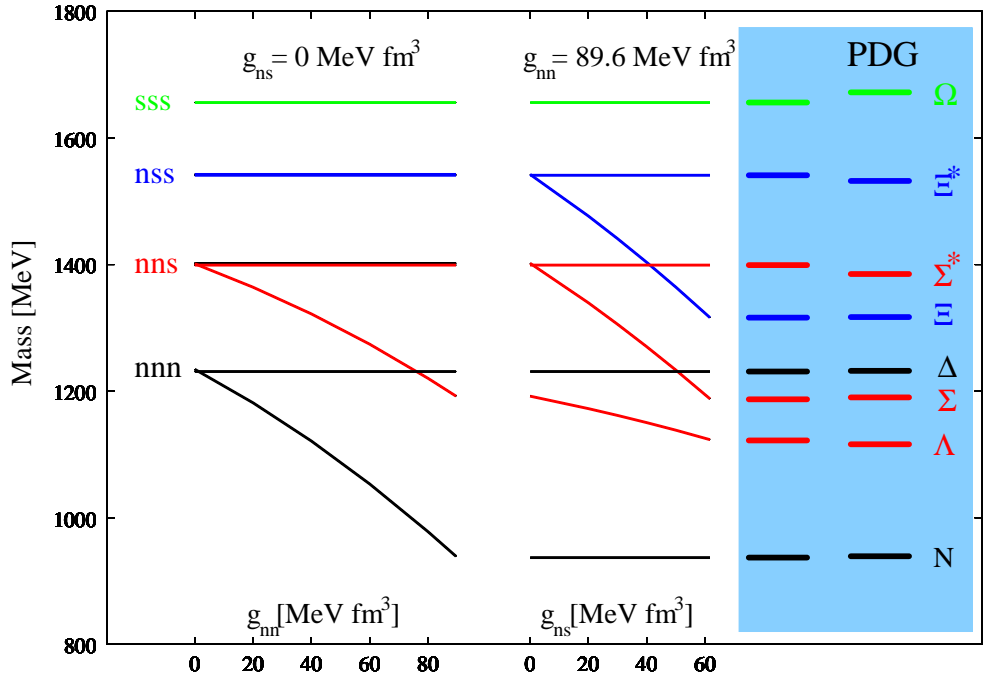},width=130mm}
  \end{center}
\caption{Generation of  the hyperfine structure of the octet and decuplet
  ground-state baryons by the instanton force in model ${\cal B}$. In the
  last column headed with 'PDG' the experimental ground-state positions
  \cite{PDG00} are shown for comparison . For a detailed explanation see
  text.}
\label{fig:groundVarM1}
\end{figure}

Now let us examine the instanton-induced hyperfine
splittings with the full relativistic dynamics of our covariant Salpeter
models. The influence
of 't~Hooft's force on the ground-state baryons in model
${\cal A}$ and model ${\cal B}$ is shown in figs.
\ref{fig:groundVarM2} and \ref{fig:groundVarM1}, respectively.

\begin{figure}[!h]
  \begin{center}
    \epsfig{file={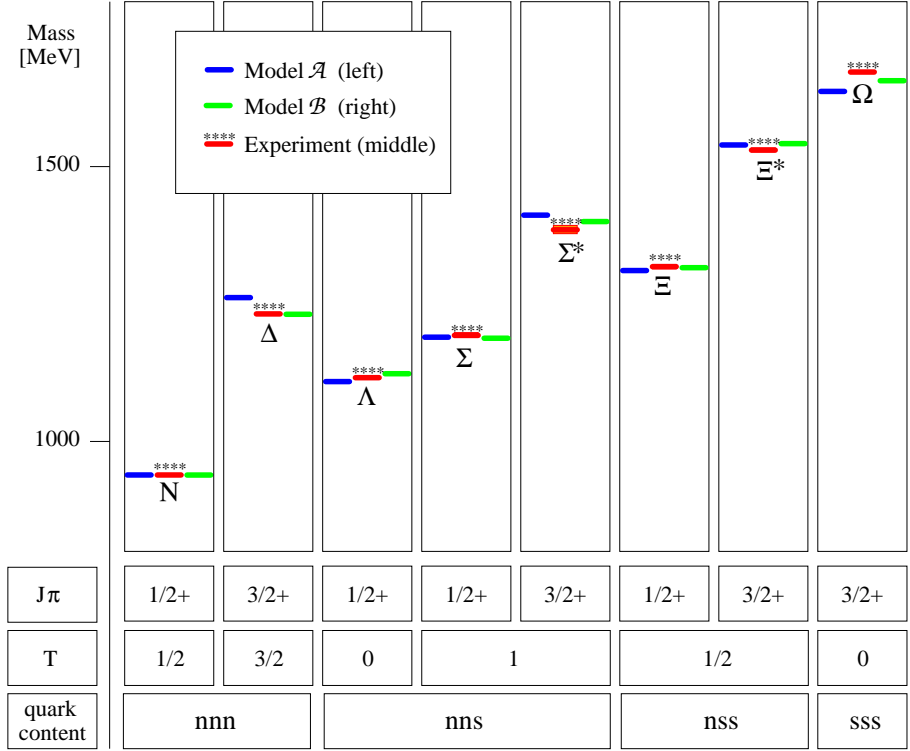},width=120mm}
  \end{center}
\caption{The resulting spin $1/2$ octet and spin $3/2$ decuplet ground-state baryons calculated in model
  ${\cal A}$ (on the left in each column) and model ${\cal B}$ (on the
  right in each column) in comparison to experimental
  findings \cite{PDG00} (middle of each column). The explicit mass values are given in table \ref{tab:groundSpec}.}
\label{fig:groundSpec}
\end{figure}

Without any residual interaction, {\it i.e.} $g_{nn}=0$ and $g_{ns}=0$, all
ground states are bound from the flavor independent three-body confinement
potential only. Therefore, the octet ground-states indeed are nearly
degenerate to the corresponding decuplet ground-states with the same number of
strange quarks. Henceforth, the decuplet ground-states remain unaffected by
't~Hooft's force. The decuplet masses calculated in models ${\cal A}$ and
${\cal B}$ with the confinement parameters and quark masses given in table
\ref{tab:ModelParam} are explicitly shown in the upper part of table
\ref{tab:groundSpec}. As discussed in the previous section, the position of
the $\Delta$-resonance in model ${\cal A}$ turns out to be slightly too
high compared to the experimental value. The same also applies to $\Sigma^*$,
whereas $\Omega$ comes out to be roughly 30 MeV too low compared to the
observed position; consequently the mass gaps between states of different
strangeness content become slightly smaller with increasing strangeness.
Nevertheless the agreement with the experimentally observed positions is still
satisfactory; see also fig. \ref{fig:groundSpec}.  In model ${\cal B}$ the
position of the $\Delta$-resonance matches the experimental value exactly and
the deviations of the remaining decuplet states to the corresponding observed
positions are smaller than in model ${\cal A}$.  Hence, the approximate
equal mass gaps due to the flavor $SU(3)$ breaking are well described in model
${\cal B}$. Altogether the positions of the decuplet ground-states are
slightly better described in model ${\cal B}$ than in model ${\cal A}$;
see also fig. \ref{fig:groundSpec} for a direct comparison of both models and
experimental data.\\ Now let us investigate the effect of 't~Hooft's force on
the octet ground-states in both models. First we consider the dependence on
the non-strange coupling $g_{nn}>0$ with the non-strange-strange coupling
still kept fixed at $g_{ns}=0$. This is shown by the mass curves in the left
part of figs. \ref{fig:groundVarM2} and \ref{fig:groundVarM1}. Due to the
non-strange scalar diquark content of the nucleon, increasing the non-strange
coupling $g_{nn}$ lowers the nucleon mass $M_N$ to its experimental value
$M_N=939$ MeV, thus generating the $\Delta-N$ mass splitting; this fixes the
coupling $g_{nn}$ in model ${\cal A}$ and model ${\cal B}$ to the values
given in table \ref{tab:ModelParam}.  At the same time, the coupling of quark
pairs to trivial spin and flavor leads also within the $\Lambda$ to an
attractive correlation of a scalar pair of non-strange quarks thus yielding a
lowering of the $\Lambda$-mass, which is not yet sufficient to match the
corresponding experimental resonance position. The remaining mass shift of
about 70 MeV in both models is expected just to come from the additional
attractive correlation of the scalar non-strange-strange diquark.  Note that
in model ${\cal A}$ as well as in model ${\cal B}$ the mass shift of
$\Lambda$ is indeed about $70\%$ weaker than that of the nucleon, in
qualitative agreement with the factor $2/3$ resulting from the static $SU(6)$
spin-flavor matrix element of 't~Hooft's force as discussed previously in the
naive model (compare also with column B of fig. \ref{fig:scemeGoundSplit}).
Of course, the two other octet states $\Sigma$ and $\Xi$ remained unaffected
so far, as they do not contain quark pairs with trivial isospin
($nn$-diquark).\\ We now turn to the $g_{ns}$ dependence of the ground-state
spectrum as shown by the mass curves in the right part of figs.
\ref{fig:groundVarM2} and \ref{fig:groundVarM1}.  With $g_{nn}$ fixed to
reproduce the $\Delta-N$ splitting in both models, an increasing coupling
$g_{ns}$ to the scalar, flavor-antisymmetric non-strange-strange diquarks has
to lower the calculated masses of the hyperons $\Lambda$, $\Sigma$ and $\Xi$
in such a way that they simultaneously fit the corresponding experimentally
observed resonance positions, thus generating the correct $\Sigma-\Lambda$,
$\Sigma^*-\Sigma$ and $\Xi^*-\Xi$ mass splittings.  This fixes the value of
$g_{ns}$ in both models.  As shown in figs. \ref{fig:groundVarM1} and
\ref{fig:groundVarM2}, the dependence of the $\Lambda$, $\Sigma$ and $\Xi$
masses on the non-strange-strange coupling $g_{ns}$ exhibits again a
qualitatively similar behavior as in the naive model discussed previously (see
for comparison column C of fig. \ref{fig:scemeGoundSplit}): with increasing
coupling $g_{ns}$, the masses of $\Sigma$ and $\Xi$ show almost the same
downward shift, whereas the additional weaker mass shift of $\Lambda$ amounts
only to $35\;\%$ and $32\;\%$ in model ${\cal A}$ and model ${\cal B}$,
respectively. This approximately agrees with the factor $1/3$ of the previous
naive estimate.  Finally, with the fixed values for $g_{ns}$ given in table
\ref{tab:ModelParam}, both models can indeed account remarkably well for the
hyperon mass splittings and provide their mass positions in very good
agreement with experiment.  The resulting masses for the octet states are
summarized in the lower part of table \ref{tab:groundSpec} and can be compared
with the corresponding
experimental values of the Particle Data Group \cite{PDG00}.  \\
Altogether, both models ${\cal A}$ and ${\cal B}$ are of the same good
quality in describing the experimentally observed positions of the octet
ground-state baryons $N$, $\Lambda$, $\Sigma$ and $\Xi$. Also the level
ordering of the octet and decuplet states is reproduced in accordance with
experimental findings, as displayed in the last two columns of figs.
\ref{fig:groundVarM1} and \ref{fig:groundVarM2}. For a direct comparison of
the results of models ${\cal A}$ and ${\cal B}$ (and the experimental
data) see fig. \ref{fig:groundSpec}.
\begin{table}[!h]
\center
\begin{tabular}{ccccc}
\hline
Ground-state&${J^\pi}$       & Exp. Mass [MeV]& Model ground-state    &  Model ground-state\\
            &                & \cite{PDG00}& in model ${\cal A}$& in model ${\cal B}$\\
\hline
$\Delta$&${\frac{3}{2}^+}$  & 1232& $\MSD(3,+,1,1260)$ & $\MSD(3,+,1,1231)$\\
$\Sigma^*$&${\frac{3}{2}^+}$& 1385& $\MSS(3,+,1,1411)$ & $\MSS(3,+,1,1399)$\\
$\Xi^*$&${\frac{3}{2}^+}$   & 1530& $\MSX(3,+,1,1539)$ & $\MSX(3,+,1,1541)$\\
$\Omega$&${\frac{3}{2}^+}$  & 1672& $\MSO(3,+,1,1636)$ & $\MSO(3,+,1,1656)$\\
\hline
$N$&${\frac{1}{2}^+}$       &  939& $\!\MSN(1,+,1,939)$  & $\!\MSN(1,+,1,939)$\\
$\Lambda$&${\frac{1}{2}^+}$ & 1116& $\MSL(1,+,1,1108)$ & $\MSL(1,+,1,1123)$\\
$\Sigma$&${\frac{1}{2}^+}$  & 1193& $\MSS(1,+,1,1190)$ & $\MSS(1,+,1,1188)$\\
$\Xi$&${\frac{1}{2}^+}$     & 1318& $\MSX(1,+,1,1310)$ & $\MSX(1,+,1,1316)$\\
\hline
\end{tabular}
\caption{Calculated decuplet (upper part) and octet (lower part) ground-state baryons in the models ${\cal A}$
  and ${\cal B}$ compared to experimental values \cite{PDG00}. Notation for
  model states as in table \ref{tab:1hwband}. 
  For a graphical presentation see fig. \ref{fig:groundSpec}.}
\label{tab:groundSpec}
\end{table}

The values of the hyperfine splittings $\Delta - N$, $\Sigma^*-\Lambda$,
$\Sigma^* -\Sigma$, $\Xi^* - \Xi$ and $\Sigma -\Lambda$ calculated in both
models are explicitly given in table \ref{tab:hyperSplit}.  
\begin{table}[!h]
\center
\begin{tabular}{cccc}
\hline
Hyperfine                 & Calculation    & Calculation    & Exp. [MeV]\\
splitting                 & of model ${\cal A}$ [MeV]& of model
                          ${\cal B}$ [MeV]\\
\hline
$\Delta - N$       & 321  & 292  &  293\\
$\Sigma^*-\Lambda$ & 303  & 276  &  269\\
$\Sigma^* -\Sigma$ & 221  & 211  &  192\\
$\Xi^* - \Xi$      & 229  & 225  &  212\\[2mm]
$\Sigma -\Lambda$  & 82   &  65  &  75\\
\hline
\end{tabular}
\caption{The hyperfine splittings between  decuplet and octet ground-state baryons in the models ${\cal A}$
  and ${\cal B}$ compared to experimental values.}
\label{tab:hyperSplit}
\end{table}
The absolute values
of the ground-state splittings can be nicely reproduced. Here the predictions
of model ${\cal B}$ are slightly better than of model ${\cal A}$, where
all splittings turn out to be slightly too big. The small discrepancies in
model ${\cal A}$ can be traced back to the decuplet states,
whose calculated absolute positions came out to be slightly too high in model, as discussed before.

Finally, we want to comment on the values of the 't~Hooft couplings $g_{nn}$
and $g_{ns}$ which have been adjusted to the experimental octet ground-state
positions.  The two model variants ${\cal A}$ and ${\cal B}$ mainly
differ in the form assumed for the spinorial Dirac structure of the linearly
rising part of the three-body confinement potential, or more precisely in the
way how the scalar and the time-like vector part are combined. In particular,
both versions yield the same non-relativistic limit.  Although we are using
for both models the same value $\lambda = 0.4$ fm for the effective range of
the regularized instanton-induced interaction, the absolute values for the
't~Hooft couplings $g_{nn}$ and $g_{ns}$ differ significantly: the couplings
in model ${\cal A}$ are about $50\%$ larger than in model ${\cal B}$
indicating a dependence of the action of 't~Hooft's force on the structure of
the confinement potential.  Nevertheless, the ratio of the two couplings
$g_{nn}$ and $g_{ns}$ turns out to be the same. Both models yield the ratio
$g_{ns}/g_{nn} = 0.69$ which again is in qualitative accordance with the crude
phenomenological estimate $g_{ns}/g_{nn}=2/3$ from the previous naive
consideration. Moreover, we should give a comment concerning the special
choice for the effective range $\lambda = 0.4$ fm of the regularized
instanton-induced four-fermion interaction. We note that there is some freedom
to choose the value for $\lambda$.  While a choice smaller than $\sim 0.3$ fm
spoils the agreement with experiment (and moreover is also much more expensive
to handle numerically), for arbitrary values $\lambda > 0.3$ fm the couplings
$g_{nn}$ and $g_{ns}$ can always be adjusted such that an equally good result
for the hyperfine ground-splittings can be achieved.  Recall from sect.
\ref{sec:eff_tHo_Lag} that 't~Hooft's force describes the interaction of quark
pairs via instantons up to a critical extension $\rho_c$ which is required to
cure the infrared problem. Thus, we expect the effective range $\lambda$ of
't~Hooft's force roughly to correspond to the critical instanton size
$\rho_c$. The value for $\rho_c$ (and thus also $\lambda$) should not be
larger than $\sim 0.5$ fm to be still within the scope of the two-loop
approximation of the $\rho$-dependent running coupling constant
(\ref{thooftcoupling}). Hence, a reasonable value $\lambda$ is somewhere in
between 0.3 and 0.5 fm and we choose $\lambda= 0.4$ fm.

\subsection{Summary for the  hyperfine structure of ground-states}
In summary, the use of 't~Hooft's residual instanton-induced interaction
within our covariant framework provides in both confinement models
${\cal A}$ and ${\cal B}$ (with parameters being fixed to reproduce the
linear $\Delta$-Regge trajectory) a good and consistent description of the
decuplet and octet ground-state baryons (see fig. \ref{fig:groundSpec} for a
comprehensive presentation of our results). It turns out that also with the
fully relativistic dynamics the $\Delta-N$, $\Sigma^*-\Sigma-\Lambda$ and
$\Xi^*-\Xi$ hyperfine splittings arise due to the attractive correlation of
quark pairs with trivial spin and antisymmetric flavor to 'scalar diquarks' in
qualitatively the same manner as one already expects from a naive static,
non-relativistic picture. The phenomenology of the hyperfine splittings
requires a ratio $g_{ns}/g_{ns}<1$ of the 't~Hooft couplings which is
compatible with the expectation from instanton physics. Our
result is at least as satisfactory as other attempts (see {\it e.g.}
ref.~\cite{CaIs86,CaRo00} and the discussion in appendix \ref{sec:OGE}) that explain the
hyperfine structure by the short range spin-spin hyperfine
part of the one-gluon-exchange.\\

Now all parameters of our model variants ${\cal A}$ and ${\cal B}$ are fixed
to the values given in table \ref{tab:ModelParam}. Calculations of all other
resonance masses are parameter-free and thus constitute true predictions.  In
particular, we can investigate next, to what extent the instanton force can
also account for the features of the non-strange and strange {\it excited}
baryons.  Therefore, we now turn to a detailed discussion of the complete
excited $N$-baryon spectrum and especially analyze how instanton effects shape
its characteristic structures. As already mentioned, the predictions for the
strange excited baryons will be discussed in a separate paper \cite{Loe01c}.

\section{The Nucleon spectrum}
\label{sec:Nuc}
We now turn to the investigation of the complete spectrum of nucleon
resonances with isospin $T=\frac{1}{2}$ and strangeness $S^*=0$, where now in
addition to the confinement force also the influence of the instanton-induced
interaction in general plays an essential role even for the excited states.
\subsection{Remarks -- Implications of 't~Hooft's force and the experimental situation}
Let us begin this discussion with some general remarks concerning the action
of 't~Hooft's force in the nucleon sector.  The influence of the instanton
force on the nucleon states is related to the content of quark pairs with
trivial spin and isospin. The positive and negative energy components of the
Salpeter amplitude ${\Phi_{J^\pi}^N}$ describing an excited flavor-octet
nucleon state with spin and parity $J^\pi$ are obtained by the embedding
map (see ref. \cite{Loe01a})
\begin{equation}
\label{embedSalp_N}
{\Phi_{J^\pi}^N} \;=\; T^{+++} {\varphi_{J^\pi}^N} \;+\; T^{---} {\varphi_{J^{-\pi}}^N}
\end{equation}
of totally $S_3$-symmetric Pauli spinors
$\varphi^N_{J^{\pi}}$ and $\varphi^N_{J^{-\pi}}$ which generally can be
decomposed into the following four different spin-flavor
$SU(6)$-configurations:
\begin{equation}
\label{NdecompPauli}
\ket{\varphi^N_{J^{\pm}}} 
=
\ket{N\;J^\pm,\; ^2 8[56]}
\;+\;
\ket{N\;J^\pm,\;^2 8[70]}
\;+\;
\ket{N\;J^\pm,\;^4 8[70]}
\;+\;
\ket{N\;J^\pm,\;^2 8[20]},
\end{equation}
with
\begin{equation}
\label{NconfigPauli}
\begin{array}{rcl}
\ket{N\;J^\pm,\; ^2 8[56]} &:=& 
\sum\limits_{L} \Bigg[ 
\ket{\psi^{L\;\pm}_{\cal S}}\tens
\frac{1}{\sqrt 2}\bigg(
\ket {\chi^\frac{1}{2}_{{\cal M}_{\cal A}}}
\tens
\ket {\phi^N_{{\cal M}_{\cal A}}}
+
\ket {\chi^\frac{1}{2}_{{\cal M}_{\cal S}}}
\tens
\ket {\phi^N_{{\cal M}_{\cal S}}}
\bigg)\Bigg]^J,\\[3mm]
\ket{N\;J^\pm,\;^2 8[70]} &:=& 
\sum\limits_{L} \Bigg[\phantom{+\;}\frac{1}{2}\; 
\ket{\psi^{L\;\pm}_{{\cal M}_{\cal A}}}\tens 
\bigg(
\ket {\chi^\frac{1}{2}_{{\cal M}_{\cal A}}}
\tens
\ket {\phi^N_{{\cal M}_{\cal S}}}
+
\ket {\chi^\frac{1}{2}_{{\cal M}_{\cal S}}}
\tens
\ket {\phi^N_{{\cal M}_{\cal A}}}
\bigg)\\[-2mm]
&& 
\phantom{\sum\limits_{L} \Bigg[}+
\frac{1}{2}\;
\ket{\psi^{L\;\pm}_{{\cal M}_{\cal S}}}\tens
\bigg(
\ket {\chi^\frac{1}{2}_{{\cal M}_{\cal A}}}
\tens
\ket {\phi^N_{{\cal M}_{\cal A}}}
-
\ket {\chi^\frac{1}{2}_{{\cal M}_{\cal S}}}
\tens
\ket {\phi^N_{{\cal M}_{\cal S}}}\bigg)\Bigg]^J,\\[3mm]
\ket{N\;J^\pm,\;^4 8[70]} &:=& 
\sum\limits_{L} \Bigg[\frac{1}{\sqrt 2} 
\bigg(
\ket{\psi^{L\;\pm}_{{\cal M}_{\cal A}}}\tens 
\ket{\chi^\frac{3}{2}_{\cal S}}\tens  
\ket{\phi^N_{{\cal M}_{\cal A}}}
-
\ket{\psi^{L\;\pm}_{{\cal M}_{\cal S}}}\tens 
\ket{\chi^\frac{3}{2}_{\cal S}}\tens  
\ket{\phi^N_{{\cal M}_{\cal S}}}
\bigg)\Bigg]^J,\\[3mm]
\ket{N\;J^\pm,\;^2 8[20]} &:=&  
\sum\limits_{L} \Bigg[\ket{\psi^{L\;\pm}_{\cal A}}\tens
\frac{1}{\sqrt 2}\bigg(
\ket {\chi^\frac{1}{2}_{{\cal M}_{\cal A}}}
\tens
\ket {\phi^N_{{\cal M}_{\cal S}}}
-
\ket {\chi^\frac{1}{2}_{{\cal M}_{\cal S}}}
\tens
\ket {\phi^N_{{\cal M}_{\cal A}}}
\bigg)\Bigg]^J.
\end{array}
\end{equation}
Here $\psi^{L\;\pm}_{R_L}$, $\chi^S_{R_S}$ and $\phi^N_{R_F}$ are the spatial,
spin and flavor wave functions with definite $S_3$-symmetries $R_L, R_S,
R_F\in\{{\cal S},{\cal M}_{\cal S}, {\cal M}_{\cal A},
{\cal A}\}$. The sum runs over possible orbital angular momenta $L$ that
can be coupled with the internal spin $S$ to the total spin $J$ as denoted by
the brackets $[\ldots]^J$. Due to the strong selection rule of the
instanton-induced interaction, its action on the different states of the
nucleon spectrum is qualitatively understandable from a simplified picture, in
which one disregards the negative energy component and the relativistic
effects from the embedding map of the Salpeter amplitudes (corresponding to
the non-relativistic limit). Then we expect all those resonances, which in the pure
confinement case ($g_{nn}=0$) are dominantly $^4 8[70]$ and $^2 8[20]$, to be
hardly influenced by 't~Hooft's force.  Consequently, these resonances should
be determined mainly by the confining kernel alone, similar to the resonances
in the $\Delta$-spectrum.  For $^4 8[70]$ states in eq. (\ref{NconfigPauli})
this selection rule is apparent from the total symmetry of the spin-function
$\chi^{3/2}_{\cal S}$. The $^2 8[20]$ states possess a totally
antisymmetric spatial wave function, which is not affected by a point-like
interaction. Moreover its spin-flavor wave function, which decomposes into
mixed symmetric spin functions $\chi^{1/2}_{{\cal M}_{\cal S}}$ and
mixed antisymmetric flavor functions $\phi^N_{{\cal M}_{\cal A}}$ and
vice versa, never contains quark pairs with trivial spin and trivial flavor
simultaneously.  However, resonances which are dominantly $^2 8[56]$, like
{\it e.g.}  the nucleon ground-state, are expected to exhibit quite strong
effects induced by the instanton force.  Furthermore, also the resonances
which mainly consist of $^2 8[70]$ states should be affected by the 't~Hooft
interaction.  Finally, it should be noted that 't~Hooft's force in general
mixes $^2 8[56]$ and $^2 8[70]$ configurations.  From these simple
non-relativistic considerations we thus expect that the majority of states is
described by the confinement force alone, while 't~Hooft's force acts in a
selective manner on a particular limited set of states.

However, one has to be careful with such naive non-relativistic
considerations.  Due to our fully covariant Salpeter approach,
relativistic effects might become essential in two ways:
\begin{enumerate}
\item From the outset (without instanton force), the initial mixing of
the four different spin-flavor $SU(6)$ configurations
(\ref{NconfigPauli}) in (\ref{NdecompPauli}), especially for the
excited states, can strongly depend on the relativistic (spin-orbit)
effects that emerge from the chosen confinement Dirac structure
in connection with the embedding map (\ref{embedSalp_N}) of the
Salpeter amplitudes. In particular, distinct Dirac
structures cause different initial intra-band splittings and level
orderings.  Consequently, the influence of the instanton-induced
interaction on the excited nucleon spectrum may be different in
our two model variants.  As we will see in the
following discussion, models ${\cal A}$ and ${\cal B}$
indeed show substantial differences concerning the effects of 't~Hooft's force
in the excited spectra, in contrast to the description of the
ground-state hyperfine structure, where
't~Hooft's force in both models works equally well.  This dependence
of the effects of the instanton force on the confinement mechanism in
fact is a purely relativistic effect, since the expressions for the
confinement kernels of both models lead to the same non-relativistic
limit and thus in this limit also to the same results.
\item Furthermore, the fully relativistic treatment of 't~Hooft's
force within our covariant Salpeter approach leads already by
itself to differences in the effects of this interaction as compared
to a non-relativistic treatment.  On the one hand the action on the
Pauli amplitudes (\ref{NconfigPauli}) is modified due to the
embedding map for the Salpeter amplitudes and on the other hand
the fully relativistic version of the instanton-induced interaction
causes effects that are {\it a priori} absent in the non-relativistic
limit.  In particular, it also has repulsive components, unlike its
non-relativistic version. In this respect, we should remark here that
the projector structure of 't~Hooft's interaction kernel actually
decomposes into two different parts. On the one hand there is the
projector onto {\it scalar diquarks} with trivial spin and
antisymmetric flavor. This is the dominant attractive part, which
survives in the non-relativistic limit. It is responsible for
the strong attraction of diquarks in the octet ground-state baryons
leading to the correct octet-decuplet ground-state splittings as
discussed in the previous subsection. But, on the other hand, there is
also a projector onto the orthogonal {\it pseudo-scalar diquarks} with
trivial spin and antisymmetric flavor. In this channel the instanton
force is repulsive. This part of the instanton force completely
vanishes in the non-relativistic limit.
\end{enumerate}
These issues in fact emphasize the importance of describing baryons in a fully
relativistic framework and accordingly it is interesting to study to what
extent our covariant approach leads to improvements in the description of the
excited state spectrum as compared to its non-relativistic version described
earlier in \cite{BBHMP90,Met92}. Moreover, the different action of 't~Hooft's
force in combination with the two different confinement Dirac structures of
models ${\cal A}$ and ${\cal B}$ will offer an additional, indirect
criterion to decide which version
provides a more realistic confinement force.\\

In the following discussion we will present a detailed investigation of
instanton-induced effects on the excited spectrum of the nucleon resonances in
both models (subsect. \ref{subsec:studying_inst_eff}). In subsect.
\ref{subsec:disc_compl_N} we shall start by comparing the complete presently known
empirical $N^*$-spectrum with our final resulting resonance positions obtained
with the coupling $g_{nn}$ adjusted in the previous section to
reproduce the experimentally measured position $M_N=939$ MeV of the
nucleon ground state. In this respect, the most striking experimental features
of the nucleon spectrum, that a realistic quark model should account for, are
\cite{PDG00}:
\begin{itemize}
\item
The low position of four states in the positive-parity $2\hbar\omega$
band, which lie quite isolated from the other states of this shell
around 2000 MeV and which form a striking pattern: The most
prominent member of this structure is the well-established lowest
isoscalar/scalar excitation, the famous Roper resonance
$N\frac{1}{2}^+(1440,\mbox{****})$ that appears even below the first
excitations of the nucleon in the negative-parity $1\hbar\omega$
band. The puzzling low position of this particular resonance has been
extensively discussed in the literature as the so-called \textit{Roper
problem} \cite{IsKa79,GlRi96,Gl00,Kre00}.  Furthermore, there are three other well-established
states $N\frac{1}{2}^+(1710,\mbox{***})$, $N\frac{3}{2}^+(1720,\mbox{****})$ and
$N\frac{5}{2}^+(1680,\mbox{****})$ which are approximately degenerate at around 1700 MeV.
\item
The hyperfine structure of five observed three- and four-star
states assigned to the negative-parity $1\hbar\omega$ shell, {\it
i.e.} the mass splitting between the two groups of almost degenerate
states $N\frac{1}{2}^-(1650,\mbox{****})$ -- $N\frac{3}{2}^-(1700,\mbox{***})$ --
$N\frac{5}{2}^-(1675,\mbox{****})$ and $N\frac{1}{2}^-(1535,\mbox{****})$ --
$N\frac{3}{2}^-(1520,\mbox{***})$.
\item
The overlap of  alternating even- and odd-parity bands and
accordingly the striking appearance of approximate ''parity doublets'':
The overlapping negative-parity $1\hbar\omega$ and positive-parity $2\hbar\omega$
shells reveal the parity doublets
\begin{center}
\begin{tabular}{lcl}
$N\frac{1}{2}^+(1710,\mbox{***})$  &--& $N\frac{1}{2}^-(1650,\mbox{****})$,\\
$N\frac{3}{2}^+(1720,\mbox{****})$ &--& $N\frac{3}{2}^-(1700,\mbox{***})$,\\
$N\frac{5}{2}^+(1680,\mbox{****})$ &--& $N\frac{5}{2}^-(1675,\mbox{****})$.
\end{tabular}
\end{center}
In the higher mass region we find {\it e.g.} the approximate
doublets
\begin{center}
\begin{tabular}{lcl}
$N\frac{7}{2}^+(1990,\mbox{**})$   &--& $N\frac{7}{2}^-(2190,\mbox{****})$,\\
$N\frac{9}{2}^+(2220,\mbox{****})$ &--& $N\frac{9}{2}^-(2250,\mbox{****})$.
\end{tabular}
\end{center}
The splittings within the parity partners are mostly within
the experimental uncertainties.
\end{itemize}
Another aspect, which is currently of high interest, is the question of the
so-called ''\textit{missing resonances}'', {\it i.e.} states that appear in
quark models but which have not been seen in $\pi N$ partial-wave analyses.
As already observed in the $\Delta$-sector, our Salpeter equation-based
quark model (and constituent quark models for baryons in general) predicts a
much richer resonance spectrum of states than has been observed so far
in scattering experiments. Most of the resonance parameters of the $N^*$ and
$\Delta$ states listed in the {\it 'Review of Particle properties'}
\cite{PDG00} stem largely from partial-wave analyses of mostly older $\pi
N\rightarrow \pi N$ scattering data.  Fortunately, the newly established
experimental electron and photon facilities at CEBAF, ELSA, {\it etc.}, make
it possible to investigate additional mechanisms of nucleon resonance
excitations with photons with considerably improved experimental accuracy.
Assuming that the ''missing'' states couple only weakly to the formation
channels in $N\pi$ scattering \cite{CaRo93,CaRo00} and thus escape from
experimental observation, the investigation of these new complementary
formation channels should lead to the discovery of some of these ''missing''
states in the near future.  Indeed, in the sector of nucleon resonances
considered here, there are already indications of three new states around 1900
MeV obtained from recent studies of photo-induced reactions with the {SAPHIR}
detector at the ELSA electron accelerator in Bonn \cite{Kle99,Wor00}. These
allow a first glimpse of the high-mass spectrum of $N^*$-resonances:
\begin{description}
\item[$\bfgrk \eta'$ {\bf photoproduction:}] A fit to the SAPHIR total and
differential cross sections for the $\eta'$ photoproduction obtained
from the reaction chain $\gamma p\rightarrow p \eta' \rightarrow p
\pi^+ \pi^- \eta \rightarrow p \pi^+\pi^-\pi^+\pi^-\pi^0$ has been
made \cite{Ploe89} assuming resonance dominance and taking only S- and
P-wave multipoles into account. The data indicate a coherent resonant
production of two $p\eta'$ partial waves, $S_{11}$ and $P_{11}$. The
extracted resonance parameters are
\begin{center}
\begin{tabular}{cccc}
partial wave & $J^\pi$ & resonance position M [MeV]   & decay width $\Gamma$ [MeV]\\
\hline 
$S_{11}$ & $\frac{1}{2}^-$ &  $1897\pm 50^{+30}_{-2}$  & $396 \pm 115^{+35}_{-45}$\\
$P_{11}$ & $\frac{1}{2}^+$ &  $1986\pm 26^{+10}_{-30}$ & $296 \pm 100^{+60}_{-10}$
\end{tabular}
\end{center}
\item[${\bfgrk\gamma} {\bfgrk p} {\bfgrk\rightarrow} {\bfgrk K^+}
{\bfgrk\Lambda}$:] Recent measurements of the $\gamma p \rightarrow
K^+ \Lambda$ total cross sections from SAPHIR \cite{Tr98} indicate for
the first time a broad structure around 1900 MeV. This structure could
not be resolved before due to low quality of the old data. An
analysis of these new and associated differential cross-section and
recoil-polarization data in the framework of an isobar model
\cite{MaBe99} suggests the existence of a broad $D_{13}$ state, where
the choice of a $D_{13}$ state is based on the agreement with
quark-model predictions \cite{CaRo94,CaRo98}. The fitted resonance
parameters are:
\begin{center}
\begin{tabular}{cccc}
partial wave & $J^\pi$ & resonance position M [MeV]   & decay width $\Gamma$ [MeV]\\
\hline 
$D_{13}$ & $\frac{3}{2}^-$ &  $1895$  & $372$
\end{tabular}
\end{center}
\end{description}
The discovery of ''missing'' states and the measurement of their resonance
positions with high accuracy provides a convincing test for the
quality and the predictive power of various constituent quark
models in order to distinguish between realistic and less realistic
quark models for the description of baryon masses. In this respect, it
is interesting if and how these indications for these three new
states fit into our models ${\cal A}$ and ${\cal B}$.\\

Before we begin our discussion organized according to the 
phenomenological issues of the excited $N^*$-spectrum listed, it is worth to
remark once more that there is no freedom\footnote{In principle there
is the freedom to choose a different effective range $\lambda$ of the
regularized 't~Hooft interaction along with a new adjusted coupling
$g_{nn}$ to readjust the correct $N-\Delta$ mass difference. However,
within the range of possible values $\lambda$, which are consistent with
a reasonable description of the ground-state baryons, the structure of
the excited nucleon spectrum shows only a fairly weak sensitivity to
the choice of $\lambda$.} left to fit the excited nucleon states. With
the five parameters $a$, $b$, $m_n$ and $g_{nn}$,
$\lambda$ fixed from the $\Delta$-spectrum and the $\Delta-N$ splitting, 
all the excited resonances of the $N^*$-spectrum are now
true predictions.
In the subsequent subsection \ref{subsec:studying_inst_eff} we will then illustrate in some more
detail, how instanton-induced effects due to 't~Hooft's
quark-quark interaction are in fact responsible for the phenomenology of the $N^*$-spectrum.

\subsection{Discussion of the complete N-spectrum}
\label{subsec:disc_compl_N}
\begin{figure}[h]
  \begin{center}
    \epsfig{file={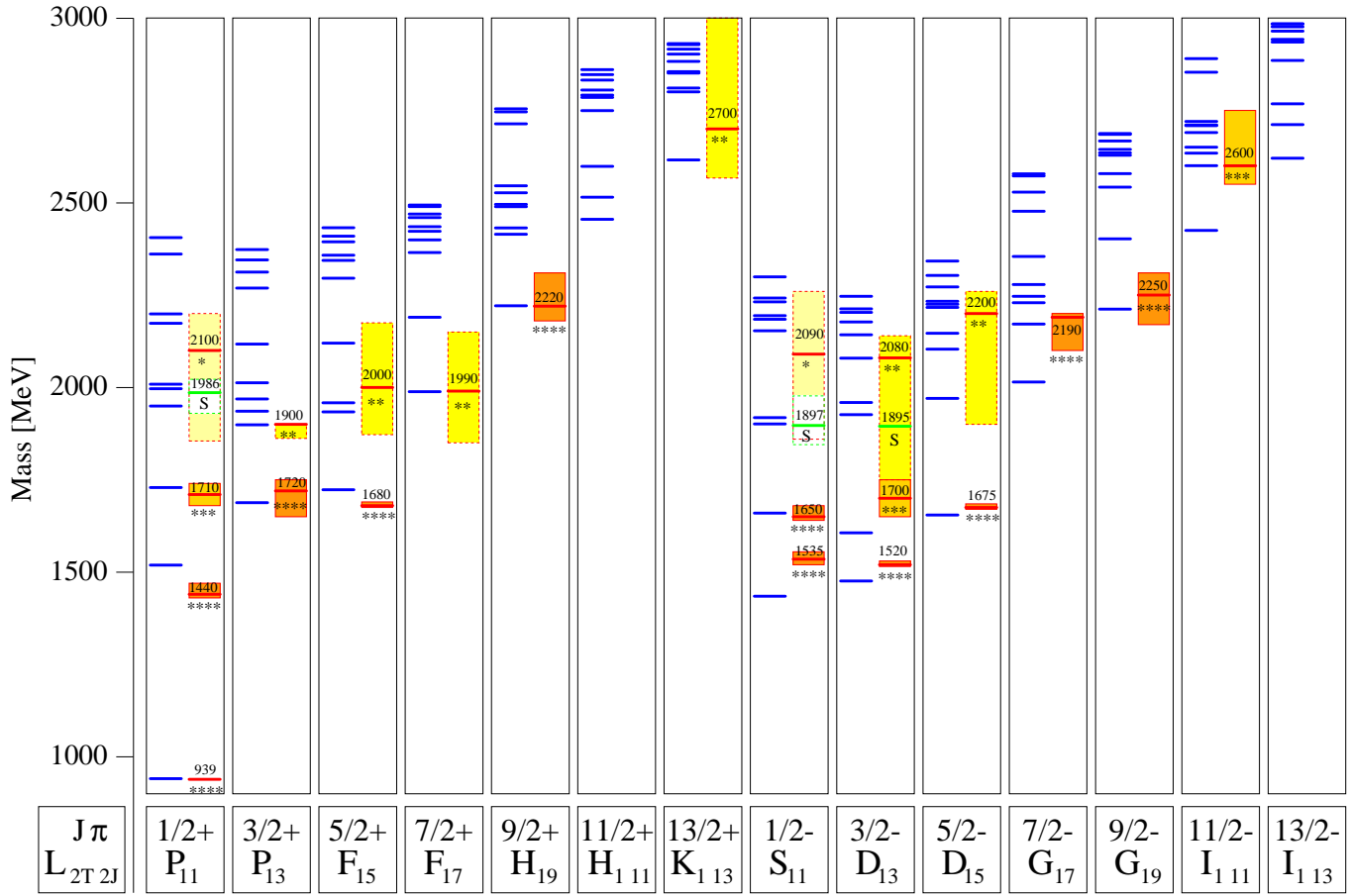},width=180mm}
  \end{center}
\caption{The calculated positive and negative parity
\textbf{N-resonance spectrum} (isospin $T=\frac{1}{2}$ and
strangeness $S^* = 0$) in \textbf{model ${\cal A}$} (left part of each column)
in comparison to the experimental spectrum taken from Particle Data Group \cite{PDG00} (right part of each
column).  The resonances are classified by the total spin
$J$ and parity $\pi$. The experimental resonance position is indicated
by a bar, the corresponding uncertainty by the shaded box, which is
darker the better a resonance is established; the status of each
resonance is additionally indicated by stars. The states labeled by 'S'
belong to new SAPHIR results \cite{Ploe89,MaBe99,Kle99,Wor00}, see text.}
\label{fig:NucM2}
\end{figure}

\begin{figure}[h]
  \begin{center}
    \epsfig{file={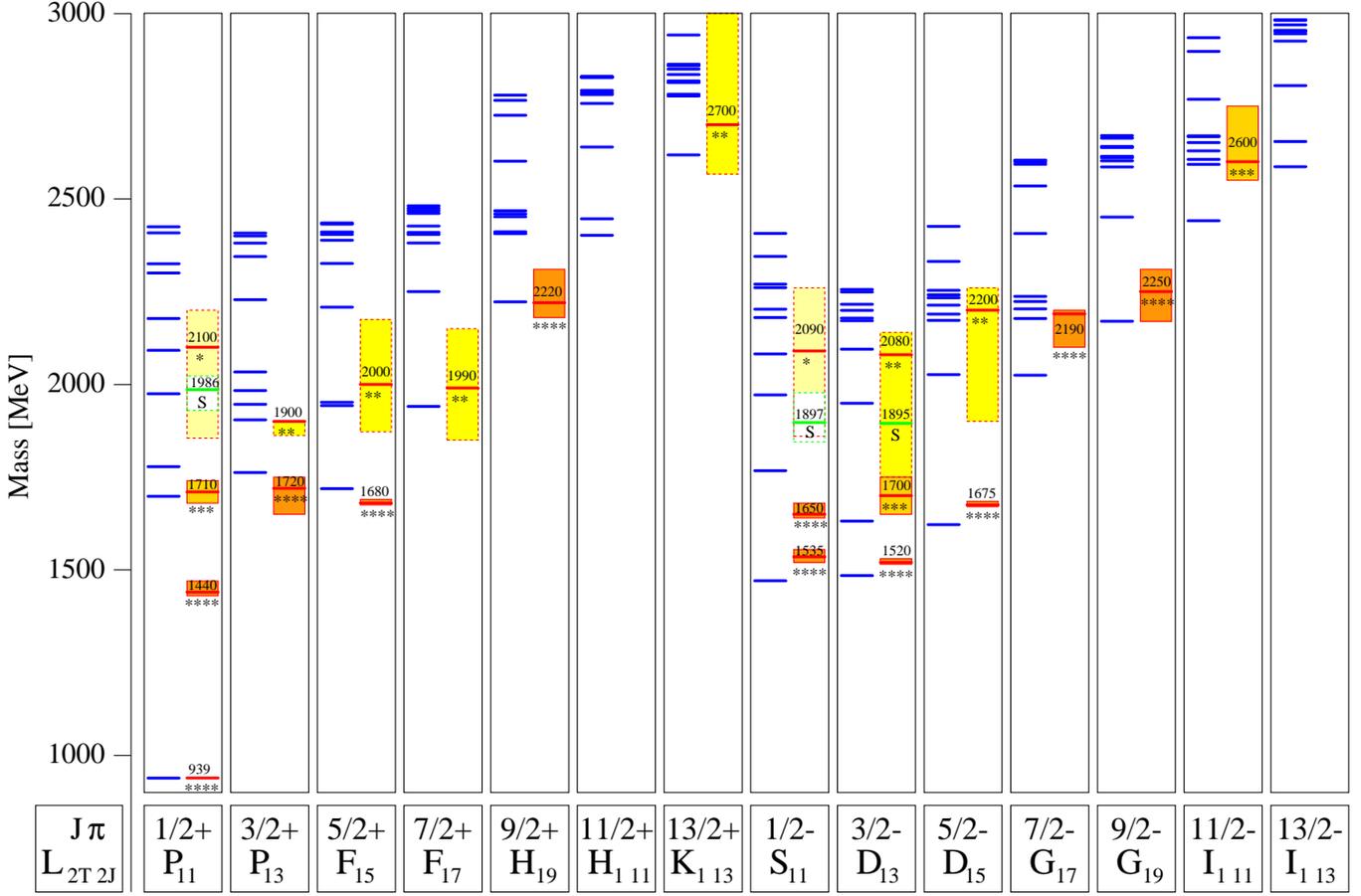},width=180mm}
  \end{center}
\caption{The calculated positive and negative parity
\textbf{N-resonance spectrum} (isospin $T=\frac{1}{2}$ and
strangeness $S^* = 0$) in \textbf{model ${\cal B}$} (left part of each column)
in comparison to the experimental spectrum taken from Particle Data Group \cite{PDG00} (right part of each
column).  The resonances are classified by the total spin
$J$ and parity $\pi$. See also caption to fig. \ref{fig:NucM2}.}
\label{fig:NucM1}
\end{figure}

Figures \ref{fig:NucM2} and \ref{fig:NucM1} show the resulting
positions of the positive- and negative-parity nucleon resonances with
total spins up to $J=\frac{13}{2}$ obtained in model
${\cal A}$ and ${\cal B}$, respectively.  These are compared with
the experimentally observed positions of all presently known
resonances of each status taken from the Particle Data Group
\cite{PDG00}.  Again, the resonances in each column are classified by
the total spin $J$ and the parity $\pi$, where left in each column the
results for at most ten excitations in model ${\cal A}$ or ${\cal B}$ are shown.  In comparison the
experimental positions \cite{PDG00} are displayed on the right in each
column with the uncertainties of the resonance positions indicated by
the shaded boxes and the rating of each resonance denoted by the
corresponding number of stars and a different shading of the error
box. In addition we also display the determined resonance positions of
the three new states that have been recently discovered by the SAPHIR
collaboration \cite{Ploe89,MaBe99,Kle99,Wor00}.  These states are indicated by the symbol 'S'.\\ In the
following, we turn to a shell-by-shell discussion of the complete
nucleon spectrum.  According to their  assignment to a particular
shell, we additionally summarized the explicit positions of
the excited model states in tables \ref{tab:N2hwband},
\ref{tab:N1hwband}, \ref{tab:N3hwband}, \ref{tab:N4hwband},
\ref{tab:N5hwband} and \ref{tab:N6hwband}.

\subsubsection{States of the 2${\bfgrk\hbar}{\bfgrk\omega}$ band}
Let us begin our discussion with the
intra-band structure of the positive-parity $2\hbar\omega$ band
including states with spins $J^\pi = \frac{1}{2}^+$, $\frac{3}{2}^+$,
$\frac{5}{2}^+$ and $\frac{7}{2}^+$.  The  predicted
positions of states belonging to this shell and their assignments to
observed resonances due to a comparison with the phenomenological
mass values are given explicitly in table \ref{tab:N2hwband}.

\begin{table}[h]
\center
\begin{tabular}{ccccccc}
\hline
Exp. state   &PW&${J^\pi}$        & Rating     & Mass range [MeV]& Model state &  Model state \\
\cite{PDG00} &  &                      &            & \cite{PDG00}    & in model ${\cal A}$& in model ${\cal B}$  \\
\hline
$N(1440)$&$P_{11}$&${\frac{1}{2}^+}$& ****     &1430-1470  &$\MSN(1,+,2,1518)$   &$\MSN(1,+,2,1698)$ \\
$N(1710)$&$P_{11}$&${\frac{1}{2}^+}$& ***      &1680-1740  &$\MSN(1,+,3,1729)$   &$\MSN(1,+,3,1778)$ \\\\
$P_{11}(1986)$&$P_{11}$&${\frac{1}{2}^+}$& SAPHIR &1930-2022 &$\begin{array}{c}
                                                               \MSN(1,+,4,1950)\\
                                                               \MSN(1,+,5,1996)\\
                                                               \;\MSN(1,+,6,2009)^\dagger
                                                               \end{array}$& $\MSN(1,+,4,1974)$ \\
$N(2100)$&$P_{11}$&${\frac{1}{2}^+}$& *         &1855 - 2200           &   &$\MSN(1,+,5,2092)$ \\

\hline
$N(1720)$&$P_{13}$&${\frac{3}{2}^+}$& ****     &1650-1750  &$\MSN(3,+,1,1688)$   &$\MSN(3,+,1,1762)$ \\
$N(1900)$&$P_{13}$&${\frac{3}{2}^+}$& **       &1862-1900  &$\MSN(3,+,2,1899)$   &$\MSN(3,+,2,1904)$ \\
         &&                         &          &           &$\MSN(3,+,3,1936)$   &$\MSN(3,+,3,1946)$ \\
         &&                         &          &           &$\MSN(3,+,4,1969)$   &$\MSN(3,+,4,1983)$ \\
         &&                         &          &           &$\MSN(3,+,5,2013)$   &$\MSN(3,+,5,2033)$ \\
\hline
$N(1680)$&$F_{15}$&${\frac{5}{2}^+}$& ****     &1675-1690  &$\MSN(5,+,1,1723)$   &$\MSN(5,+,1,1718)$ \\\\
$N(2000)$&$F_{15}$&${\frac{5}{2}^+}$& **       &1872-2175  &$\begin{array}{c}
                                                              \MSN(5,+,2,1934)\\
                                                              \MSN(5,+,3,1959)
                                                             \end{array}$
                                                           &$\begin{array}{c}
                                                              \MSN(5,+,2,1943)\\ 
                                                              \MSN(5,+,3,1952)
                                                             \end{array}$\\
\hline
$N(1990)$&$F_{17}$&${\frac{7}{2}^+}$& **       &1850-2150  &$\MSN(7,+,1,1989)$   &$\MSN(7,+,1,1941)$ \\
\hline
\end{tabular}
\caption{Calculated positions of nucleon states assigned to the positive parity $2\hbar\omega$ shell
  in comparison to the corresponding experimental mass values taken from
  \cite{PDG00}. Notation as introduced in the caption of table \ref{tab:1hwband}.
  The resonance $P_{11}(1986)$ denoted by 'SAPHIR' (in the column 'Rating') corresponds to the
  new $\eta'$-photoproduction result from the SAPHIR collaboration
  \cite{Ploe89}.
  $^\dagger$ The predicted state $\MSN(1,+,6,2009)$ in model ${\cal A}$ actually 
  belongs to the $4\hbar\omega$ shell but is lowered into the region of the $2\hbar\omega$
  band due to the instanton force, see fig. \ref{fig:Model2N+gvar} 
  in subsect. \ref{subsec:studying_inst_eff}.}
\label{tab:N2hwband}
\end{table}
Indeed we find in the $\frac{1}{2}^+$, $\frac{3}{2}^+$ and
$\frac{5}{2}^+$-sectors for both models ${\cal A}$ and ${\cal B}$ a
selective lowering of exactly four states, well separated from the
remaining bulk of states which is centered around 2000 MeV. As we will
illustrate in the next subsect. \ref{subsec:studying_inst_eff}, this is
indeed a consequence of the strongly attractive action of the
instanton-induced residual interaction in these dominantly $^2 8[56,+]$ or $^2
8[70,+]$ states.\\ Concerning these states let us first focus on the results
of model ${\cal A}$. Figure \ref{fig:NucM2} impressively shows that in
model ${\cal A}$ these four separated states fit rather well into the
experimentally observed pattern of splittings, thus yielding a unique
one-to-one correspondence between our model states and the observed
resonances.  In particular, we can account for the puzzling low position of the
Roper resonance $N\frac{1}{2}^+(1440,\mbox{****})$: the calculated
position at 1518 MeV is only 78 MeV too high compared to the
phenomenologically determined mass value. The discrepancy to the upper edge of
the uncertainty range even amounts to only 48 MeV.  We should note here that
more recent analyses \cite{MaSa92,VrDyLe00} even determine a slightly higher
Roper resonance position at $1462\pm 80$ MeV and $1479\pm 80$ MeV,
respectively.  We also obtain a very satisfactory description of the other
three states that are grouped around 1700 MeV: in the
$N\frac{1}{2}^+$ sector the second radial excitation after the
Roper state predicted at 1729 MeV fits exactly into the uncertainty range of the
$N\frac{1}{2}^+(1710,\mbox{***})$.  The same is the case in the
$N\frac{3}{2}^+$ sector, where the first excitation predicted at 1688 MeV
nicely agrees with the well-established four-star state
$N\frac{3}{2}^+(1720,\mbox{****})$. In the $N\frac{5}{2}^+$ sector our
prediction for the first excited state at 1723 MeV slightly overestimates the
observed position of the $N\frac{5}{2}^+(1680,\mbox{****})$. We would like to
mention that a selective lowering of these states is also found in a
non-relativistic treatment, see \cite{BBHMP90,Met92}. However these calculations
could only qualitatively describe these splittings: the Roper resonance
and the other three positive-parity excited states tend to be too massive by
about 200-250 MeV. In this respect, we thus find the predictions of our fully
covariant model ${\cal A}$ being of considerably better quality,
emphasizing the necessity of a fully relativistic framework to describe this
striking structure quantitatively. Looking at fig. \ref{fig:NucM1}, the
corresponding predictions for these four states in model ${\cal B}$ confirm
this statement: even though model ${\cal B}$ has the same non-relativistic
limit as model ${\cal A}$ the fully relativistic treatment within our
Salpeter framework yields a totally different result.  Although we likewise
observe in the $N\frac{1}{2}^+$, $N\frac{3}{2}^+$ and $N\frac{5}{2}^+$ sectors
a separation of four states from the rest of the $2\hbar\omega$ band, model
${\cal B}$ obviously strongly fails in the description of the low position
of the Roper resonance. Instead, we find in the $N\frac{1}{2}^+$ sector
two slightly split states at 1698 MeV and 1778 MeV that lie near the
experimentally observed resonance $N\frac{1}{2}^+(1710,\mbox{***})$.  On the
other hand, the predictions of the two other resonances in the
$N\frac{3}{2}^+$ and $N\frac{5}{2}^+$ sectors reproduce the states
$N\frac{3}{2}^+(1720,\mbox{****})$ and $N\frac{5}{2}^+(1680,\mbox{****})$
rather satisfactorily. It is worth noting that the difference to model
${\cal A}$ concerning the Roper resonance is caused by a quite
different influence of the instanton-induced interaction within the
confinement versions ${\cal A}$ and ${\cal B}$. It has its origin in the
different mixing of the spin-flavor $SU(6)$ contributions (\ref{NconfigPauli})
to the embedded Pauli spinors and moreover in a different level ordering
within the intra-band structure. This originates from various relativistic effects that are
induced by the two confinement Dirac structures of model ${\cal A}$
and ${\cal B}$. Also these features shall be clarified in some more detail
in the course of the next subsect. \ref{subsec:studying_inst_eff}. Here let
us note that this striking difference between the two models concerning the
Roper resonance strongly supports model ${\cal A}$ to provide the
more realistic confinement version in combination with 't~Hooft's force as
residual interaction.\\ As already mentioned, apart from these four separated,
low-lying resonances, the remaining bulk of states predicted in the
$2\hbar\omega$ shell is clustered around 2 GeV. In both models their
mean mass corresponds nicely with the ranges of possible values of the
three two-star states $N\frac{3}{2}^+(1900,\mbox{**})$,
$N\frac{5}{2}^+(2000,\mbox{**})$ and $N\frac{7}{2}^+(1990,\mbox{**})$ observed
in this resonance region. Obviously, both models predict a substantial number
of ''missing'' states in this region and thus the assignment of our model
states to observed resonances by a comparison of the masses alone is in
general not unique.  Let us discuss the situation in each spin sector
from $\frac{1}{2}^+$ to $\frac{7}{2}^+$ in turn:\\
In the $N\frac{7}{2}^+$ sector the assignment is still unambiguous: both
models predict only a single state in accordance with the single resonance
$N\frac{7}{2}^+(1990,\mbox{**})$ seen in this mass range for the $F_{17}$
partial wave. The predicted mass values at 1989 MeV in model ${\cal A}$ and
at 1941 MeV in model ${\cal B}$ both agree with the observed position of
this two-star resonance. Moreover, it is interesting to note that the Salpeter 
amplitude of this first excited $N\frac{7}{2}^+$ state exhibits an
almost pure $^4 8 [70]$ configuration ($> 99\%$) in both models and by no
means is influenced by 't~Hooft's force.  Hence, this state is determined by
the confinement potential alone. In the $N\frac{5}{2}^+$ sector there are
predictions of two further nearly degenerate states in this higher resonance
region at 1934 and 1959 MeV in model ${\cal A}$ and at 1943 and 1952 MeV
in model ${\cal B}$. The calculated masses correspond rather well to the
$N\frac{5}{2}^+(2000,\mbox{**})$. In the $N\frac{3}{2}^+$ sector we find even
a group of four other states in the $2\hbar\omega$ band which all
lie in the range between $\sim 1900$ and $\sim 2000$ MeV. In both models, the
first of these states (predicted at 1899 MeV in ${\cal A}$ and at 1904 MeV
in ${\cal B}$) fits close to the reported position of the resonance
$N\frac{3}{2}^+(1900, **)$. Thus, the gap of roughly 200
MeV between the resonances
$N\frac{3}{2}^+(1720,\mbox{****})$ and $N\frac{3}{2}^+(1900,\mbox{**})$ is
fairly well reproduced in both models and none of the three remaining
''missing'' states of this shell lies in between this gap as partly
predicted in other quark models, as {\it e.g.} in the
collective quark model of Bijker, Iachello and Leviatan \cite{BIL94,BIL00},
which is based on a spectrum-generating algebra. In the $N\frac{1}{2}^+$
sector, the results of model ${\cal A}$ and ${\cal B}$ again differ
significantly.  In (the less realistic) model ${\cal B}$, the two remaining
states that complete the $2\hbar\omega$ shell in this sector, are predicted at
1974 MeV and 2092 MeV. The first state agrees fairly well with the position of
the new discovered resonance $P_{11}(1986)$ extracted by a recent analysis of
$\eta'$-photoproduction at the SAPHIR detector \cite{Ploe89}.  The second state fits
the average value of the weakly established one-star resonance
$N\frac{1}{2}^+(2100,\mbox{*})$.  Model ${\cal A}$, however, predicts a
cluster of three close states at 1950 MeV, 1996 MeV and 2009 MeV. We
should note that the first two predictions correspond to two (of four)
remaining states that one usually expects to be assigned to the $2\hbar\omega$
shell.  The third one is lowered from the $4\hbar\omega$ band into this
resonance region due to a strong attraction of the instanton force for this
state (for an illustration see fig. \ref{fig:Model2N+gvar} in the next
subsect. \ref{subsec:studying_inst_eff}).  All three predicted masses
nicely fit within the uncertainty range of the recently discovered 'SAPHIR
resonance' $P_{11}(1986)$, which overlaps with the quite large error range of
the one-star $N\frac{1}{2}^+(2100,\mbox{*})$.  To decide which of these three
model states has to be assigned to this new resonance due to possible
different couplings of these states to the $N\eta'$ decay channel would
require the calculation of quasi-two-body decays of baryons into different
meson-baryon final states within our covariant Bethe-Salpeter
framework.  Unfortunately this information is not yet available but it will be
a principal objective of our investigations in the near future.  For the
moment it is interesting to emphasize that, similar to the $N\frac{3}{2}^+$
sector, there are no predicted ''missing'' states in between the new 'SAPHIR resonance'
$P_{11}(1986)$ and the established $N\frac{1}{2}^+(1710,\mbox{***})$
resonance. Also here, our model ${\cal A}$ differs significantly from other
constituent quark models \cite{CaIs86,BIL94,BIL00}. In this respect, the possible discovery of
new (so far ''missing'' and undiscovered) baryon states at CEBAF, ELSA and
elsewhere just in this $N^*$ resonance region is highly interesting and will
provide an additional helpful criterion to distinguish between the different
constituent quark models presently discussed in the literature \cite{CaRo00}. 
\subsubsection{States of the 1${\bfgrk\hbar}{\bfgrk\omega}$ band}
We now turn to the description of those states which, in the language of
the harmonic oscillator basis, belong to the negative-parity $1\hbar\omega$
band.  Both models predict (as usual in constituent quark models
for baryons) five states with spins $J^\pi=\frac{1}{2}^-$, $\frac{3}{2}^-$ and
$\frac{5}{2}^-$ that can be uniquely identified with the five observed,
well-established four- and three-star resonances listed in the baryon summary
table of the Particle Data Group \cite{PDG00}. A comparison of the
predicted masses with the corresponding empirical resonance
positions is given in table \ref{tab:N1hwband}.

\begin{table}[h]
\center
\begin{tabular}{ccccccc}
\hline
Exp. state   &PW&${J^\pi}$        & Rating     & Mass range [MeV]& Model state &  Model state \\
\cite{PDG00} &  &                      &            & \cite{PDG00}    & in model ${\cal A}$& in model ${\cal B}$  \\
\hline
$N(1535)$&$S_{11}$&${\frac{1}{2}^-}$& ****     &1520-1555  &$\MSN(1,-,1,1435)$   &$\MSN(1,-,1,1470)$ \\
$N(1650)$&$S_{11}$&${\frac{1}{2}^-}$& ****     &1640-1680  &$\MSN(1,-,2,1660)$   &$\MSN(1,-,2,1767)$ \\
\hline
$N(1520)$&$D_{13}$&${\frac{3}{2}^-}$& ****     &1515-1530  &$\MSN(3,-,1,1476)$   &$\MSN(3,-,1,1485)$ \\
$N(1700)$&$D_{13}$&${\frac{3}{2}^-}$& ***      &1650-1750  &$\MSN(3,-,2,1606)$   &$\MSN(3,-,2,1631)$ \\
\hline
$N(1675)$&$D_{15}$&${\frac{5}{2}^-}$& ****     &1670-1685  &$\MSN(5,-,1,1655)$   &$\MSN(5,-,1,1622)$ \\
\hline
\end{tabular}
\caption{Calculated positions of all nucleon states assigned to the negative parity $1\hbar\omega$ shell
  in comparison to the corresponding experimental mass values taken from
  \cite{PDG00}. Notation as in table \ref{tab:1hwband}.}
\label{tab:N1hwband}
\end{table}
In both models, the single $1\hbar\omega$ state predicted in the
$N\frac{5}{2}^-$ sector is an approximately pure $^4 8 [70]$ state ($>99\%$).
It thus remains totally unaffected by the instanton force similar to the first
positive parity excitation in the $N\frac{7}{2}^+$ sector. Hence also this
state is determined by the three-quark confinement kernel alone. In model
${\cal A}$ this state is predicted at 1655 MeV, very close to the
experimental position of the four-star resonance $N\frac{5}{2}^-(1675,\mbox{****})$.
Model ${\cal B}$ slightly underestimates the empirical mass value by
roughly 50 MeV.  Also the mass splitting between the two resonances
$N\frac{3}{2}^-(1520,\mbox{****})$ and $N\frac{3}{2}^-(1720,\mbox{***})$ in the
$N\frac{3}{2}^-$ sector is equally well reproduced in both models.  
However, the calculated positions appear
slightly too low compared to the two
observed resonances in this sector. As already observed in the
positive-parity $2\hbar\omega$ shell, the main difference between the two
models again shows up  in the sector with spin $1/2$: model
${\cal B}$, which already strongly failed in describing the 
low-lying Roper resonance, yields in
the  negative-parity $N\frac{1}{2}^-$ sector a much too large
splitting of the two $1\hbar\omega$ states. The position of the four-star
resonance $N\frac{1}{2}^-(1535,\mbox{****})$ is underestimated, while at the same
time the calculated mass corresponding to $N\frac{1}{2}^-(1650,\mbox{****})$
appears much too high. Again, the situation is better described in model ${\cal A}$:
although the position of the $N\frac{1}{2}^-(1535,\mbox{****})$ is still predicted
100 MeV too low, the predicted mass of the second
excitation at 1660 MeV nicely agrees with the empirical resonance position of the
$N\frac{1}{2}^-(1650,\mbox{****})$, even within its uncertainty range.
Altogether, we thus obtain a rather well predicted pattern of hyperfine splittings of
the $1\hbar\omega$ band in model ${\cal A}$, which thus again is of better quality
than in model ${\cal B}$. As already observed in the positive-parity sector, the
differences between model ${\cal A}$ and model ${\cal B}$ have their
origin in a different influence of the instanton-induced interaction. 
Again this can be traced back to a difference in relativistic effects that
stem from the embedding map of the Salpeter amplitudes in combination
with the two distinct confinement Dirac structures. For a more detailed
discussion of this issue we again refer to subsect. \ref{subsec:studying_inst_eff}.

\subsubsection{Relative arrangement of the 2${\bfgrk\hbar}{\bfgrk\omega}$ and 1${\bfgrk\hbar}{\bfgrk\omega}$ bands -- approximate
parity doublets in the second resonance region} 
\begin{figure}[h]
  \begin{center}
    \epsfig{file={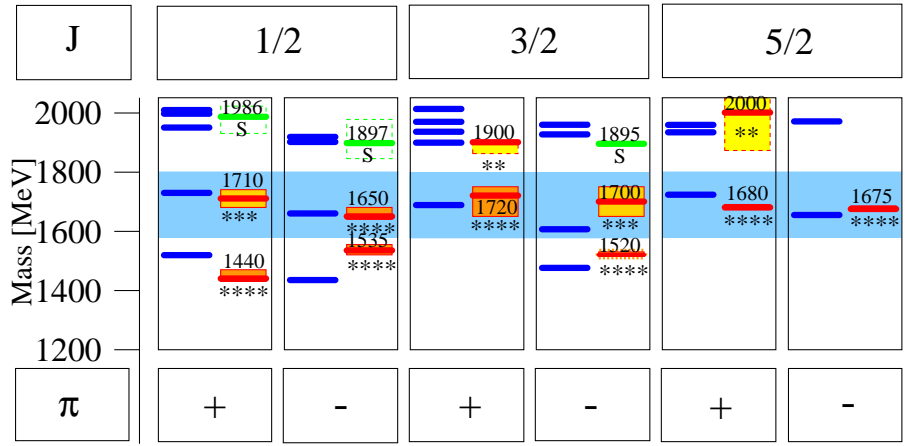},width=120mm}
  \end{center}
\caption{Approximate parity doublets due to the overlap of the
positive-parity $2\hbar\omega$ states and the negative-parity
$1\hbar\omega$ states in the second resonance region around $\sim
1700$ MeV (shaded regions) as predicted in model ${\cal A}$. The
left part in each column shows our calculation which is compared with
the experiment depicted in the right part.  For each spin
$J=\frac{1}{2}$, $\frac{3}{2}$ and $\frac{5}{2}$ the left column shows
the even and the right column the odd parity.  For an explicit
illustration how these doublet structures are generated by 't~Hooft's
instanton-induced interaction see fig. \ref{fig:PDgvar1hw2hw3hw} and
the corresponding detailed discussion of instanton effects in
subsect. \ref{subsec:studying_inst_eff}.}
\label{fig:PD1hw2hw}
\end{figure}

After discussing the
individual hyperfine structures of the positive-parity $2\hbar\omega$
band and the negative-parity $1\hbar\omega$ band separately, let us
now compare the relative positions of the positive and negative parity
states involved. We restrict this discussion to model ${\cal A}$,
which so far led to consistently better results, especially in the
sectors with spin $J=\frac{1}{2}$.  Due to the selective lowering of a
particular set of states of the $2\hbar\omega$ band relative to the
other states we can indeed reproduce the striking overlap of states of
the two shells with opposite parity.  The situation is displayed in
fig. \ref{fig:PD1hw2hw}, where now in contrast to fig.
\ref{fig:NucM2} the states with the same total spin, but opposite
parity are directly displayed side by side.

As discussed, model ${\cal A}$ nicely accounts for the low
position of the Roper resonance.  Unfortunately, due to the
slightly too low predicted position of the first negative-parity
$N\frac{1}{2}^-$ state, which still appears below the calculated
position of the Roper resonance, the striking inversion of
the ordering of positive- and negative-parity states in the spin-$1/2$ 
sector cannot be achieved. On the other hand, figure
\ref{fig:PD1hw2hw} impressively demonstrates that our model
${\cal A}$ indeed reproduces the remarkable appearance of the
lowest approximate ''parity doublets'' of the nucleon spectrum in
the second resonance region around $\sim 1700$ MeV (shaded regions in fig.
\ref{fig:PD1hw2hw}) remarkably well; see also table \ref{tab:PD_1hw_2hw}.
Apart from the slightly too low calculated position of the
$N\frac{3}{2}^-(1700,\mbox{***})$ even the experimentally observed splittings
between the parity partners in general are fairly well described.
\begin{table}[!h]
\begin{center}
\begin{tabular}{c}
\hline 
\textbf{approximate parity doublets around 1700 MeV}\\[2mm]
$\begin{array}{rlcl}
&'2\hbar\omega',\quad \pi=+ && '1\hbar\omega',\quad \pi=-\\
\hline
&N\frac{1}{2}^+(1710,\mbox{***})  &-& N\frac{1}{2}^-(1650,\mbox{****})\\[1mm]
{\rm exp.}:&{ 1680-1740\;{\rm MeV} }        & & {1640-1680\;{\rm MeV}}\\
{\rm model}\; {\cal A}:& 1729 \;{\rm MeV} & & 1660\;{\rm MeV}\\
\hline 
&N\frac{3}{2}^+(1720,\mbox{****}) &-& N\frac{3}{2}^-(1700,\mbox{***})\\[1mm]
{\rm exp.}:&{1650-1750\;{\rm MeV}}        & & {1650-1750\;{\rm MeV}}\\
{\rm model}\; {\cal A}:& 1688 \;{\rm MeV} & & 1606 \;{\rm MeV}\\
\hline
&N\frac{5}{2}^+(1680,\mbox{****}) &-& N\frac{5}{2}^-(1675,\mbox{****})\\[1mm]
{\rm exp.}:&{1675-1690\;{\rm MeV}}        & & {1670-1685\;{\rm MeV}}\\
{\rm model}\; {\cal A}:& 1723 \;{\rm MeV} & & 1655  \;{\rm MeV}\\
\end{array}$\\
\hline
\end{tabular}
\end{center}
\caption{Approximate parity doublets in the second resonance region in model ${\cal A}$.}
\label{tab:PD_1hw_2hw}
\end{table}

\subsubsection{Beyond the 2${\bfgrk\hbar}{\bfgrk\omega}$ band}
Finally let us have a look at the experimentally still rather poorly explored
high energy part of the nucleon spectrum, {\it i.e.} at the states that belong
to the $3\hbar\omega$, $4\hbar\omega$, $5\hbar\omega$ and $6\hbar\omega$ bands
with observed total spins up to $J=\frac{13}{2}$.\\

We start with the states of the $3\hbar\omega$ and $4\hbar\omega$ bands
and their relative alignments.
The predicted positions in both models for the lightest few states assigned to these shells
are summarized and compared to the experimental mass values in table
\ref{tab:N3hwband} and \ref{tab:N4hwband}, respectively.

\begin{table}[!h]
\center
\begin{tabular}{ccccccc}
\hline
Exp. state   &PW&${J^\pi}$        & Rating     & Mass range [MeV]& Model state &  Model state \\
\cite{PDG00} &  &                      &            & \cite{PDG00}    & in model ${\cal A}$& in model ${\cal B}$  \\
\hline
$S_{11}(1897)$&$S_{11}$&${\frac{1}{2}^-}$& SAPHIR     &1845-1977 &$\begin{array}{c}
                                                                    \MSN(1,-,3,1901)\\
                                                                    \MSN(1,-,4,1918)
                                                                   \end{array}$
                                                                 &$\MSN(1,-,3,1971)$\\[5mm]
$N(2090)$&$S_{11}$&${\frac{1}{2}^-}$& *     &1860-2260  &$\begin{array}{c}
                                                           \MSN(1,-,5,2153)\\
                                                           \MSN(1,-,6,2185)\\ 
                                                           \MSN(1,-,7,2194)\\
                                                           \MSN(1,-,8,2232)\\
                                                           \MSN(1,-,9,2242)
                                                          \end{array}$ 
                                                        &$\begin{array}{c}
                                                           \MSN(1,-,4,2082)\\
                                                           \MSN(1,-,5,2180)\\
                                                           \MSN(1,-,6,2203)\\
                                                           \MSN(1,-,7,2261)
                                                          \end{array}$\\[5mm]
         &        &                 &       &           &&$\begin{array}{c}
                                                             \MSN(1,-,8,2270)\\
                                                             \MSN(1,-,9,2345)
                                                            \end{array}$\\
\hline
$D_{13}(1895)$&$D_{13}$&${\frac{3}{2}^-}$& SAPHIR     &$\approx 1895$ &$\begin{array}{c}
                                                                         \MSN(3,-,3,1926)\\
                                                                         \MSN(3,-,4,1959)
                                                                        \end{array}$&$\MSN(3,-,3,1949)$\\[5mm]
$N(2080)$&$D_{13}$&${\frac{3}{2}^-}$& **     &1750-2140  &$\MSN(3,-,5,2079)$&$\MSN(3,-,4,2095)$\\[5mm]
         &        &                 &        &           &$\begin{array}{c}   
                                                            \MSN(3,-,6,2143)\\
                                                            \MSN(3,-,7,2177)\\
                                                            \MSN(3,-,8,2203)\\
                                                            \MSN(3,-,9,2213)\\
                                                            \;\MSN(3,-,10,2247)
                                                           \end{array}$
                                                         &$\begin{array}{c}
                                                            \MSN(3,-,5,2172)\\
                                                            \MSN(3,-,6,2179)\\
                                                            \MSN(3,-,7,2200)\\
                                                            \MSN(3,-,8,2216)\\
                                                            \MSN(3,-,9,2249)\\
                                                            \;\MSN(3,-,10,2256)
                                                           \end{array}$\\
\hline
         &        &                 &        &          &$\MSN(5,-,2,1970)$&$\MSN(5,-,2,2026)$\\[5mm]
$N(2200)$&$D_{15}$&${\frac{5}{2}^-}$& **     &1900-2260 &$\begin{array}{c}
                                                           \MSN(5,-,3,2104)\\
                                                           \MSN(5,-,4,2147)\\
                                                           \MSN(5,-,5,2217)\\
                                                           \MSN(5,-,6,2225)\\
                                                           \MSN(5,-,7,2233)
                                                          \end{array}$
                                                        &$\begin{array}{c}
                                                           \MSN(5,-,3,2173)\\
                                                           \MSN(5,-,4,2189)\\
                                                           \MSN(5,-,5,2213)\\
                                                           \MSN(5,-,6,2233)\\
                                                           \MSN(5,-,7,2241)\\
                                                           \MSN(5,-,8,2254)
                                                          \end{array}$\\[5mm]
         &        &                 &        &          &$\begin{array}{c}
                                                           \MSN(5,-,8,2272)\\
                                                           \MSN(5,-,9,2303)
                                                          \end{array}$ & $\MSN(5,-,9,2331)$\\
\hline
         &        &                 &        &          &$\MSN(7,-,1,2015)$&$\MSN(7,-,1,2024)$\\
$N(2190)$&$G_{17}$&${\frac{7}{2}^-}$& ****   &2100-2200 &$\MSN(7,-,2,2171)$&$\MSN(7,-,2,2177)$\\[5mm]
         &        &                 &        &          &$\begin{array}{c}
                                                           \MSN(7,-,3,2229)\\
                                                           \MSN(7,-,4,2247)\\
                                                           \MSN(7,-,5,2279)
                                                          \end{array}$
                                                        &$\begin{array}{c}
                                                           \MSN(7,-,3,2203)\\
                                                           \MSN(7,-,4,2223)\\
                                                           \MSN(7,-,5,2237)
                                                          \end{array}$\\
\hline
$N(2250)$&$G_{19}$&${\frac{9}{2}^-}$& ****   &2170-2310 &$\MSN(9,-,1,2212)$&$\MSN(9,-,1,2170)$\\
\hline
\end{tabular}
\caption{Calculated positions for the lightest few negative-parity
nucleon states in the $3\hbar\omega$ shell in comparison to the
corresponding experimental mass values taken from \cite{PDG00}. Notation as in
table \ref{tab:1hwband}. The resonances denoted by 'SAPHIR' (in the column 'Rating') corresponds to the
new photoproduction results from the SAPHIR collaboration, see text.}
\label{tab:N3hwband}
\end{table}

\begin{table}[!h]
\center
\begin{tabular}{ccccccc}
\hline
Exp. state   &PW&${J^\pi}$        & Rating     & Mass range [MeV]& Model state &  Model state \\
\cite{PDG00} &  &                      &            & \cite{PDG00}    & in model ${\cal A}$& in model ${\cal B}$  \\
\hline
$N(2100)$&$P_{11}$&${\frac{1}{2}^+}$&*     &1850-2200&$\begin{array}{c}
                                                        \MSN(1,+,6,2009)\\
                                                        \MSN(1,+,7,2174)\\
                                                        \MSN(1,+,8,2199)
                                                       \end{array}$&$\MSN(1,+,6,2177)$\\  
\\
         &        &                 &      &          &$\MSN(1,+,9,2361)$&$\MSN(1,+,7,2300)$\\
         &        &                 &      &          &&$\MSN(1,+,8,2325)$\\
         &        &                 &      &          &&$\MSN(1,+,9,2408)$\\
\hline
         &$P_{13}$&${\frac{3}{2}^+}$&      &           &$\MSN(3,+,6,2117)$&$\MSN(3,+,6,2228)$\\
         &        &                 &      &           &$\MSN(3,+,7,2269)$&$\MSN(3,+,7,2345)$\\
         &        &                 &      &           &$\MSN(3,+,8,2312)$&$\MSN(3,+,8,2381)$\\
\hline
         &$F_{15}$&${\frac{5}{2}^+}$&      &           &$\MSN(5,+,4,2120)$&$\MSN(5,+,4,2208)$\\
         &        &                 &      &           &$\MSN(5,+,5,2296)$&$\MSN(5,+,5,2326)$\\
         &        &                 &      &           &$\MSN(5,+,6,2344)$&$\MSN(5,+,6,2389)$\\
\hline
         &$F_{17}$&${\frac{7}{2}^+}$&      &           &$\MSN(7,+,2,2190)$&$\MSN(7,+,2,2250)$\\
         &        &                 &      &           &$\MSN(7,+,3,2365)$&$\MSN(7,+,3,2381)$\\
         &        &                 &      &           &$\MSN(7,+,4,2399)$&$\MSN(7,+,4,2404)$\\
\hline
$N(2220)$&$H_{19}$&${\frac{9}{2}^+}$&****  &2180-2310  &$\MSN(9,+,1,2221)$&$\MSN(9,+,1,2221)$\\
\\
         &        &                 &      &           &$\MSN(9,+,2,2415)$&$\MSN(9,+,2,2406)$\\
         &        &                 &      &           &$\MSN(9,+,3,2432)$&$\MSN(9,+,3,2411)$\\
\hline
         &$H_{1\;11}$&${\frac{11}{2}^+}$&  &           &$\MSN(11,+,1,2455)$&$\MSN(11,+,1,2402)$\\
         &           &                  &  &           &$\MSN(11,+,2,2515)$&$\MSN(11,+,2,2446)$\\
\hline
\end{tabular}
\caption{Calculated positions for the lightest few positive-parity
nucleon states in the $4\hbar\omega$ shell in comparison to the
corresponding experimental mass values taken from \cite{PDG00}. Notation as in
table \ref{tab:1hwband}.}
\label{tab:N4hwband}
\end{table}

The Particle Data Group \cite{PDG00} lists all in all five
negative-parity resonances in the energy region between roughly 2000
and 2300 MeV with $J^\pi = \frac{1}{2}^-$ to $\frac{9}{2}^-$. In
the notation of the oscillator shell model these should be assigned to the
$3\hbar\omega$ band: There are the well-established
resonances of the four-star category, which are the lowest observed
orbital excitations in the $N\frac{7}{2}^-$ and $N\frac{9}{2}^-$
sector ({\it i.e.}  states of Regge-trajectory-type sequences),
the $N\frac{7}{2}^-(2190,\mbox{****})$ and the $N\frac{9}{2}^-(2250
,\mbox{****})$. Also there is in each of the lower spin
sectors $J^\pi = \frac{1}{2}^-$, $\frac{3}{2}^-$ and $\frac{5}{2}^-$
some evidence for a radially excited resonance in this mass
region. These are the weakly established one- and two-star resonances
$N\frac{1}{2}^-(2090,\mbox{*})$, $N\frac{3}{2}^-(2080,\mbox{**})$ and
$N\frac{5}{2}^-(2200,\mbox{**})$ which have quite big uncertainties
of several hundred MeVs.  All these states form a shell structure with
a mean mass of about 2200 MeV. Concerning the
$N\frac{9}{2}^-(2250,\mbox{****})$ it is worth noting that the PDG 
states in the corresponding sector with the same total spin
$J=\frac{9}{2}$ but with opposite positive parity the first excited
resonance $N\frac{9}{2}^+(2220, \mbox{****})$ at almost the same mass value,
{\it i.e.} nearly degenerate with the $N\frac{9}{2}^-(2250
,\mbox{****})$. Due to its high spin $J=\frac{9}{2}$, this 
well-established four-star resonance in the positive-parity sector has to
be a member of the $4\hbar\omega$ shell.  The appearance of this
approximate parity doublet $N\frac{9}{2}^+(2220,\mbox{****})$ --
$N\frac{9}{2}^-(2250,\mbox{****})$ is a very striking feature in the
high energy part of the nucleon spectrum, and we should note here that 
(non-relativistic or 'relativized') constituent quarks models
which use one-gluon-exchange as residual interaction \cite{CaIs86} generally
cannot account for this well-established structure.\\ In addition
there are the new indications of resonant structures recently observed
in the $S_{11}$ and $D_{13}$ partial waves of the photoproduction
measurements of $\gamma p\rightarrow p\eta'$ and $\gamma p\rightarrow
K^+\Lambda$ with the SAPHIR detector at ELSA in Bonn: the
$S_{11}(1897)$ and $D_{13}(1895)$, which are labeled by the symbol 'S'
in figs. \ref{fig:NucM2} and \ref{fig:NucM1}. Both states should
belong to the $3\hbar\omega$ band, but their positions at about 1900
MeV are comparatively low, namely more than 200 MeV below the other
observed states in the $3\hbar\omega$ shell quoted by the PDG. 
These new resonances are nearly degenerate with the upper part
of the positive parity $2\hbar\omega$ shell.

Before comparing these experimentally found structures with our
predicted spectrum in this energy region in detail, it is 
instructive to discuss first the predicted intra-band structures of
the $3\hbar\omega$ shell and the corresponding implications of the
instanton force in general.  As can be seen in figs. \ref{fig:NucM2}
and \ref{fig:NucM1}, in both models the bulk of states
appears in a region between $\sim 2100$ and $\sim 2300$
MeV, which fairly well agrees with the range of possible mass values of the
resonances presently stated by the PDG. The predicted spectrum
of states in this shell is even richer as that of $2\hbar\omega$ band discussed
before. Furthermore,  in both models we again observe a
particular set of states, which is selectively lowered
relative to the rest of the states in the $\sim 2200$ MeV region.
Consequently, also the $3\hbar\omega$ band splits into two well
separated parts, in a similar manner as observed in the $2\hbar\omega$
band: again, this lowering is due to the attractive action of the
instanton-induced interaction in these states as will become more
apparent by the more detailed investigation of these instanton-induced
effects in the next subsection (see figs. \ref{fig:Model2N-gvar} and
\ref{fig:Mod1aNgvar} for an illustration of these effects in model
${\cal A}$ and model ${\cal B}$, respectively).  The lowering of
these particular $3\hbar\omega$ states occurs for the
total spins $J^\pi=\frac{1}{2}^-$, $\frac{3}{2}^-$, $\frac{5}{2}^-$
and $\frac{7}{2}^-$ and is absent for the single state of this band
predicted in the sector with spin $J=\frac{9}{2}^-$.  It is worth to
mention here that the Salpeter amplitudes of all these
states in fact commonly exhibit a strong mixture of dominant $^2
8[56]$ and $^2 8[70]$ configurations.  The shift due to 't~Hooft's
force hence leads to a structure, which then lies in a region between
roughly 1900 and 2000 MeV in model ${\cal A}$ and around 2000 MeV
in model ${\cal B}$.  This substructure lies fairly
well in between the other members of the $3\hbar\omega$ shell and the upper
part of the $1\hbar\omega$ shell and hence it overlaps with the upper
part of the positive parity $2\hbar\omega$ band.  The arrangement
of the $3\hbar\omega$ and $2\hbar\omega$ bands is quite similar
to that of the $2\hbar\omega$ and $1\hbar\omega$ shells, leading
there to the appearance of approximate parity doublets discussed
above. In fact, similar predicted alignments even appear between the
higher alternating even- and odd-parity bands, {\it e.g.} between the
lowest states of the $4\hbar\omega$ band and the upper part of the
$3\hbar\omega$ band.

\begin{figure}[!h]
  \begin{center}
    \epsfig{file={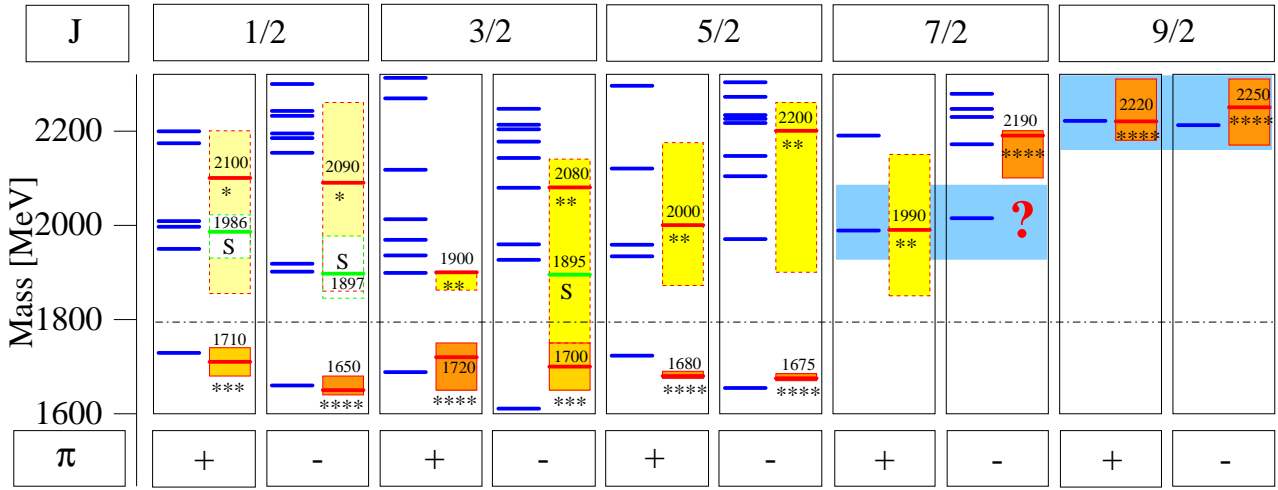},width=170mm}
  \end{center}
\caption{Approximate parity doublets due to the overlap of the
  negative-parity $3\hbar\omega$ shell with the positive-parity
  $2\hbar\omega$ shell and $4\hbar\omega$ shell, respectively, as predicted in
  model ${\cal A}$.  For an explicit illustration how these doublet
  structures are generated due to 't~Hooft's instanton-induced interaction see
  the figs. \ref{fig:PDgvar1hw2hw3hw} and \ref{fig:PDregge} and the
  corresponding detailed discussion of instanton effects in subsect.
  \ref{subsec:studying_inst_eff}.}
\label{fig:PD2hw3hw4hw}
\end{figure}

Now let us discuss in detail to what extent these predicted intra-band
structures along with the corresponding positioning of the alternating
even- and odd-parity bands ({\it i.e.} $4\hbar\omega \leftrightarrow
3\hbar\omega$ and $2\hbar\omega\leftrightarrow 3\hbar\omega$) are
indeed realized in the empirical nucleon mass spectrum. In the
$N\frac{9}{2}^-$ sector both models predict only a single state
belonging to the upper part of the $3\hbar\omega$ shell, which is the
lowest excitation in this sector. The calculated masses at 2212 MeV
and 2170 MeV in model ${\cal A}$ and model ${\cal B}$,
respectively, match the single observed four-star resonance
$N\frac{9}{2}^-(2250,\mbox{****})$ in the $G_{19}$ partial wave quite
well. Similar to the lowest predicted states in the $N\frac{7}{2}^+$
and $N\frac{5}{2}^-$ sector also this state shows an almost pure $^4
8[70]$ configuration and thus remains totally unaffected by the
instanton-induced interaction. In the corresponding $N\frac{9}{2}^+$
sector with the same spin but with opposite positive parity both
models indeed can even account for the low position of the
$N\frac{9}{2}^+(2220)$. Both predict for this state the same
mass value of 2221 MeV.  Consequently, our two alternative models can
reproduce the striking parity-doublet structure
$N\frac{9}{2}^+(2220,\mbox{****})$--$N\frac{9}{2}^-(2250,\mbox{****})$
in excellent agreement with the well-established experimental
findings; see also fig. \ref{fig:PD2hw3hw4hw}.  In this respect we
should mention that in contrast to the negative-parity $3\hbar\omega$
state this positive-parity $4\hbar\omega$ state is strongly influenced
by 't~Hooft's force, since in both models its spin-flavor part is
dominantly a $^2 8[56]$ configuration with an additional strong
admixture of $^2 8[70]$.  Consequently, this state is significantly
lowered (together with a group of other states belonging to the
$4\hbar\omega$ band). It is remarkable that the 't~Hooft coupling
$g_{nn}$ (as fixed by the $\Delta-N$-splitting) is just the
right size to produce this (almost degenerate) parity doublet.

In the $N\frac{7}{2}^-$ sector both model variants predict the first
excitation roughly 170 MeV too light compared to the four-star
resonance $N\frac{7}{2}^-(2190,\mbox{****})$: model ${\cal A}$
predicts this state at 2015 MeV and model ${\cal B}$ at 2024 MeV.
In both models this state is strongly lowered by 't~Hooft's force,
since in both model variants the states contain a dominant $^2 8[70]$
configuration ($\sim 65\%$) with an additional quite strong admixture
of a $^2 8[56]$ contribution ($\sim 32\%$). The second excited state
in this sector, however is hardly influenced by 't~Hooft's force,
since it is dominantly $^4 8 [70]$ in both models. Thus, it appears in
the upper part of the $3\hbar\omega$ band and matches the
$N\frac{7}{2}^-(2190,\mbox{****})$ quite well: model ${\cal A}$
predicts the position at 2171 MeV and model ${\cal B}$ at 2177 MeV
in nice agreement with the empirical mass value.  It is quite
interesting to speculate whether the too low predicted position of the
first excitation is really a shortcoming of our model or if indeed the
first excited resonance in $N\frac{7}{2}^-$ should appear below
$N\frac{7}{2}^-(2190,\mbox{****})$ at about 2015 MeV.  Let us discuss
this alternative interpretation on the basis of our spectroscopic
results in the $3\hbar\omega$ shell in general. We should mention that
a collective model of baryon masses developed by Bijker, Iachello and
Leviatan \cite{BIL94,BIL00} shows a similar result in the
$N\frac{7}{2}^-$ sector. But in this model the nucleon states of the
$3\hbar\omega$ band seem in general to be predicted too light. In our
models, however, this is not the case, since a comparatively low
position is observed only for a group of altogether six states of the
$3\hbar\omega$ shell with spins $J^\pi$ from $\frac{1}{2}^-$ to
$\frac{7}{2}^-$. This group is lowered with respect to the other
states of this shell due to the selective action of 't~Hooft's force.
The lowest excitation predicted in $N\frac{7}{2}^-$ is just the state
with the highest spin $J=\frac{7}{2}$ in this group . Hence,
concerning the rather low predicted position of the first excited
$N\frac{7}{2}^-$ state the question is, how realistic this well
separated part of the $3\hbar\omega$ shell in fact is.  In this
respect, the analysis of the new SAPHIR data on $\eta'$ and kaon
photoproduction, which show evidence for resonances in the $S_{11}$
and $D_{13}$ partial waves, respectively, is very fortunate. Indeed,
the determined resonance positions strongly support this
structure. Let us restrict this discussion to the more realistic model
${\cal A}$: the resonance position extracted from the $S_{11}$
partial wave of $\gamma p \rightarrow p \eta'$ at 1897 MeV (with the
error range of 1845 MeV to 1977 MeV) excellently agrees with the two
predicted $N\frac{1}{2}^-$ states of model ${\cal A}$ in this
region at 1901 and 1918 MeV.  Also the determined resonance position
extracted from the $D_{13}$ partial wave of $\gamma p \rightarrow K^+
\Lambda$ at almost the same mass value of 1895 MeV (unfortunately
there is no error assigned to this value) fairly agrees with the
prediction of the two close lying $N\frac{3}{2}^-$ states in model
${\cal A}$ at 1926 and 1959 MeV. In the $N\frac{5}{2}^-$ our model
${\cal A}$ predicts a single low-lying state belonging to the same
structure at 1970 MeV which still is ''missing'' and finally in the
$N\frac{7}{2}^-$ sector we correspondingly have the low predicted
state in question at 2015 MeV which indeed should appear far below the
$N\frac{7}{2}^-(2190,\mbox{****})$, if we take this instanton-induced
structure seriously. An experimentally proven existence of such
comparatively low-lying states in the $N\frac{5}{2}^-$ and
$N\frac{7}{2}^-$ sector might become a sensitive test for our model
${\cal A}$ and especially for the residual 't~Hooft interaction
used in this model. In this respect, we should note here that the
prediction of such well separated states in $N\frac{5}{2}^-$ and
$N\frac{7}{2}^-$ seems to depend strongly on the residual quark-quark
interaction employed: other quark models which use a different
residual force (as for instance one-gluon-exchange \cite{CaIs86}) do
not predict such states. Moreover, it is worth to anticipate already
here that the corresponding lowest $\Lambda$-state in the
$\Lambda\frac{7}{2}^-$ sector (see our subsequent paper \cite{Loe01c})
is strongly lowered by the same instanton-induced effect and thus
nicely explains the comparatively low position of the
$\Lambda\frac{7}{2}^-(2100,\mbox{****})$.  \\ Figure
\ref{fig:PD2hw3hw4hw} shows the relative arrangements of parts of the
negative-parity $3\hbar\omega$ band and parts of the positive-parity
$2\hbar\omega$ and $4\hbar\omega$ bands in model ${\cal A}$.
In contrast to fig. \ref{fig:NucM2} the sectors with the same total
spin $J$ and opposite parity are now directly displayed side by side.
In addition, also the mass region around 1700 MeV with the previously
discussed approximate parity doublets of the $2\hbar\omega$ and
$1\hbar\omega$ bands is shown for comparison.  The figure nicely
illustrates the relatively strong overlap of the predicted structures
of the negative-parity $3\hbar\omega$ and the positive-parity
$2\hbar\omega$ shell. Experimental evidence supporting this comes from
the comparatively low positions of the newly discovered 'SAPHIR
resonances'.  The predicted overlap of the $3\hbar\omega$ and
$4\hbar\omega$ shell nicely reproduces the observed parity doublet
pattern in the $N\frac{9}{2}^\pm$ sectors.  Referring back to our
discussion of a possibly low-lying state in the
$N\frac{7}{2}^-$ sector, it is very interesting that the lowest state
calculated in this sector at 2015 MeV (in model ${\cal A}$)
together with the experimentally observed lowest positive-parity
excitation $N\frac{7}{2}^+(1990,\mbox{**})$ ($\hat=$ 1988 MeV in model
${\cal A}$) fits much better into the scheme of approximate parity
doublets than the lowest state observed, up to now the
$N\frac{7}{2}^-(2190,\mbox{****})$, does.\\

Finally, let us compare our results with the highest mass states of
the experimental nucleon spectrum, {\it i.e.} the lowest excitations
in the sectors $N\frac{11}{2}^-$ and $N\frac{13}{2}^+$ reported by the
Particle Data Group \cite{PDG00}. These are the three- and two-star
resonances $N\frac{11}{2}^-(2600,\mbox{***})$ and
$N\frac{13}{2}^+(2700,\mbox{**})$, which in the notation of the
oscillator model should belong to the negative-parity $5\hbar\omega$
and the positive-parity $6\hbar\omega$ shell, respectively. The
positions predicted for the lightest few $5\hbar\omega$ and
$6\hbar\omega$ states in both models are given in tables
\ref{tab:N5hwband} and \ref{tab:N6hwband}, respectively.

\begin{figure}[!h]
\begin{center}
\epsfig{file={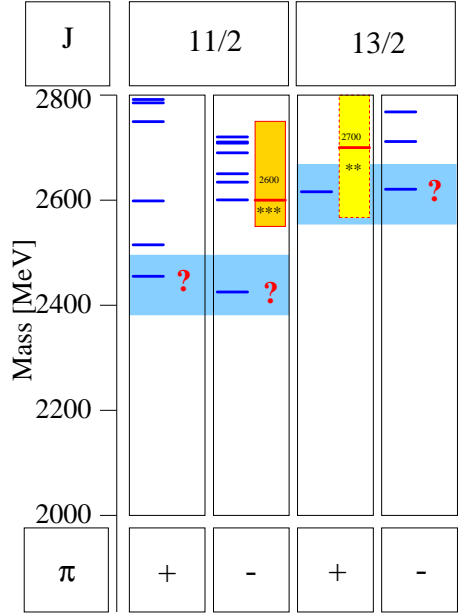},width=60mm}
\end{center}
\caption{Approximate parity doublets in $N\frac{11}{2}^\pm$
and $N\frac{13}{2}^\pm$ (in model ${\cal A}$).}
\label{fig:PD4hw5hw6hw}
\end{figure}

In the $N\frac{11}{2}^-$ sector the situation is very similar to that
found in the $N\frac{7}{2}^-$ sector. Again the lowest excitation is
strongly influenced by 't~Hooft's force and thus is strongly lowered
relative to other states in this sector. Consequently, the first
excited state is predicted roughly 160-175 MeV too low compared to the
position of the resonance $N\frac{11}{2}^-(2600,\mbox{***})$: model
${\cal A}$ predicts the first excitation at 2425 MeV and model
${\cal B}$ at 2441 MeV.  The higher excitations, however, again fit
nicely within the range of possible values of the
$N\frac{11}{2}^-(2600,\mbox{***})$ (see also table
\ref{tab:N5hwband}). Again, we can just speculate, if this is a
shortcoming of our models or if there is really a state below the
$N\frac{11}{2}^-(2600,\mbox{***})$. But once again it is quite
interesting to note that the predicted first excitation in this sector
forms an approximate parity doublet structure together with the first
excitation predicted in the $N\frac{11}{2}^+$ sector which belongs to
the positive-parity $4\hbar\omega$ band.  Unfortunately, no resonance
has been seen experimentally in $N\frac{11}{2}^+$ hitherto. Model
${\cal A}$ predicts the first excitation in $N\frac{11}{2}^+$ at
2455 MeV and model ${\cal B}$ at 2402 MeV (see also table
\ref{tab:N4hwband}). A graphical illustration of this approximate
spin-$11/2$ parity doublet in model ${\cal A}$ is given in the left
part of fig.  \ref{fig:PD4hw5hw6hw}.

The lowest excitation $N\frac{13}{2}^+(2700,\mbox{**})$ observed in
the $N\frac{13}{2}^+$ sector is the highest spin of a resonance in the
nucleon sector measured at all. This resonance is the highest lying
member of the positive-parity $N$-Regge trajectory (see discussion
below).  Due to a lowering by 't~Hooft's force, both models predict
the first excited state in the $N\frac{13}{2}^+$ sector well isolated,
{\it i.e.}  roughly 160-180 MeV below the other excited states. The
predicted masses are close to the measured position and lie within the
range of possible values for this two-star resonance: model
${\cal A}$ yields 2616 MeV and model ${\cal B}$ 2619 MeV (see
also table \ref{tab:N6hwband}). Again we find this lowest excited
model state in $N\frac{13}{2}^+$ belonging to a parity doublet
structure together with the corresponding lowest excitation in the
sector $N\frac{13}{2}^-$ with the same spin but negative parity. This
negative-parity state belongs to the $5\hbar\omega$ shell, and model
${\cal A}$ and model ${\cal B}$ predict the position at 2621 MeV
and 2587 MeV, respectively (see also table \ref{tab:N5hwband}). The
right part of fig. \ref{fig:PD4hw5hw6hw} depicts this parity doublet
structure in model ${\cal A}$. Unfortunately, the first excitation
of $N\frac{13}{2}^-$ has not been discovered yet.
\begin{table}[!h]
\center
\begin{tabular}{ccccccc}
\hline
Exp. state   &PW&${J^\pi}$        & Rating     & Mass range [MeV]& Model state &  Model state \\
\cite{PDG00} &  &                      &            & \cite{PDG00}    & in model ${\cal A}$& in model ${\cal B}$  \\
\hline
         &$G_{19}$&${\frac{9}{2}^-}$&      &          &$\MSN(9,-,2,2402)$&$\MSN(9,-,2,2451)$\\
         &        &                 &      &          &$\MSN(9,-,3,2542)$&$\MSN(9,-,3,2586)$\\
         &        &                 &      &          &$\MSN(9,-,4,2579)$&$\MSN(9,-,4,2602)$\\
\hline
         &        &                 &      &          &$\MSN(11,-,1,2425)$&$\MSN(11,-,1,2441)$\\\\
$N(2600)$&$I_{1\;11}$&${\frac{11}{2}^-}$&***&2550-2750&$\MSN(11,-,2,2600)$&$\MSN(11,-,2,2593)$\\
         &        &                 &      &          &$\MSN(11,-,3,2634)$&$\MSN(11,-,3,2606)$\\
         &        &                 &      &          &$\MSN(11,-,4,2650)$&$\MSN(11,-,4,2629)$\\  
\hline
         &$I_{1\;13}$&${\frac{13}{2}^-}$&      &      &$\MSN(13,-,1,2621)$&$\MSN(13,-,1,2587)$\\
         &           &                 &      &       &$\MSN(13,-,2,2712)$&$\MSN(13,-,2,2655)$\\
\hline
\end{tabular}
\caption{Calculated positions for the lightest few negative-parity
nucleon states in the $5\hbar\omega$ shell with $J\geq\frac{9}{2}$ in comparison to the
corresponding experimental mass values taken from \cite{PDG00}. Notation as in
table \ref{tab:1hwband}.}
\label{tab:N5hwband}
\end{table} 
\begin{table}[!h]
\center
\begin{tabular}{ccccccc}
\hline
Exp. state   &PW&${J^\pi}$        & Rating     & Mass range [MeV]& Model state &  Model state \\
\cite{PDG00} &  &                      &            & \cite{PDG00}    & in model ${\cal A}$& in model ${\cal B}$  \\
\hline 
         &$H_{1\;11}$&${\frac{11}{2}^+}$&  &          &$\MSN(11,+,3,2598)$&$\MSN(11,+,3,2640)$\\
         &        &                 &      &          &$\MSN(11,+,4,2749)$&$\MSN(11,+,4,2757)$\\
         &        &                 &      &          &$\MSN(11,+,5,2785)$&$\MSN(11,+,5,2781)$\\
\hline
$N(2700)$&$K_{1\;13}$&${\frac{13}{2}^+}$&**&2567-3100&$\MSN(13,+,1,2616)$&$\MSN(13,+,1,2619)$\\\\
         &        &                 &      &          &$\MSN(13,+,2,2800)$&$\MSN(13,+,2,2777)$\\
         &        &                 &      &          &$\MSN(13,+,3,2811)$&$\MSN(13,+,3,2782)$\\  
\hline
\end{tabular}
\caption{Calculated positions for the lightest few positive-parity
nucleon states in the $6\hbar\omega$ shell with $J\geq\frac{11}{2}$ in comparison to the
corresponding experimental mass values taken from \cite{PDG00}. Notation as in
table \ref{tab:1hwband}.}
\label{tab:N6hwband}
\end{table}

To summarize our investigations concerning the relative alignment of the
alternating even- and odd-parity bands in the excited nucleon spectrum, let us
finally stress the most striking feature of our model that could be exposed in
the course of this discussion and becomes immediately evident from
figs. \ref{fig:PD2hw3hw4hw} and \ref{fig:PD4hw5hw6hw}. It is the systematical
occurrence of approximately degenerate states with the same spin
and opposite parity. On the one hand our models nicely reproduce the
approximate doublet structures which in fact are observed experimentally. On the other hand,
these doublets also appear, where the present experimental situation either seems to deviate
from such a parity doublet structure, or members of these doublets are
''missing''. In fact, we predict for really all the lowest excitations with spins
from $J=\frac{5}{2}$ to $J=\frac{13}{2}$ such a pattern of approximate parity
doublets. As already partly indicated in the previous discussion, this
systematics originates from 't~Hooft's force: the two members
of the doublet belong to two adjacent shells with opposite parity and 
't~Hooft's force lowers one of these 
strongly, whereas the other one remains totally unaffected. In this
way, both states become approximately degenerate in energy. For a more detailed
discussion of this scenario in the context of a detailed investigation of instanton-induced effects we refer
to subsect. \ref{subsec:studying_inst_eff}.

\subsubsection{The positive-parity $N$-Regge trajectory}
Finally, let us conclude the discussion of the nucleon
spectrum with the states $N\frac{1}{2}^+(939,\mbox{****})$,
$N\frac{5}{2}^+(1680,\mbox{****})$, $N\frac{9}{2}^+(2220,\mbox{****})$
and $N\frac{13}{2}^+(2700,\mbox{**})$ of the positive-parity $N$-Regge 
trajectory, which belong to a sequence of three-quark states
with increasing separations of the quarks.  Figures
\ref{fig:NucM2} and \ref{fig:NucM1} show that all states of the
trajectory are fairly well described in both confinement models
${\cal A}$ and ${\cal B}$; see also table \ref{tab:n_regge}, in
which the calculated positions of Regge states are explicitly summarized
and compared to the experimental mass values.
This result shows that the string-like, flavor-independent confinement force of both
models (with parameters fixed to account for the
positive-parity $\Delta$-trajectory up to highest orbital excitations)
works equally well also for nucleon excitations with increasing
quark separations.  But here one should note the substantial difference to the
states of the $\Delta$-Regge trajectory: in contrast to the
$\Delta$-trajectory, all states of the $N$-trajectory are quite strongly
lowered due to the action of the instanton-induced force.  The nucleon
ground state $N(939)$ shows the largest downward mass shift and with
increasing mass of the Regge states the mass shift decreases. In
this respect it is very interesting to investigate to what extent 't~Hooft's
force actually influences the Regge behavior $M^2\sim J$.

\begin{figure}[!h]
  \begin{center}
    \input{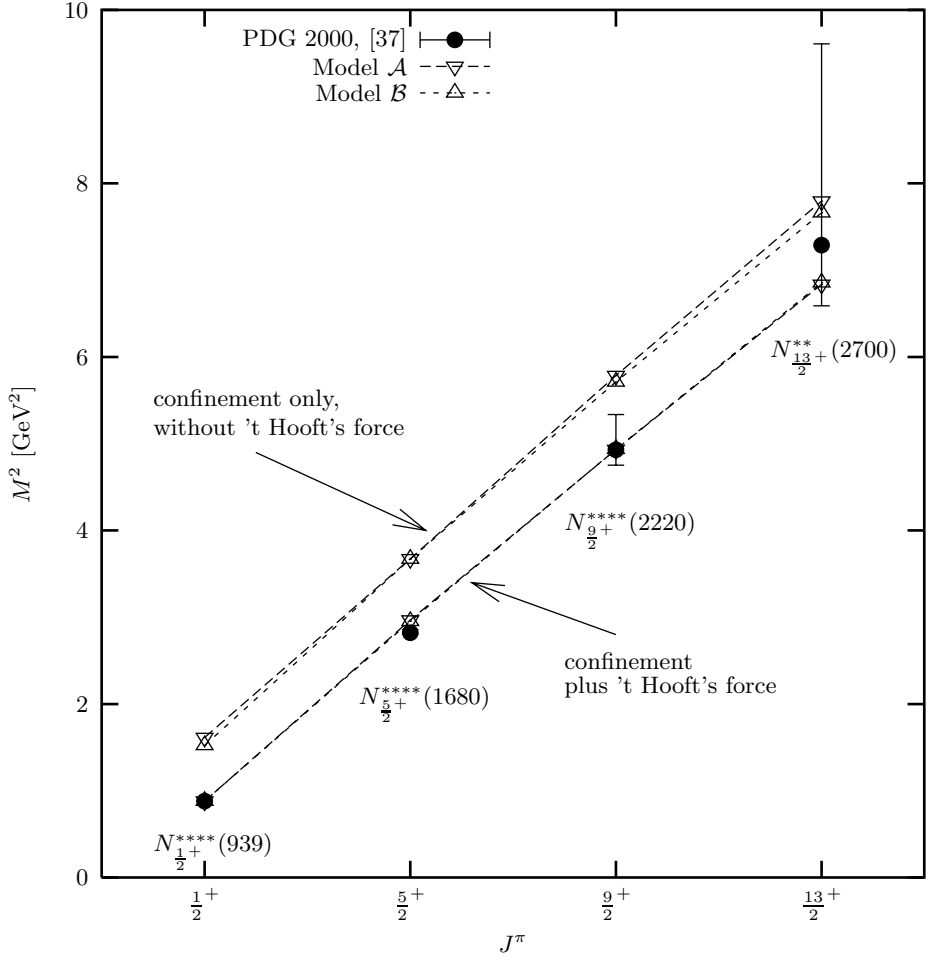}
  \end{center}
\caption{Chew-Frautschi plot ($M^2$ {\it vs.} $J$) of the
  positive-parity $N$-Regge trajectory $N\frac{1}{2}^+$,
  $N\frac{5}{2}^+$, $N\frac{9}{2}^+$, $N\frac{13}{2}^+$, $\ldots$, in the
  models ${\cal A}$ and ${\cal B}$ (lower curves) compared to
  experimental masses from the Particle Data Group (see \cite{PDG00}).
  The upper curves show the trajectories without influence of 't~Hooft's
  instanton-induced interaction. Note the  remarkable fact that both models exhibit
  the correct linear Regge characteristic $M^2 \sim J$ with almost the
  same slope in both cases, {\it i.e.} without and even with the contribution of
  the instanton force.  Hence, adjusting the 't~Hooft coupling $g_{nn}$ to
  describe just the ground state correctly in both models, the whole model trajectory fits
  the empirical Regge behavior in excellent quantitative agreement. See
  table \ref{tab:n_regge} for explicit values.}
\label{fig:n_regge}
\end{figure}

Figure \ref{fig:n_regge} shows the Chew-Frautschi plot
($M^2$ {\it versus} $J$) of the positive-parity $N$-Regge trajectory
obtained in model ${\cal A}$ and ${\cal B}$, respectively.
To illustrate the influence of 't~Hooft's instanton-induced interaction
we additionally displayed the trajectory as given by the
confinement forces of models ${\cal A}$ and ${\cal B}$
alone. Thus, neglecting 't~Hooft's force, both confinement variants yield
the characteristic Regge behavior $M^2\propto J$, as one would
expect from the results for the $\Delta$-Regge trajectory discussed before.
In particular, the slope is approximately the same as in the
$\Delta$-sector. It  is quite astonishing that the effect of the
instanton force on the states of the trajectory indeed is such that it
maintains the linear Regge characteristic $M^2\propto J$ with
almost the same slope, in nice agreement with phenomenology.
Hence, adjusting the effective 't~Hooft coupling $g_{nn}$ to describe
just the ground state ({\it i.e.} the $\Delta-N$ splitting) correctly, the whole model trajectory shows an
equally large downward shift in the square of the baryon masses
$M^2$ thus leading to an excellent quantitative agreement with the
empirical linear $N$-Regge trajectory. This nontrivial
compatibility of the instanton-induced effects with the observed linear Regge 
characteristic in the nucleon spectrum is a quite remarkable
and an interesting feature of 't~Hooft's force. Moreover, figure \ref{fig:n_regge}
convincingly demonstrates once more that even in the highest mass regions of the 
nucleon spectrum, instanton-induced effects are crucial for the correct description
of resonance positions. 
\begin{table}[!h]
\center
\begin{tabular}{cccccc}
\hline
Regge         & Rating & $J^\pi$           & exp. Mass [MeV] & Mass [MeV]         & Mass [MeV]\\
state               &        &                   &                 &  Model ${\cal A}$& Model ${\cal B}$\\
\hline
$N(939)$       & ****   & $\frac{1}{2}^+$   & 939             & 939              &  939 \\ 
$N(1680)$      & ****   & $\frac{5}{2}^+$   & 1675-1690       & 1723             & 1718 \\
$N(2220)$      & ****   & $\frac{9}{2}^+$   & 2180-2310       & 2221             & 2221 \\
$N(2700)$      & **     & $\frac{13}{2}^+$  & 2567-3100       & 2616             & 2619\\
\hline
\end{tabular}
\caption{Position of states belonging to the positive-parity $N$-Regge 
trajectory calculated in the models ${\cal A}$ and
${\cal B}$ in comparison to the experimental resonance positions \cite{PDG00}.
For a graphical presentation see fig. \ref{fig:n_regge}.}
\label{tab:n_regge}
\end{table}

\subsection{A study of instanton effects in the excited nucleon spectrum}
\label{subsec:studying_inst_eff}
So far we presented a detailed discussion of the predicted nucleon
spectrum in both confinement models ${\cal A}$ and ${\cal B}$
with the strength of the instanton-induced interaction fixed to
match the observed position of the nucleon ground state
$N(939)$.  We investigated the calculated structures of the single
positive- and negative-parity shells and compared these to the
hitherto observed experimental patterns.  In fact, we found a remarkably good
agreement between the predictions of our model version ${\cal A}$
and the empirical nucleon mass spectrum in the lower resonance regions as
well as for the higher and highest mass regions including orbital
excitations up to $J=\frac{13}{2}$ (Regge
trajectory). In particular, all the striking hyperfine splittings within
the intra-band structures and accordingly the arrangements between the
alternating even- and odd-parity shells could be nicely reproduced in
at least qualitative but mostly even in completely quantitative
agreement with the experimental findings. In particular, we clearly reproduced
\begin{itemize}
\item
the important downshift of the Roper resonance and  three other low-lying
states of the $2\hbar\omega$ shell around 1700 MeV,
\item 
the hyperfine structure of the five states in the $1\hbar\omega$ band,
\item
the fine structure of the $3\hbar\omega$ band as far as it is indicated by the new
SAPHIR photoproduction results,
\item
the approximate parity doublets due to partial overlapping 
of shells: $1\hbar\omega\leftrightarrow 2\hbar\omega$,
$2\hbar\omega\leftrightarrow 3\hbar\omega$, 
$3\hbar\omega\leftrightarrow 4\hbar\omega$,
$4\hbar\omega\leftrightarrow 5\hbar\omega$ and
$5\hbar\omega\leftrightarrow 6\hbar\omega$.
\end{itemize}
In the discussion of the foregoing subsection we already mentioned
that in our fully covariant model
${\cal A}$ all these striking features of the excited nucleon
spectrum arise due to 't~Hooft's instanton-induced quark-quark
interaction. To supplement this
discussion we shall now illustrate in some more detail that the
effects of this residual interaction are in fact responsible for
generating all these structures simultaneously.  In this respect it is
instructive to study how the 't~Hooft interaction affects the energy
levels when, starting from the case with confinement only, it is
switched on and its strength is gradually increased.  It is our aim to
demonstrate in this way that prominent features of the spectrum like the
$\Delta-N$ ground-state splitting and the conspicuous low position of
the Roper resonance can be put along with all the other observed
phenomena into a somewhat wider perspective, {\it i.e.} that instanton-induced effects 
indeed dominate the fine structure of the {\it whole}
nucleon spectrum.

Another issue that has to be clarified in this
context is the difference between the predictions of model
${\cal A}$ and model ${\cal B}$. Although model ${\cal B}$
could also account for the most features  discussed, it strongly
failed concerning the low position of the Roper resonance in the
$N\frac{1}{2}^+$ sector and moreover predicted a much too large
splitting of the two lowest $S_{11}$ resonances in the corresponding
$N\frac{1}{2}^-$ sector with negative parity.  As already mentioned,
this indicates that the action of 't~Hooft's force might
strongly depend on the Dirac structures chosen for the three-body
confinement kernel and on the corresponding relativistic effects induced
in combination with the embedding map of the Salpeter
amplitudes. Also this aspect shall be analyzed here in some more
detail.\\

To study the effects of 't~Hooft's force in model ${\cal A}$ we display the
dependence of the nucleon masses on the effective 't~Hooft coupling $g_{nn}$
in figs. \ref{fig:Model2N+gvar} and \ref{fig:Model2N-gvar} for the
positive and negative parity sector, respectively. The corresponding results
of model ${\cal B}$ are shown in fig. \ref{fig:Mod1aNgvar}. In each
sector with spin $J$ and parity $\pi$ the leftmost spectrum in each column
shows the spectrum obtained with the three-body confinement kernel alone. The curves show the
change of the spectrum as a function of the 't~Hooft coupling $g_{nn}$:
Starting from the case with confinement only the coupling is gradually
increased and finally is fixed to reproduce the $\Delta-N$ splitting (right
spectrum in each column). The rightmost spectrum shows for comparison the experimental resonance
positions as before.
\subsubsection{Pure confinement spectra}
Let us first consider the spectra in figs. \ref{fig:Model2N+gvar},
\ref{fig:Model2N-gvar} and \ref{fig:Mod1aNgvar} as they are determined by the
two different three-body confinement kernels alone, {\it i.e.}  in the case
$g_{nn}=0$ (leftmost spectrum in each column): Without 't~Hooft's force the linear confining
interaction arranges the nucleon spectrum into a sequence of alternating even-
and odd-parity shells.  As one expects, the situation is quite similar to the
spectrum of $\Delta$-resonances. The states of these rather narrow bands are
clustered around a common mean value, which, however, fairly agrees only
with the upper parts of the rather wide spread band structures observed
experimentally.\\In fact some of these experimentally observed resonances even
are already well described by single model states in both confinement
versions. This, for instance, is the case for the three resonances
$N\frac{5}{2}^-(1675,\mbox{****})$, $N\frac{7}{2}^+(1990,\mbox{**})$ and
$N\frac{9}{2}^-(2250,\mbox{****})$. These are single states with maximum total
spin $J$ in the $1\hbar\omega$, $2\hbar\omega$ and $3\hbar\omega$ shell,
respectively.  In general, in the naive oscillator shell model, the involved
relative orbital angular momenta in the $N\hbar\omega$ shell can
be maximally combined to a total orbital angular momentum $L_{\rm max}=
N$. To achieve the maximal total spin $J_{\rm max}(N)$ in this shell,
$L_{\rm max}= N$ has to be coupled with the symmetric $S=\frac{3}{2}$ internal
spin-quartet function to get $J_{\rm max}(N)= N + \frac{3}{2}$.  Thus, we find
for the different $N\hbar\omega$ multiplets of the oscillator shell model the
maximum total spins as displayed in table \ref{tab:max_tot_spins}, where the
corresponding states have internal spin $S=\frac{3}{2}$ and are just the
lowest excited states in the sectors $J_{\rm max}^\pi(N)$.
\begin{table}[!h]
\begin{center}
\begin{tabular}{|c||c|c|c|c|c|c|}
\hline
Shell $N\hbar\omega$    &$1\hbar\omega$ &$2\hbar\omega$ &$3\hbar\omega$ &$4\hbar\omega$  &$5\hbar\omega$ & $6\hbar\omega$\\
\hline
$J_{\rm max}^\pi(N)= N + \frac{3}{2}$ &$\frac{5}{2}^-$&$\frac{7}{2}^+$&$\frac{9}{2}^-$&$\frac{11}{2}^+$&$\frac{13}{2}^-$&$\frac{15}{2}^+$\\
\hline
\end{tabular}
\caption{Maximum total spins of states arising in the different oscillator shells.}
\label{tab:max_tot_spins}
\end{center}
\end{table}

Analyzing the spin-flavor $SU(6)$ configurations of the lowest excited states
with spin and parity $J^\pi=\frac{5}{2}^-$, $\frac{7}{2}^+$, $\frac{9}{2}^-$,
$\frac{11}{2}^+$ and $\frac{13}{2}^-$ in our models ${\cal A}$ and
${\cal B}$, we find that they indeed have in common to be almost pure $^4
8[70]$ configurations ($>99 \%$), see table
\ref{tab:ConfMixunaffected}. Consequently, due to the selection rules of
't~Hooft's force, these states are expected to be hardly influenced by the
residual interaction and thus indeed should be almost determined by the
confining force alone. In fact, switching on 't~Hooft's force, these states,
{\it i.e.} their masses (as shown in figs.
\ref{fig:Model2N+gvar}, \ref{fig:Model2N-gvar} and \ref{fig:Mod1aNgvar}) as well as
their Salpeter amplitudes (table \ref{tab:ConfMixunaffected}) are by no
means affected with increasing strength $g_{nn}>0$, in nice accordance with
the observed resonances
$N\frac{5}{2}^-(1675,\mbox{****})$, $N\frac{7}{2}^+(1990,\mbox{**})$ and
$N\frac{9}{2}^-(2250,\mbox{****})$ in these sectors.
\begin{table}[!h]
\begin{center}
\begin{tabular}{cc}
{\large \textbf{Model} ${\cal A}$:}
& Mass and configuration mixing \\
& with \underline{and} without 't~Hooft's force\\
\begin{tabular}{|c|c||cc|} 
\hline
Shell&
$J_{\rm max}^\pi$ &
Exp. state&Rating\\[1mm]
&&{$\Delta M$ [MeV]}&\\
\hline
\hline
$1\hbar\omega$&$\frac{5}{2}^-$&N(1675)&****\\
              &               &{1670-1685}       &\\
\hline
\hline
$2\hbar\omega$&$\frac{7}{2}^+$&N(1990)& **\\
              &               &{1850-2150}       &\\
\hline
\hline
$3\hbar\omega$&$\frac{9}{2}^-$&N(2250)& ****\\
              &               &{2170-2310}       &\\
\hline
\hline
$4\hbar\omega$&$\frac{11}{2}^+$&''missing''&\\
              &               &&\\
\phantom{$4\hbar\omega$}&\phantom{$\frac{11}{2}^+$}&''missing''&\\
              &               &&\\
\hline
\hline
$5\hbar\omega$&$\frac{13}{2}^-$&''missing''&\\
              &               &&\\
\phantom{$5\hbar\omega$}&\phantom{$\frac{13}{2}^+$}&''missing''&\\
              &               &&\\
\hline
\end{tabular}&
\begin{tabular}{|c|c|cccc|}
\hline
Mass & 
pos. & 
$\!\!{}^2 8 [56]\!\!$&
$\!\!{}^2 8 [70]\!\!$&
$\!\!{}^4 8 [70]\!\!$&
$\!\!{}^2 8 [20]\!\!$\\[1mm]
$[$MeV$]$& 
neg. & 
$\!\!{}^2 8 [56]\!\!$&
$\!\!{}^2 8 [70]\!\!$&
$\!\!{}^4 8 [70]\!\!$&
$\!\!{}^2 8 [20]\!\!$\\
\hline
\hline
1655 &   98.4 & 0.0 & 0.0 &  {\bf \underline{98.4}} &      0.0\\
\phantom{$\frac{5}{2}^-$} &    1.6 & 0.1 & 0.3  &      1.0                   &      0.1\\
\hline
\hline
1989 &   98.1 & 0.0 & 0.0 &  {\bf \underline{98.1}} &      0.0\\
\phantom{$\frac{7}{2}^-$}     &    1.9 & 0.2 & 0.2  &      1.4                   &     0.2\\
\hline
\hline
2212 &   98.2 & 0.0 & 0.0 &  {\bf \underline{98.2}} &      0.0\\
\phantom{$\frac{9}{2}^-$}     &    1.8 & 0.1 & 0.3  &      1.2                   &     0.2\\
\hline
\hline
2455 & 98.0 & 0.0 & 0.0 &  {\bf \underline{97.9}} & 0.0      \\
\phantom{$\frac{11}{2}^+$}     & 2.0     &0.2 &0.2  &1.5                    &0.2    \\
\hline
2515 & 97.7 & 0.0 & 0.0 &  {\bf \underline{97.7}} & 0.0      \\
\phantom{$\frac{11}{2}^+$}     & 2.3     &0.1 &0.4  &1.8                    &0.1    \\
\hline
\hline
2621 & 98.3 & 0.0 & 0.0 &  {\bf \underline{98.3}} & 0.0     \\
\phantom{$\frac{13}{2}^-$}     & 1.7    &0.0 & 0.3 & 1.2                    &0.2    \\
\hline
2712 &98.0 & 0.0 & 0.0 &  {\bf \underline{98.0}} & 0.0     \\
\phantom{$\frac{13}{2}^-$}     & 2.0    &0.2 & 0.1 & 1.6                   & 0.2   \\
\hline
\end{tabular}\\
{\large \textbf{Model} ${\cal B}$:}
& Mass and configuration mixing \\
& with \underline{and} without 't~Hooft's force\\
\begin{tabular}{|c|c||cc|} 
\hline
Shell&
$J_{\rm max}^\pi$ &
Exp. state&Rating\\[1mm]
&&{$\Delta M$ [MeV]}&\\
\hline
\hline
$1\hbar\omega$&$\frac{5}{2}^-$&N(1675)&****\\
              &               &{1670-1685}       &\\
\hline
\hline
$2\hbar\omega$&$\frac{7}{2}^+$&N(1990)& **\\
              &               &{1850-2150}       &\\
\hline
\hline
$3\hbar\omega$&$\frac{9}{2}^-$&N(2250)& ****\\
              &               &{2170-2310}       &\\
\hline
\hline
$4\hbar\omega$&$\frac{11}{2}^+$&''missing''&\\
              &               &&\\
\phantom{$4\hbar\omega$}&\phantom{$\frac{11}{2}^+$}&''missing''&\\
              &               &&\\
\hline
\hline
$5\hbar\omega$&$\frac{13}{2}^-$&''missing''&\\
              &               &&\\
\phantom{$5\hbar\omega$}&\phantom{$\frac{13}{2}^+$}&''missing''&\\
              &               &&\\
\hline
\end{tabular}&
\begin{tabular}{|c|c|cccc|}
\hline
Mass & 
pos. & 
$\!\!{}^2 8 [56]\!\!$&
$\!\!{}^2 8 [70]\!\!$&
$\!\!{}^4 8 [70]\!\!$&
$\!\!{}^2 8 [20]\!\!$\\[1mm]
$[$MeV$]$& 
neg. & 
$\!\!{}^2 8 [56]\!\!$&
$\!\!{}^2 8 [70]\!\!$&
$\!\!{}^4 8 [70]\!\!$&
$\!\!{}^2 8 [20]\!\!$\\
\hline
\hline
1622 &   99.9 & 0.0 & 0.0 &  {\bf \underline{99.9}} &      0.0\\
\phantom{$\frac{5}{2}^-$} &    0.1 & 0.0 & 0.0  &      0.1                   &      0.0\\
\hline
\hline
1941 &   99.9 & 0.0 & 0.0 &  {\bf \underline{99.9}} &      0.0\\
\phantom{$\frac{7}{2}^+$}     &    0.1 & 0.0 & 0.0  &      0.1                   &     0.0\\
\hline
\hline
2170 &   99.9 & 0.0 & 0.0 &  {\bf \underline{99.9}} &      0.0\\
\phantom{$\frac{9}{2}^-$}     &    0.1 & 0.0 & 0.0  &      0.1                   &     0.0\\
\hline
\hline
2402 & 99.8 & 0.0 & 0.0 &  {\bf \underline{99.8}} & 0.0      \\
\phantom{$\frac{11}{2}^+$}     & 0.2     &0.0 &0.0  &0.1                    &0.0    \\
\hline
2446 & 99.7 & 0.0 & 0.0 &  {\bf \underline{99.7}} & 0.0      \\
\phantom{$\frac{11}{2}^+$}     & 0.3     &0.0 &0.0  &0.2                    &0.0    \\
\hline
\hline
2587 & 99.8 & 0.0 & 0.0 &  {\bf \underline{99.8}} & 0.0     \\
\phantom{$\frac{13}{2}^-$}     & 0.2    &0.0 & 0.0 & 0.2                    &0.0    \\
\hline
2655 &99.7 & 0.0 & 0.0 &  {\bf \underline{99.7}} & 0.0     \\
\phantom{$\frac{13}{2}^-$}     & 0.3    &0.0 & 0.0 & 0.3                   & 0.0   \\
\hline
\end{tabular}
\end{tabular}
\end{center}
\caption{Masses and configuration mixing of states in each $N\hbar\omega$
  shell with maximal total spin $J_{\rm max}(N)=N+\frac{3}{2}$  ($N=1,\ldots, 5,$
compare table \ref{tab:max_tot_spins}).  For each different contribution to the Salpeter amplitude the
corresponding Salpeter norm is given in $\%$. In each row the upper line
shows the positive and the lower line the negative energy contribution. 
These states remain totally unaffected by 't~Hooft's force, {\it i.e.} the values without and with 't~Hooft's force (right) are
exactly the same. See text for a more detailed explanation.}
\label{tab:ConfMixunaffected}
\end{table}
On the one hand, this result indicates that the confinement force, whose
parameters have been fixed on the phenomenology of the $\Delta$-spectrum,
works equally well also in the nucleon spectrum concerning the arrangement and
positions of the shells. On the other hand it shows that the strong selection
rules of 't~Hooft's interaction are really consistent with the phenomenology
of these three
particular nucleon resonances.\\
Concerning the structures of the shells in the pure confinement spectrum it should
be noted that model ${\cal A}$ exhibits somewhat larger intra-band
splittings than model ${\cal B}$, due to slightly bigger relativistic
spin-orbit effects in model ${\cal A}$ (here compare to the discussion of the
$\Delta$-spectrum). However, unlike the $\Delta$-spectrum, the multiplet
structure of alternating positive and negative parity bands is obviously
strongly broken in the phenomenological nucleon spectrum, and the intra-band
structures generated by the two different confinement kernels by no means can
account for the quite large observed hyperfine splittings in the shells.  This
is most apparent from the experimentally rather well-established structure
of the lower part of the $2\hbar\omega$ shell, where (in case of the low-lying
Roper resonance) the deviations between the pure confinement spectra and the
empirical mass spectrum are even of the same order of magnitude as the
$\Delta-N$ ground-state splitting, which also cannot be accounted for in the
case $g_{nn}=0$. Similar large deviations are also found in higher mass
regions, as {\it e.g.} in  case of the $N\frac{9}{2}^+(2220,\mbox{****})$.  Thus,
apart from the  $\Delta-N$ ground-state hyperfine splitting, also the shortcomings
in the excited pure confinement spectrum strongly indicate missing residual
spin-spin interactions in the nucleon sector.\\ Hence, the question arises, to
what extent 't~Hooft's instanton-induced interaction, which already could
nicely explain the octet-decuplet ground-state splittings, can {\it simultaneously} also
account for the striking  mass splittings observed in the excited nucleon
spectrum.  That this is indeed the case in our model version ${\cal A}$ is
already evident from the previous discussion of the complete nucleon spectrum
but will now be  even more convincingly demonstrated when, starting from the
case with confinement only, the 't~Hooft coupling $g_{nn}$ is turned on and
then is increased gradually until the model ground-state fits the experimental
position of $N\frac{1}{2}^+(939)$. In this respect let us first focus on the
effects in the positive parity spectrum of model ${\cal A}$ shown in fig.
\ref{fig:Model2N+gvar}.

\subsubsection{Instanton-induced effects in model ${\cal A}$ -- positive-parity spectrum}
In the $2\hbar\omega$ band one finds a selective lowering of exactly
four states relative to the other states as required by the
experimental findings and it is quite impressive how the instanton-induced 
effects even shape the pattern of these four model states to
come into considerably good agreement with the observed (well-established) 
pattern at the value of the coupling, where the $\Delta-N$ splitting is reproduced.
\begin{figure}[h]
 \begin{center} 
 \epsfig{file={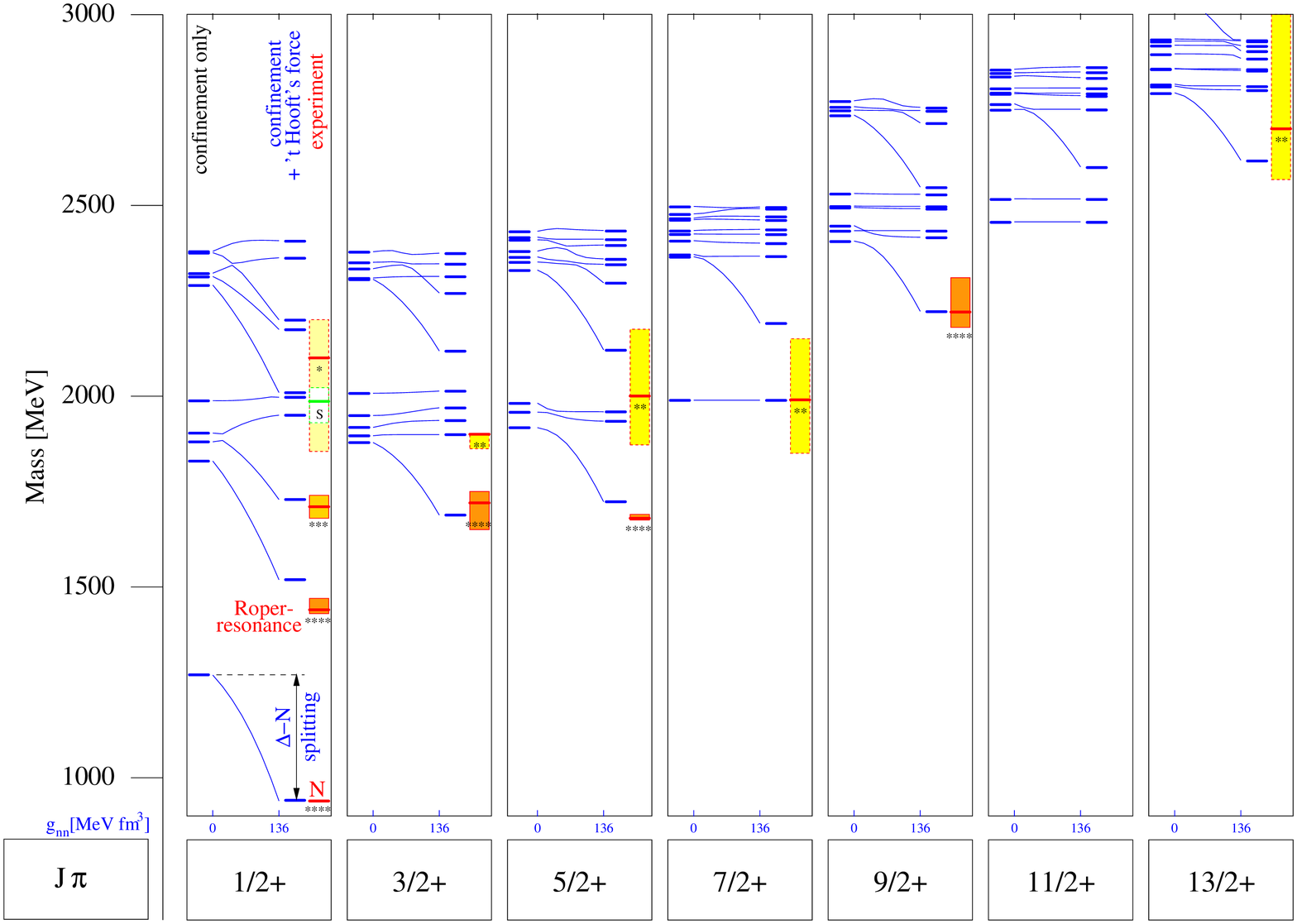},width=180mm}
 \end{center}
\caption{The effect of the instanton-induced interaction on the
  positive-parity nucleon resonances in model ${\cal A}$. In each column
  the leftmost spectrum is determined by confinement alone. The curves
  illustrate the variation of the spectrum with increasing 't~Hooft coupling
  $g_{nn}$ which finally is fixed to reproduce the correct
  $N-\Delta$-splitting.  As can be seen, the 't~Hooft interaction can indeed
  account for essential structures ({\it e.g.}  the {\bf Roper resonance}) of
  the experimental spectrum, which is shown on the right in each column.}
\label{fig:Model2N+gvar}
\end{figure}
In addition to the graphical representation of this effect in fig.
\ref{fig:Model2N+gvar} we demonstrate the influence of 't~Hooft's
force on the $2\hbar\omega$ states (in comparison to the ground-state)
also in table \ref{tab:ConfMixA2hw} by displaying the explicit
mass values as well as the configuration mixing of these states for
the cases without ($g_{nn}=0$ MeV fm$^3$) and with ($g_{nn}=136$ MeV
fm$^3$) 't~Hooft's force.  In fact, one of the states in the
ground-state sector $N\frac{1}{2}^+$, namely the lowest excitation in
the pure confinement spectrum at 1830 MeV, is lowered even strongly
enough (by the same amount of roughly 300 MeV like the ground-state)
to match almost the conspicuous low position of the Roper
resonance $N\frac{1}{2}^+(1440,\mbox{****})$.  In this respect it is
very interesting to note that, similar to the ground-state, this state
shows an almost pure $^2 8[56]$ spin-flavor $SU(6)$ configuration in the
case without 't~Hooft interaction. But also when 't~Hooft's force is
switched on with a strength to bring the ground-state roughly 300 MeV
down to its experimental value, both the ground-state and the Roper 
state still exhibit almost the same configuration mixing. The final
configuration mixing of both states  differs from the pure
confinement case essentially by a $7-8\%$ admixture of a $^2 8[70]$ configuration.
Thus, the behavior of the Roper state in model ${\cal A}$
under the influence 't~Hooft's force is very similar to that of the
ground-state. The third excited $N\frac{1}{2}^+$ state of the pure
confinement spectrum has a dominant $^2 8[70]$ configuration. In
comparison to the Roper state this state is only moderately
lowered by 't~Hooft's force by 175 MeV and hence becomes the second
excited state matching nicely the position of the
$N\frac{1}{2}^+(1710)$. Also each of the lowest states in the
$N\frac{3}{2}^+$ and $N\frac{5}{2}^+$ sectors exhibits in the pure
confinement case a dominant $^2 8[56]$ configuration, however, with
rather big admixtures of the other configuration.  These states show a
common downward shift of roughly 190 MeV somewhat smaller than the
Roper resonance but quite similar to the second excitation in
the $N\frac{1}{2}^+$ sector. Hence the equally big downward mass shift of
all three states is just of the right size to reproduce the nearly
degenerate structure of the three resonances
$N\frac{1}{2}^+(1710,\mbox{***})$, $N\frac{3}{2}^+(1720,\mbox{****})$
and $N\frac{5}{2}^+(1680,\mbox{****})$ around 1700 MeV.

\begin{table}[!h]
\begin{center}
\begin{tabular}{ccc}
&\textbf{without 't~Hooft's force} & \textbf{with 't~Hooft's force}\\
\begin{tabular}{|c|}
\hline
$J$\\[1mm]
\\
\hline
\hline
$\frac{1}{2}$\\[1.2mm]\\
\\\\
\\\\
\\\\
\\\\
\hline
\hline
$\frac{3}{2}$\\[0.7mm]\\
\\\\
\\\\
\\\\
\\\\
\hline
\hline
$\frac{5}{2}$\\[0.4mm]\\
\\\\
\\\\
\hline
\hline
$\frac{7}{2}$\\[-0.1mm]\\
\hline
\end{tabular}
&\hspace*{-4mm}
\begin{tabular}{|c|c|cccc|}
\hline
Mass & 
pos. & 
$\!\!{}^2 8 [56]\!\!$&
$\!\!{}^2 8 [70]\!\!$&
$\!\!{}^4 8 [70]\!\!$&
$\!\!{}^2 8 [20]\!\!$\\[1mm]
$[$MeV$]$& 
neg. & 
$\!\!{}^2 8 [56]\!\!$&
$\!\!{}^2 8 [70]\!\!$&
$\!\!{}^4 8 [70]\!\!$&
$\!\!{}^2 8 [20]\!\!$\\
\hline
\hline
1270 & 98.5  & {\bf \underline{98.5}} & 0.0  & 0.0  & 0.0\\
     & 1.5   & 0.3 & 0.6 &   0.5                  &0.0\\
\hline
\hline
1830 & 98.2   & {\bf \underline{97.5}}  & 0.4  & 0.2 &0.1\\
     & 1.8    &  0.4                    & 0.7 & 0.7 & 0.0\\
\hline
1880 & 97.9   &0.0  &13.7  &  {\bf \underline{71.8}} &12.4\\
     & 2.1   &0.0  & 0.3 & 1.8                    &0.0\\
\hline
1903 & 98.1   &0.6  & {\bf \underline{75.9}} & 19.4  &2.2\\
     & 1.9   &0.4  & 0.7 & 0.8                    &0.1\\
\hline
1988 & 97.7   &0.1  &8.4  & 6.4  &{\bf \underline{82.9}}\\
     &  2.3  & 0.0 & 0.8 & 1.0                    &0.5\\
\hline           
\hline
1878 & 98.0 & {\bf \underline{53.0}}   & 18.7 & 23.7 & 2.5\\
     &  2.0  & 0.4 & 0.4 & 1.2                    &0.0\\
\hline
1896 & 98.3  &0.5  &0.7  &  {\bf\underline{93.7}} &3.3\\
     &  1.7  &0.1  &0.3  &  1.2                   &0.1\\
\hline
1918 &97.9    &6.3  &24.6  &  {\bf \underline{67.0}} &0.0\\
     &2.1    & 0.1 & 0.6 & 1.4                    &0.1\\
\hline
1949 &97.9    &37.6  &{\bf \underline{44.0}}  &7.5   &8.8\\
     & 2.1   & 0.4 & 1.0 & 0.7                    &0.0\\
\hline
2007 & 97.8   &0.7  &9.8  &  4.3 & {\bf \underline{83.0}} \\
     & 2.2   &0.0  &0.8  & 1.0                    &0.4\\
\hline           
\hline
1917 & 98.0   &{\bf \underline{61.7}}  &2.8  &33.5   &0.0\\
     &  2.0   &0.8  & 0.6 & 0.5                    &0.1\\
\hline
1957 & 97.9   &20.8  &20.8  &  {\bf \underline{56.3}} &0.0\\
     & 2.1   & 0.2 & 0.7 & 0.6                    &0.6\\
\hline
1981 & 98.0   &15.5  &{\bf \underline{74.3}}  & 8.1  &0.0\\
     &  2.0  &0.6  &0.4  &0.7                     &0.3\\
\hline           
\hline
1989 & 98.1   &0.0  &0.0  &  {\bf \underline{98.1}} &0.0\\
     &  1.9  &0.2  & 0.2 & 1.4                    &0.2\\
\hline
\end{tabular}
&\hspace*{-4mm}
\begin{tabular}{|c|c|cccc|}
\hline
Mass & 
pos. & 
$\!\!{}^2 8 [56]\!\!$&
$\!\!{}^2 8 [70]\!\!$&
$\!\!{}^4 8 [70]\!\!$&
$\!\!{}^2 8 [20]\!\!$\\[1mm]
$[$MeV$]$& 
neg. & 
$\!\!{}^2 8 [56]\!\!$&
$\!\!{}^2 8 [70]\!\!$&
$\!\!{}^4 8 [70]\!\!$&
$\!\!{}^2 8 [20]\!\!$\\
\hline
\hline
939 & 96.3   &{\bf \underline{90.4}}  &5.8  & 0.0  &0.0\\
     & 3.7   & 1.3  &1.8  & 0.6                    &0.0\\

\hline
\hline
1518 & 97.2   & {\bf \underline{90.8}} &6.1  & 0.1  &0.2\\
     &  2.8  & 0.8  & 1.3  & 0.6                    &0.0\\
\hline
1729 & 97.9   & 15.5 & {\bf \underline{81.4}} & 0.7  &0.3\\
     &  2.1  & 0.6 & 0.8 &  0.6                   &0.1\\
\hline
1950 & 97.9   & 0.9 & 0.9 &  {\bf \underline{88.6}} &7.4\\
     & 2.1    & 0.0   & 0.2 & 1.8 & 0.1\\
\hline
1996 & 97.7   & 34.2 & 5.9 & 6.8 & {\bf \underline{50.8}} \\
     & 2.3    &  0.2 &  0.6& 1.3 & 0.3\\
\hline           
\hline
1688 & 97.0   & {\bf \underline{66.6}} & 29.9  &0.5   &0.1\\
     & 3.0   & 1.1  & 0.7 & 1.3                    &0.0\\
\hline
1899 & 98.1   &0.2  & 0.1 &  {\bf \underline{93.9}} &4.0\\
     & 1.9   & 0.2 & 0.5 & 1.1                    &0.1\\
\hline
1936 & 97.8   &6.5  & 14.6 &  {\bf \underline{67.1}} &9.7\\
     &  2.2  & 0.2 & 0.8 & 1.2                    &0.0\\
\hline
1969 & 98.1   & 20.2  & {\bf \underline{56.0}} &21.6   &0.3\\
     & 1.9   &0.2  & 0.8 & 1.0                    &0.1\\
\hline
2013 & 97.8   &0.8  &1.1  & 12.9  &{\bf \underline{82.9}}\\
     & 2.2   &0.0  & 0.9 & 1.0                    &0.3\\
\hline           
\hline
1723 & 96.8   &{\bf \underline{65.9}}  &30.6  &0.3   &0.0\\
     &  3.2  &1.0  & 1.2 & 0.9                    &0.2\\
\hline
1934 & 97.8   &13.1  &25.7  &  {\bf \underline{59.0}} &0.0\\
     &  2.2  & 1.1 &0.6  & 0.5                    &0.0\\
\hline
1959 & 97.9   & 16.0 &{\bf \underline{43.3}}  &38.5   &0.0\\
     &  2.1  & 0.3 &0.6  & 0.5                    &0.8\\
\hline           
\hline
1989 &  98.1  &0.0  & 0.0 &  {\bf \underline{98.1}} &0.0\\
     &  1.9  & 0.2 & 0.2 & 1.4                    &0.2\\
\hline
\end{tabular}
\end{tabular}
\end{center}
\caption{Masses and configuration mixing for the nucleon ground-state and the
excited states of the positive-parity $2\hbar\omega$ shell without (left) and
with (right) 't~Hooft's 
force in model ${\cal A}$. For each contribution to the
Salpeter amplitude the corresponding Salpeter norm is
given in $\%$. In each row the upper line shows the positive and the lower line
the negative energy contributions. Dominant contributions are bold printed and
underlined.}
\label{tab:ConfMixA2hw}
\end{table}

Note however that the formation of such a well separated shell substructure,
which is lowered relative to the bulk of the other
states, is not just restricted to the $2\hbar\omega$ band but in general can
be found also in the other even parity shells.  In the $4\hbar\omega$ and
$6\hbar\omega$ band the downward shift of these structures amounts to roughly 180
MeV when $g_{nn}$ is fixed by the correct ground-state position. Again the
states of these well separated sub-shells have dominant $^2 8[56]$
contributions ($55 \%$ - $70 \%$) with additional (in general quite big)
admixtures of $^2 8[70]$ configurations ($30\%$ - $45 \%$).  In the
$4\hbar\omega$ shell this downward shift is just the right size to
achieve the correct description of the well-established Regge state
$N\frac{9}{2}^+(2220,\mbox{****})$ and in the $6\hbar\omega$ shell the shift
isolates the first excitation of the $N\frac{13}{2}^+$ sector, which is
assigned to the Regge state $N\frac{13}{2}^+(2700,\mbox{**})$. Notice that all
excited members of the positive parity $N$-Regge trajectory, {\it i.e.}
each of the lowest excitations in the $N\frac{5}{2}^+$, $N\frac{9}{2}^+$ and
$N\frac{13}{2}^+$ sectors, belong to these sub-shells of the
$2\hbar\omega$, $4\hbar\omega$ and $6\hbar\omega$ band, respectively, and
remember the  remarkable feature that their mass shifts, along with the lowering of
the ground-state, conform excellently with the observed linear Regge 
characteristic of the trajectory states $N\frac{1}{2}^+(939,\mbox{****})$,
$N\frac{5}{2}^+(1680,\mbox{****})$, $N\frac{9}{2}^+(2220,\mbox{****})$ and
$N\frac{13}{2}^+(2700,\mbox{**})$ as discussed in the previous subsection (see
fig. \ref{fig:n_regge}). In summary we thus find that along with the
$\Delta-N$ ground-state splitting really \textit{all} essential hyperfine
structures of the positive-parity $N^*$-spectrum, which the confinement kernel
of model ${\cal A}$ alone cannot account for, are in fact consistently and
simultaneously generated by the instanton-induced 't~Hooft interaction.
Apart from the success of 't~Hooft's force in generating the hyperfine structure of
ground-state baryons, these results thus give a further strong evidence for our
conjecture that the instanton-induced short distance interaction in fact might
play the dominant role for the structure of the whole nucleon spectrum.

\subsubsection{Instanton-induced effects in model ${\cal A}$ -- negative-parity spectrum}
\begin{figure}[h]
  \begin{center}
    \epsfig{file={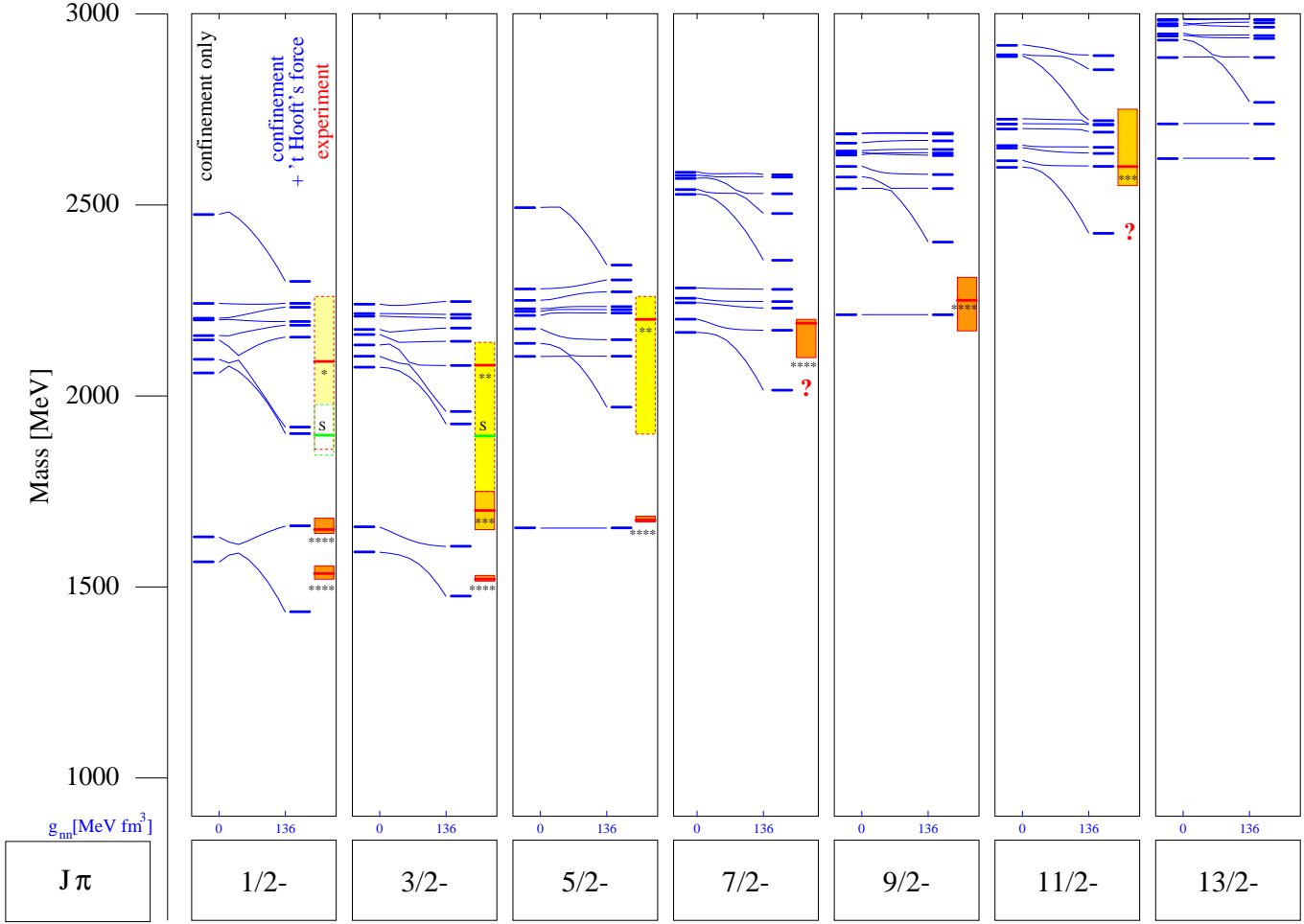},width=180mm}
  \end{center}
\caption{Instanton-induced splittings of negative parity nucleon
states in model ${\cal A}$. In each column the leftmost spectrum is
obtained with the confinement force alone. The curves show the change
of the spectrum with increasing 't~Hooft coupling $g_{nn}$. The
right spectrum finally displays the spectrum with $g_{nn}$ fixed to
reproduce the right nucleon ground-state position at 939 MeV. For
comparison the experimental resonance positions are displayed
rightmost in each column.}
\label{fig:Model2N-gvar}
\end{figure}

We now turn to the instanton-induced effects in the negative parity
nucleon spectrum predicted by model ${\cal A}$. Figure
\ref{fig:Model2N-gvar} shows that the instanton force causes similar
effects in the odd-parity $1\hbar\omega$, $3\hbar\omega $,
$5\hbar\omega $ bands. 
Again, in each shell a multiplet of particular
states shows a downward level shift of roughly 180-200 MeV, thus
generating substructures of nearly degenerate states that are well
separated from the rest of the shell. Analyzing the spin-flavor
content of the corresponding Salpeter amplitudes, we find that all these states
have dominant contributions of $^2 8[70]$ configurations with
additional large admixtures of $^2 8[56]$. 

In the $1\hbar\omega$ band
this mechanism generates the level splittings of the five
experimentally well-established states.  The single state of this
shell in the $N\frac{5}{2}^-$ sector, which is almost a pure $^4 8[70]$
configuration, remains totally unaffected, as previously mentioned.
Note that moderate spin-orbit effects of the confinement kernel ${\cal A}$
already break the degeneracy of the states in the pure confinement 
spectrum and cause a slight mixture of the
$^2 8[70]$ and $^4 8[70]$ configurations in the two states of
the $N\frac{1}{2}^-$ and $N\frac{3}{2}^-$ sectors,
respectively.  In both sectors the mass splitting of $\sim 65$ MeV is such
that the state with dominant $^2 8[70]$ configuration lies above the
state with dominant $^4 8[70]$ configuration. The action of 't~Hooft's
force just reverses this initial level ordering and moreover causes a
significant admixture of the $^2 8[56]$ configuration (roughly $10\%$)
to the lowest states in $N\frac{1}{2}^-$ and $N\frac{3}{2}^-$.  This
scenario is documented in table \ref{tab:ConfMixA1hw}, where we
displayed the masses as well as the explicit configuration mixings of
the $1\hbar\omega$ states for the cases without and with 't~Hooft's
interaction. Note that in the $N\frac{1}{2}^-$ sector the
relativistic version of the instanton force acts repulsive on the
dominantly $^4 8[70]$ state due to the repulsive
part of 't~Hooft's force acting in the pseudo-scalar diquark sector.
In this way the position of $N\frac{1}{2}^-(1650)$ is nicely reproduced.
Unfortunately, the lowering of the other state in this sector is
slightly too strong, such that it shifts the first predicted excitation
below the experimentally observed $N\frac{1}{2}^-(1535)$ and thus
below the first predicted state with the opposite
positive parity assigned to the Roper resonance.  Nonetheless,
the splitting between the upper multiplet
$N\frac{1}{2}^-(1650)-N\frac{3}{2}^-(1700)-N\frac{5}{2}^-(1675)$ and
the lower multiplet $N\frac{1}{2}^-(1535)-N\frac{3}{2}^-(1520)$ is
all in all reasonably well reproduced in this manner.
\begin{table}[!h]
\begin{center}
\begin{tabular}{ccc}
&\textbf{without 't~Hooft's force} & \textbf{with 't~Hooft's force}\\
\begin{tabular}{|c|}
\hline
$J$\\[1mm]
\\
\hline
\hline
$\frac{1}{2}$\\
\\
\\
\\
\hline
\hline
$\frac{3}{2}$\\
\\
\\
\\
\hline
\hline
$\frac{5}{2}$\\
\\
\hline
\end{tabular}
&\hspace*{-4mm}
\begin{tabular}{|c|c|cccc|}
\hline
Mass & 
pos. & 
$\!\!{}^2 8 [56]\!\!$&
$\!\!{}^2 8 [70]\!\!$&
$\!\!{}^4 8 [70]\!\!$&
$\!\!{}^2 8 [20]\!\!$\\[1mm]
$[$MeV$]$& 
neg. & 
$\!\!{}^2 8 [56]\!\!$&
$\!\!{}^2 8 [70]\!\!$&
$\!\!{}^4 8 [70]\!\!$&
$\!\!{}^2 8 [20]\!\!$\\
\hline
\hline
1566 &   98.3 & 0.0 & 18.1 &  {\bf \underline{80.2    }} &      0.0\\
     &    1.7 & 0.0 & 0.1  &      1.1                    &      0.5\\
\hline
1631 &   98.1 & 0.0 & {\bf \underline{80.0    }} &  18.1 &      0.0\\
     &    1.9 & 0.7 & 0.4  &      0.5                    &      0.4\\
\hline           
\hline
1592 &   98.2 & 0.0 & 24.7 &  {\bf \underline{73.5    }} &      0.0\\
     &    1.8 & 0.0 & 0.4  &      0.8                    &      0.4\\
\hline
1657 &   98.4 & 0.0 & {\bf \underline{73.6    }} &  24.8 &      0.0\\
     &    1.6 & 0.2 & 0.4  &      0.8                    &      0.2\\
\hline
\hline
1655 &   98.4 & 0.0 & 0.0 &  {\bf \underline{98.4    }} &      0.0\\
     &    1.6 & 0.1 & 0.3  &      1.0                   &     0.1\\
\hline
\end{tabular}
&\hspace*{-4mm}
\begin{tabular}{|c|c|cccc|}
\hline
Mass & 
pos. & 
$\!\!{}^2 8 [56]\!\!$&
$\!\!{}^2 8 [70]\!\!$&
$\!\!{}^4 8 [70]\!\!$&
$\!\!{}^2 8 [20]\!\!$\\[1mm]
$[$MeV$]$& 
neg. & 
$\!\!{}^2 8 [56]\!\!$&
$\!\!{}^2 8 [70]\!\!$&
$\!\!{}^4 8 [70]\!\!$&
$\!\!{}^2 8 [20]\!\!$\\
\hline
\hline
1435 &   96.0 & 10.6 & {\bf \underline{83.2}}& 2.2  &      0.0\\
     &    4.0 & 2.4 & 0.8  &      0.1                    &      0.6\\
\hline
1660 &   98.5 & 0.4 & 3.0 & {\bf \underline{94.6}} &      0.6\\
     &    1.5 & 0.1 & 0.0  &      1.2                    &    0.2\\
\hline           
\hline
1476 &   97.3 & 9.0 & {\bf \underline{87.6}} & 0.7 &      0.0\\
     &    2.7 & 0.6 & 0.9                        & 0.9 &      0.3\\
\hline
1696 &   98.2 & 0.5 &0.5 & {\bf \underline{97.2}} &     0.0\\
     &    1.8 & 0.2 & 0.5  &      0.7             &      0.3\\
\hline
\hline
1655 &   98.4 & 0.0 & 0.0 &  {\bf \underline{98.4    }} &      0.0\\
     &    1.6 & 0.1 & 0.3  &      1.0                   &     0.2\\
\hline
\end{tabular}
\end{tabular}
\end{center}
\caption{Masses and configuration mixing for the states of the
negative-parity $1\hbar\omega$ shell without (left) and with (right)
't~Hooft's force in model ${\cal A}$. For each contribution to the
Salpeter amplitude the corresponding Salpeter norm is
given in $\%$. In each row the upper line shows the positive and the
lower line the negative energy contributions. Dominant contributions are bold printed and
underlined.}
\label{tab:ConfMixA1hw}
\end{table}

In the $3\hbar\omega$ shell region the first evidences for new
resonances in the $N\frac{1}{2}^-$ and $N\frac{3}{2}^-$ sectors around
1900 MeV from recent SAPHIR photoproduction results in the channels
$p\gamma\rightarrow p\eta'$ and $p\gamma\rightarrow K \Lambda$,
respectively, might indeed be interpreted as first experimental
indications of such a comparatively low-lying structure of the
$3\hbar\omega$ band: With $g_{nn}$ adjusted to the nucleon
ground-state, 't~Hooft's force separates two close states in both
sectors whose positions then nicely agree with the resonance positions
reported. As illustrated, also the lowest model excitations in the
$N\frac{5}{2}^-$ and $N\frac{7}{2}^-$ sector belong to this well
separated structure. Presently, there is still no experimental
evidence for such a comparatively low-lying state in the
$N\frac{5}{2}^-$ sector.  Concerning the first excited
$N\frac{7}{2}^-$ state we now see that just because of the downward
shift of roughly 180 MeV, its predicted position turns out to be 170
MeV too light compared to the position of the first observed
excitation in this sector, {\it i.e.} the four star state
$N\frac{7}{2}^-(2190)$.  However, the second excitation in
$N\frac{7}{2}^-$ along with the other $3\hbar\omega$ states in this
sector in fact is hardly influenced by the instanton force and thus
fits the $N\frac{7}{2}^-(2190)$ quite well. Hence, the explanation of
the comparatively low position of the two new SAPHIR resonances
$S_{11}(1897)$ and $D_{13}(1895)$ as instanton-induced effect requires
the confirmation of the two states in the $N\frac{5}{2}^-$ and
$N\frac{7}{2}^-$ sectors at roughly 2 GeV even below the hitherto
observed first excitation $N\frac{7}{2}^-(2190)$ in order to complete
this predicted multiplet.  The same effect as in the $N\frac{7}{2}^-$
sector can also be seen in the $N\frac{11}{2}^-$ sector where the
lowest predicted excitation belongs to the strongly lowered multiplet
structure of the $5\hbar\omega$ band.  Again, the downward mass shift
of roughly 190 MeV due to the instanton force brings the first
calculated excitation in $N\frac{11}{2}^-$ about 175 MeV below the
first observed excitation $N\frac{11}{2}^-(2600)$. But the second
predicted state matches the $N\frac{11}{2}^-(2600)$ exactly, as it
remains unaffected by 't~Hooft's force.  Hence, the residual 't~Hooft
interaction again leads to a state below the $N\frac{11}{2}^-(2600)$
at 2425 MeV, which is the state with the highest total spin in the
lowered multiplet of the $5\hbar\omega$ shell.  Note that the two
lowest predicted states of $N\frac{7}{2}^-$ and $N\frac{11}{2}^-$
belong to the same negative parity Regge trajectory and the lowering
of the states of this trajectory is quite analogous to the lowering of
the positive-parity trajectory discussed previously. There it was just
this instanton-induced downward mass shift which brought all states of
this trajectory simultaneously into excellent agreement with the
experiment.

\subsubsection{Explanation for approximate parity doublets}
\begin{figure}[h]
  \begin{center}
    \epsfig{file={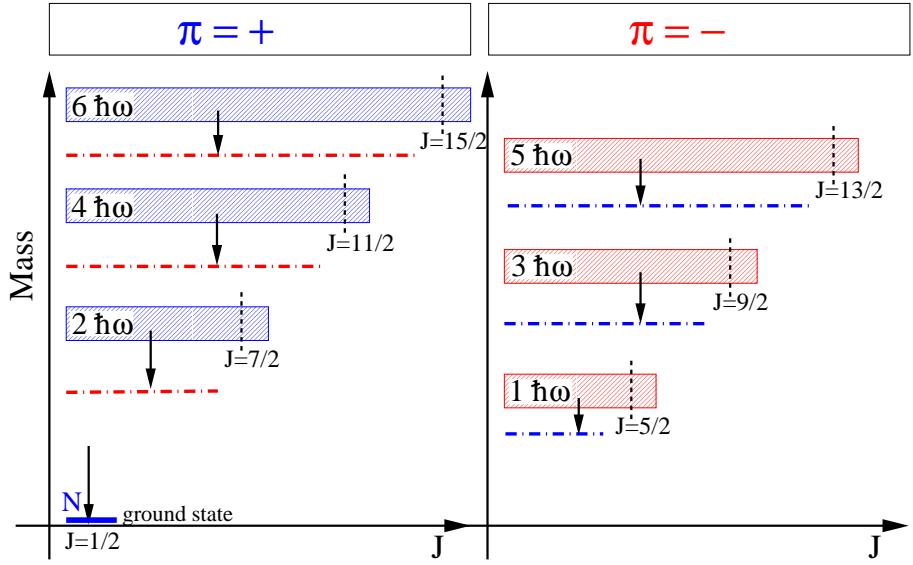},width=120mm}
  \end{center}
\caption{Schematical illustration of the explanation for the appearance approximate parity
  doublets in the nucleon spectrum due to instanton-induced effects in our
  relativistic Salpeter equation based model. For a detailed explanation
  see text.}
\label{fig:PDscheme}
\end{figure}

Finally let us focus in the context of instanton effects (in model
${\cal A}$) on the appearance of more or less mass degenerate
nucleon states of equal total spin and opposite parity. Comparing
the illustrations of instanton-induced effects for the positive and
negative parity nucleon spectrum in figs. \ref{fig:Model2N+gvar}
and \ref{fig:Model2N-gvar}, respectively, we find in fact a simple
explicit foundation of the occurrence of near parity doublets in the
excited $N^*$-spectrum due to 't~Hooft's force.  In this respect let
us consider the global arrangements of the band structures in the
nucleon spectrum formed within the framework of our model: Without
the instanton-induced interaction, the three-body confinement kernel alone
arranges the nucleon spectrum into even and odd parity shells, where
the odd parity bands lie in between the even parity bands (and vice
versa). As demonstrated, the instanton-induced interaction then splits
each of these shells into two well separated parts due to
its strong selection rules: the lower part, namely a particular subset
of states, which are mixtures of dominantly  $^2 8[56]$ and $^2
8[70]$ configurations, is significantly lowered with respect to the
upper part, {\it i.e.} the remaining bulk of states which is hardly influenced by
the residual interaction. This situation is schematically demonstrated
in fig. \ref{fig:PDscheme}.

Let us in general consider a $N\hbar\omega$ shell with $N\geq 2$ in the
positive ($N$ even) or negative ($N$ odd) parity sector, respectively. When
the 't~Hooft coupling $g_{nn}$ is adjusted to fit the nucleon ground state, the
lowering of the shell substructure is such that it comes
to lie approximately in between the unshifted parts of the adjacent
$N\hbar\omega$ and $(N-2)\hbar\omega$ bands with the same parity. Accordingly
it becomes positioned nearly degenerate with the corresponding unshifted part
of the $(N-1)\hbar\omega$ shell with the opposite parity.  This nearly
degenerate arrangement of the upper/lower part of a shell with the lower/upper
part of a shell with opposite parity consequently leads to the near parity
doublets in our model, which we already discussed in detail in the foregoing
subsection (see figs.  \ref{fig:PD1hw2hw} and \ref{fig:PD2hw3hw4hw}).

\begin{figure}[h]
\begin{center}
  \epsfig{file={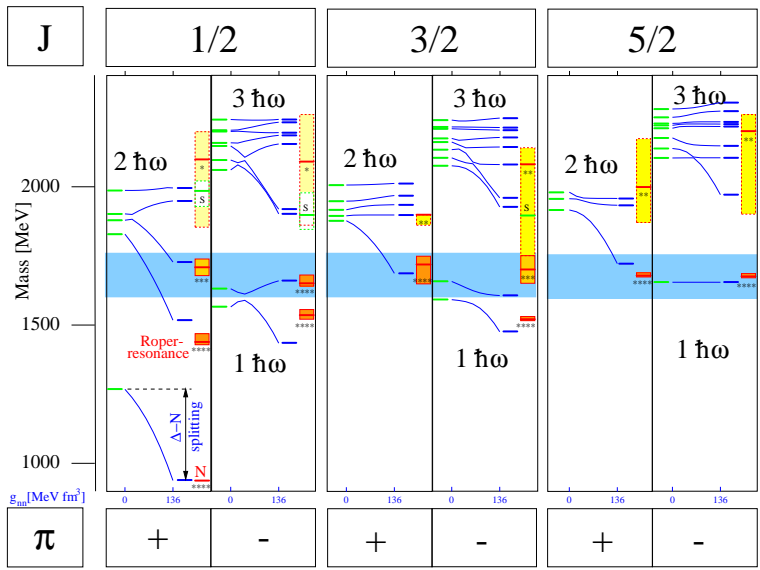},width=100mm}
\end{center}
\caption{Generation of approximate parity doublets due to the overlap of the
  positive-parity $2\hbar\omega$ and the negative-parity $1\hbar\omega$ shells
induced by 't~Hooft's force (in model ${\cal A}$).}
\label{fig:PDgvar1hw2hw3hw}
\end{figure}

The positioning of the positive-parity $2\hbar\omega$ shell and the
negative-parity $1\hbar\omega$ shell due to instanton effects, which leads to
approximate doublets in the second resonance region of the excited nucleon
spectrum around 1700 MeV, is once more explicitly illustrated in fig.
\ref{fig:PDgvar1hw2hw3hw}; we therefore combined the columns of fig.
\ref{fig:Model2N+gvar} (positive parity) and fig. \ref{fig:Model2N-gvar}
(negative parity) corresponding to the spins $J=\frac{1}{2}$, $\frac{3}{2}$
and $\frac{5}{2}$ in such a way that the sectors with same spin and opposite
parity are directly displayed side by side.
We thus see that in the framework of our covariant quark model ${\cal A}$
the instanton-induced effects indeed suggest a quite simple and appealing
explanation for these experimentally observed approximate parity doublets in
the second resonance region.

\begin{figure}[h]
  \begin{center}
    \epsfig{file={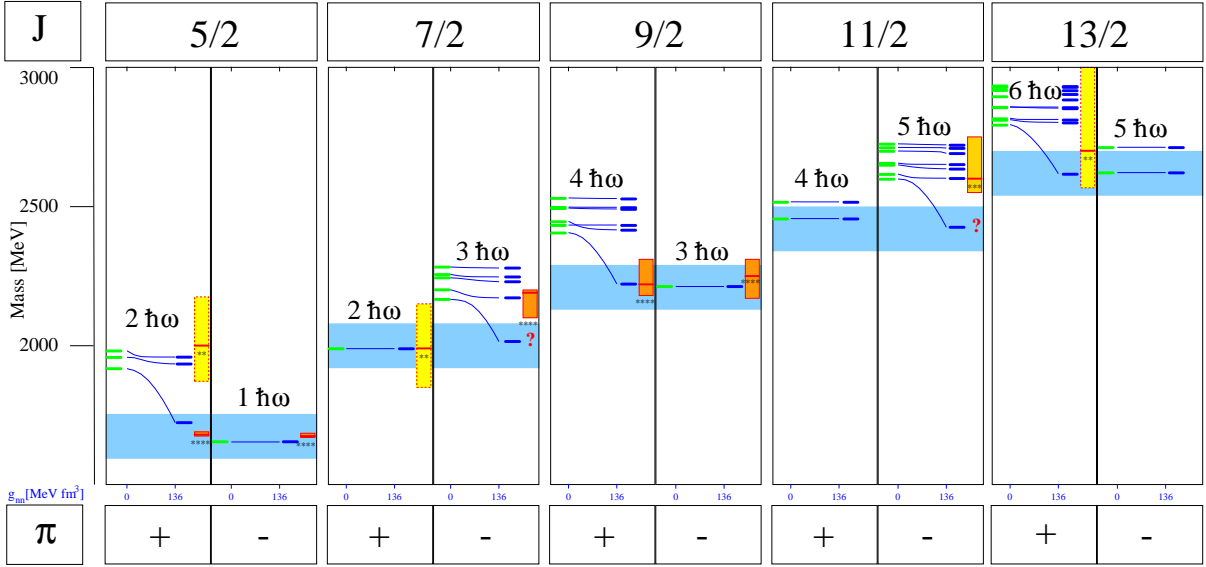},width=160mm}
  \end{center}
\caption{Instanton-induced generation of approximate parity doublets of lowest
  lying states (Regge-trajectory-type states) in the sectors with $J$ from
  $\frac{5}{2}$ to $\frac{13}{2}$ (in model ${\cal A}$).}
\label{fig:PDregge}
\end{figure}

Moreover, this mechanism works even for the higher mass regions of
the nucleon spectrum: The general arrangement of shell multiplets as indicated
schematically in fig. \ref{fig:PDscheme} is most prominent for the
lowest-lying\footnote{Note that these states are rather isolated and
  furthermore experimentally the most simply (or even the only) accessible
  resonances especially of the higher oscillator shells.}  states of the
sectors with $J^\pi$ from $\frac{5}{2}^\pm$ to $\frac{13}{2}^\pm$
(Regge-trajectory-type states): From table \ref{tab:max_tot_spins} we see that
the lowest states of the sectors $J^\pi=\frac{5}{2}^-$, $\frac{7}{2}^+$,
$\frac{9}{2}^-$, $\frac{11}{2}^+$, and $\frac{13}{2}^-$ are just the states
with maximum total spin $J_{\rm max}(N)=N+\frac{3}{2}$ of the $N\hbar\omega$
shells with $N=$1, 2, 3, 4, and 5.  As discussed earlier these states are not
influenced by 't~Hooft's force since they necessarily contain a totally
symmetric spin-quartet ($S=\frac{3}{2}$) wave function to achieve this maximal
total spin. But in the corresponding sectors with the same total spin $J$ but
opposite parity we then always find exactly one state of the higher lying $N+1$ band
that is selectively lowered by 't~Hooft's force and thus becomes the lowest
excitation in this opposite-parity sector fairly isolated from the other
states. The arrangement of these states in model ${\cal A}$ in comparison
with the present experimental situation is shown in fig. \ref{fig:PDregge},
where again the sectors with the same spin and opposite parity are displayed
side by side.
It is quite astonishing that the strength $g_{nn}$ as fixed on
the ground-state is just the right size to form in this way
{\it systematically} patterns of approximate parity doublets for {\it
all} lowest excitations in the sectors $J=\frac{5}{2}$ to
$J=\frac{13}{2}$.  In the $N\frac{5}{2}^\pm$ and $N\frac{9}{2}^\pm$ sectors
this scenario is nicely confirmed by the well-established parity
doublet structures
$N\frac{5}{2}^+(1680,\mbox{****})-N\frac{5}{2}^-(1675,\mbox{****})$
and
$N\frac{9}{2}^+(2220,\mbox{****})-N\frac{9}{2}^-(2250,\mbox{****})$,
respectively.  In the $N\frac{7}{2}$ sector, however, the present experimental findings of
the lowest excitations seem to deviate from such a parity doubling
structure due to the rather high measured resonance position of the
$N\frac{7}{2}^-(2190,\mbox{****})$ with respect to the
$N\frac{7}{2}^+(1990,\mbox{**})$. Therefore, a new investigation of this
situation and especially the presently stated position of the first
$N\frac{7}{2}^-$ excitation is highly desirable. In fact, the
interesting energy range around 2000 MeV is accessible by the new
photo- and electro-production experiments {\it e.g.} with the Crystal Barrel 
detector at ELSA (University of Bonn, Germany) or with
the CLAS detector at CEBAF (Jefferson Lab, USA), so that an early
clarification of the situation looks promising. But also an
exploration of the higher spin states with $J=\frac{11}{2}$ and
$J=\frac{13}{2}$ by future experiments on high baryon excitations is
desirable. The verification of the position of the first excited
$N\frac{11}{2}^-$ state, which presently differs significantly from
our prediction, as well as the discovery of the still ''missing''
first (two) excitation(s) in the sectors $N\frac{11}{2}^+$ and
$N\frac{13}{2}^-$ is essential to decide whether this striking feature
of systematical parity doublets induced by instanton
effects is indeed a realistic global structure exhibited
also by the experimental nucleon spectrum.

To conclude this discussion of approximate parity doublets, it is worthwhile
to mention that a further test of this scenario for parity
doublets based on instanton effects should be possible by the measurement of
electromagnetic ($p\gamma^*\rightarrow N^*$) transition form factors into each
member of the doublets shown in fig. \ref{fig:PDregge}.  Such a measurement
provides a deeper insight into the structure of the corresponding resonances.
In this respect it is crucial that one member of each doublet, namely that
which belongs to the maximum total spin $J_{\rm max}(N)=N+\frac{3}{2}$ of the
corresponding $N\hbar\omega$ shell, is by no means influenced by 't~Hooft's
force whereas its partner with opposite parity is strongly affected by
this residual interaction. Therefore, both members of the doublet have
significantly different internal structures that should be manifest
in systematically different shape of the two corresponding magnetic multipole
transition form factors:
the member of each parity doublet affected by 't~Hooft's force exhibits a
rather strong scalar-diquark correlation and thus its structure should be more
compact, in comparison to its unaffected doublet partner whose structure is
expected to be rather soft. Consequently for larger momentum transfers $Q^2$ the transition form
factors $p\gamma^*\rightarrow N^*$ for the lowest excitation in
$N^*\frac{5}{2}^-$, $N^*\frac{7}{2}^+$, $N^*\frac{9}{2}^-$,
$N^*\frac{11}{2}^+$, and $N^*\frac{13}{2}^-$ should systematically decrease
faster than those of opposite parity \cite{Kr01}.

\subsubsection{Instanton-induced effects in model ${\cal B}$ -- differences to model ${\cal A}$}
\begin{figure}[!h]
  \begin{center}
    \epsfig{file={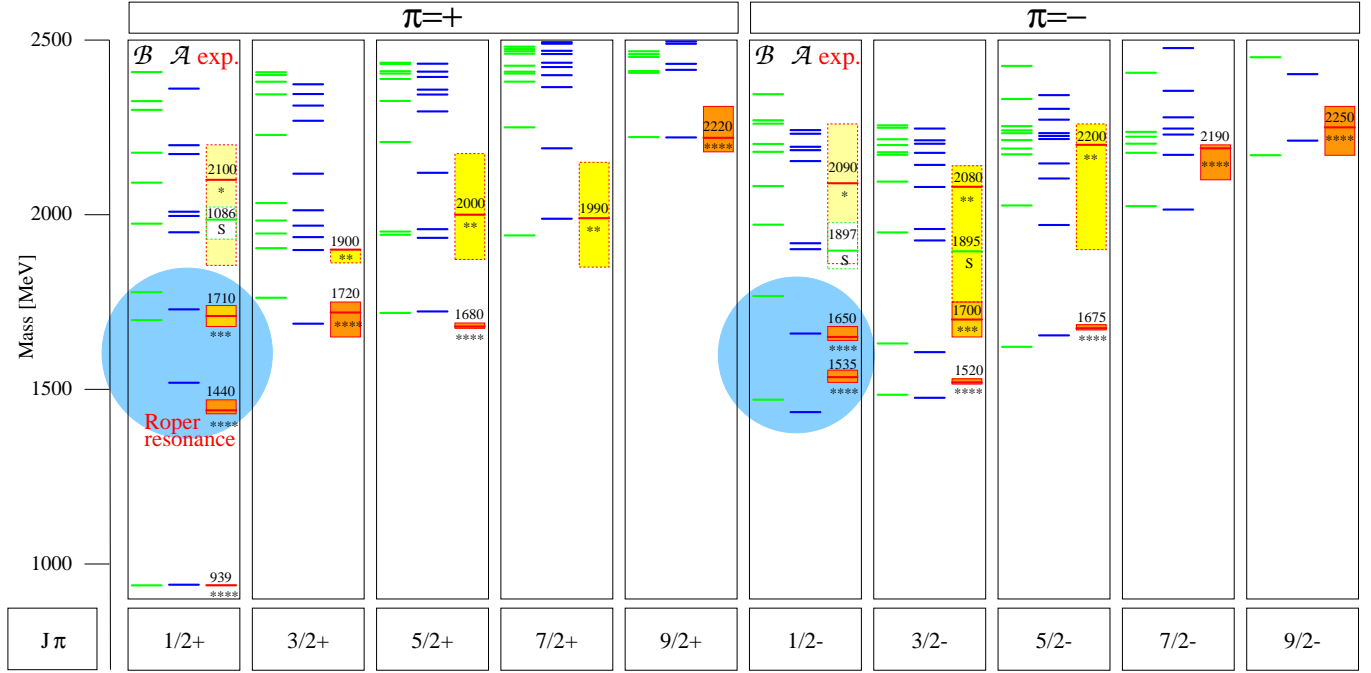},width=180mm}
  \end{center}
\caption{Comparison of the $N^*$-spectrum in model ${\cal A}$ (middle) and
${\cal B}$ (left); see text.}
\label{fig:NucM1M2}
\end{figure}

In our earlier discussion of the complete nucleon spectrum we found
some significant differences between the models ${\cal A}$ and
${\cal B}$: although also model ${\cal B}$ is able to describe
uniformly most of the structures of the excited $N^*$ spectrum up to
highest excitations similar to model ${\cal A}$ ({\it e.g.} the
positive parity N-Regge trajectory), it strongly failed, in
contrast to model ${\cal A}$, in describing one of the most
prominent and interesting features of the nucleon spectrum, namely the
Roper resonance in the $N\frac{1}{2}^+$ sector. Moreover, model
${\cal B}$ predicted a much too large hyperfine splitting between
the first two $N\frac{1}{2}^-$ excitations in the $1\hbar\omega$
shell; see fig. \ref{fig:NucM1M2}, where the results of both models
can be directly compared.  Thus, apart from further rather small
differences, the most distinct deviations between model
${\cal A}$ and ${\cal B}$ show up in the sectors with total spin
$J=\frac{1}{2}$. Since the equally good results of both models
concerning the description of the complete presently known
$\Delta$-spectrum did not allow to favor one of the two
phenomenological confinement potentials, the quite different results
concerning the Roper resonance provide a suitable indirect
criterion that strongly supports version ${\cal A}$ to provide the
more realistic confinement potential in combination with 't~Hooft's
force as residual interaction.

Studying now the effects of 't~Hooft's force on the energy levels of the $N^*$
spectrum in model ${\cal B}$ and comparing these with the corresponding
results of model ${\cal A}$ presented in the foregoing subsection, we want
to show that the differences between the models indeed have their origin
in a different influence of instanton-induced effects due to different
relativistic effects of the confinement kernels in both models:\\
In both models ${\cal A}$ and ${\cal B}$ the three-quark confinement
kernel is provided with a combination of scalar ($\Id\tens\Id\tens\Id$) and
time-like vector ($\Id\tens\gamma^0\tens\gamma^0+$ cycl. perm.) Dirac
structures.  Both versions just differ (apart from different model parameters)
in the way how these two spinorial Dirac structures are
combined in the linearly rising part of the confinement kernel.
Since both models have the {\it same non-relativistic limit} and hence
yield in this limit the same results, the
large differences between model ${\cal A}$ and ${\cal B}$ concerning the
Roper resonance in $N^*\frac{1}{2}^+$ and the hyperfine splittings of
the $1\hbar\omega$ states in $N^*\frac{1}{2}^+$ in fact can be considered as pure
relativistic effects in our Salpeter equation-based quark model.
This shall be clarified more precisely in the course of this discussion.\\

\begin{figure}[!h]
  \begin{center}
    \epsfig{file={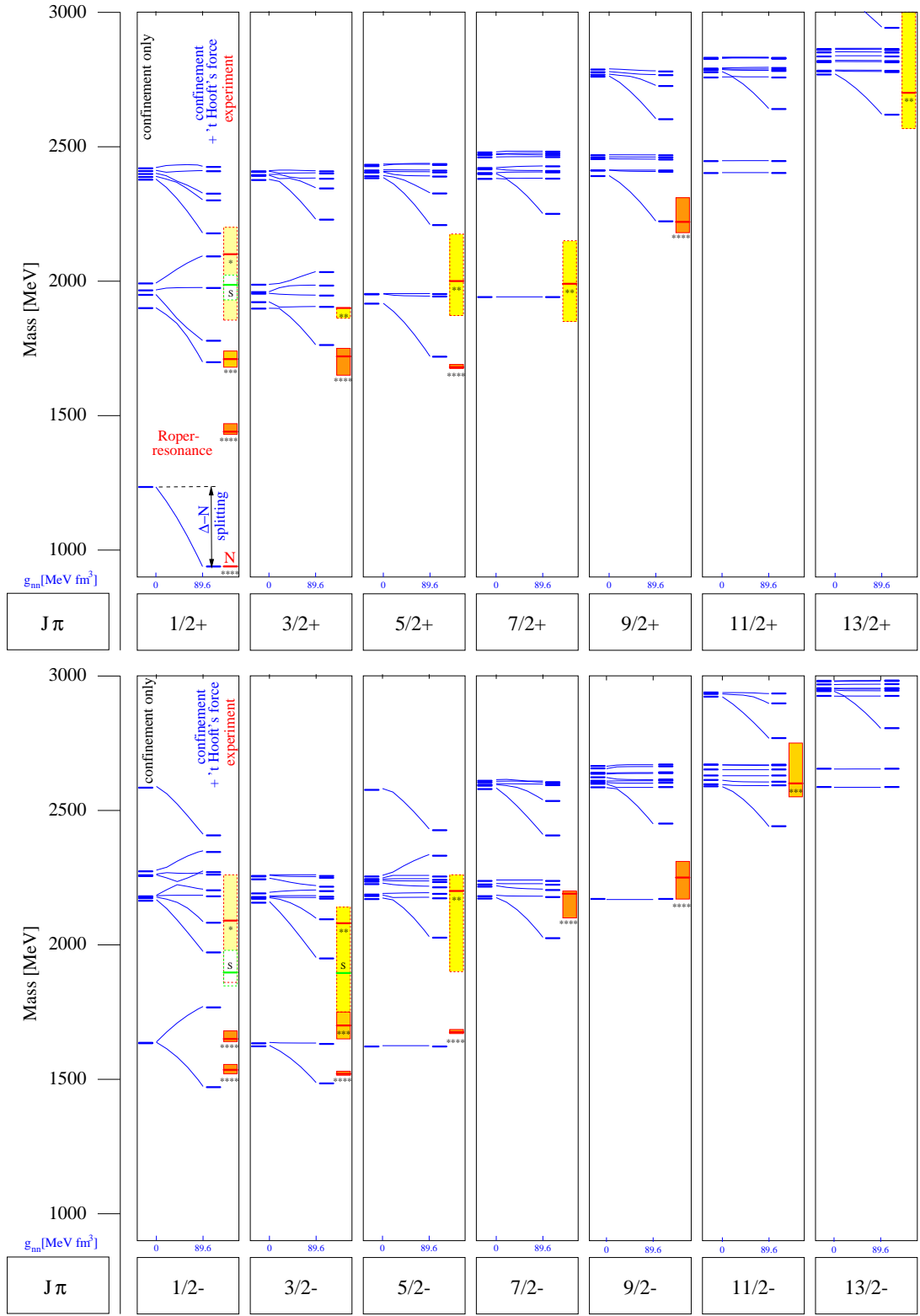},width=144mm}
  \end{center}
\caption{Instanton-induced effects in the positive (ahead) and
  negative parity (below) nucleon spectrum of model ${\cal B}$. The figures
  show the variation of the $N^*$ mass levels with increasing effective 't~Hooft
  coupling $g_{nn}$ analogous to the figs. \ref{fig:Model2N+gvar} and
  \ref{fig:Model2N-gvar} corresponding to model ${\cal A}$. Note that in
  contrast to model ${\cal A}$ the residual 't~Hooft interaction here
  cannot account for the Roper resonance $N\frac{1}{2}^+(1440)$ and
  moreover strongly fails in the description of $N\frac{1}{2}^-(1650)$!}
\label{fig:Mod1aNgvar}
\end{figure}

In analogy to figs. \ref{fig:Model2N+gvar} and \ref{fig:Model2N-gvar} of
model ${\cal A}$, fig. \ref{fig:Mod1aNgvar} shows the influence of 
't~Hooft's force on the energy levels of the excited positive- and negative-parity
nucleon spectrum in model ${\cal B}$. 
Before analyzing in detail the differences to model ${\cal A}$ let us
briefly discuss those effects that both models have in common.  Considering
globally the influence of 't~Hooft's force on the $N^*$-spectrum in model
${\cal B}$ we indeed find a lot of effects similar to those in model
${\cal A}$. In each shell we again find the systematics of well separated
multiplets of states which show dominant contributions of the $2^8[56]$ and
$2^8[70]$ configurations or their mixtures.  These shell substructures are
rather strongly lowered relative to the majority of states in the
corresponding bands that are hardly affected by 't~Hooft's force.  As in model
${\cal A}$ this systematics leads again to the occurrence of more or less
mass degenerate states of equal spin and opposite parity.  We do not want to
discuss this aspect again in detail here and just remark that this
instanton-induced mechanism of generating approximate parity doublets works
slightly better in model ${\cal A}$.  As before, the instanton interaction
for several model states induces downward mass shifts again of the right size
to reproduce the correct experimentally determined resonance positions, if the
't~Hooft coupling $g_{nn}$ is fixed such that the model ground-state matches
the experimental mass. In the positive parity spectrum these are for instance
the states assigned to the Regge trajectory, {\it i.e.} besides the ground-state the
three states $N\frac{5}{2}^+(1680,\mbox{****})$,
$N\frac{9}{2}^+(2220,\mbox{****})$ and $N\frac{13}{2}^+(2700,\mbox{**})$. In
the negative parity spectrum we again find the lowest predicted states in the
$N\frac{7}{2}^-$ and $N\frac{11}{2}^-$ roughly 160 -- 170 MeV below the
observed first excitations $N\frac{7}{2}^-(2190)$ and $N\frac{11}{2}^-(2600)$
stated by the PDG \cite{PDG00}.  This is caused in the same way as in model
${\cal A}$ just because of the strong downward mass shift of the first
excited states in these sectors by roughly 150 MeV.

\begin{figure}[!h]
  \begin{center}
    \epsfig{file={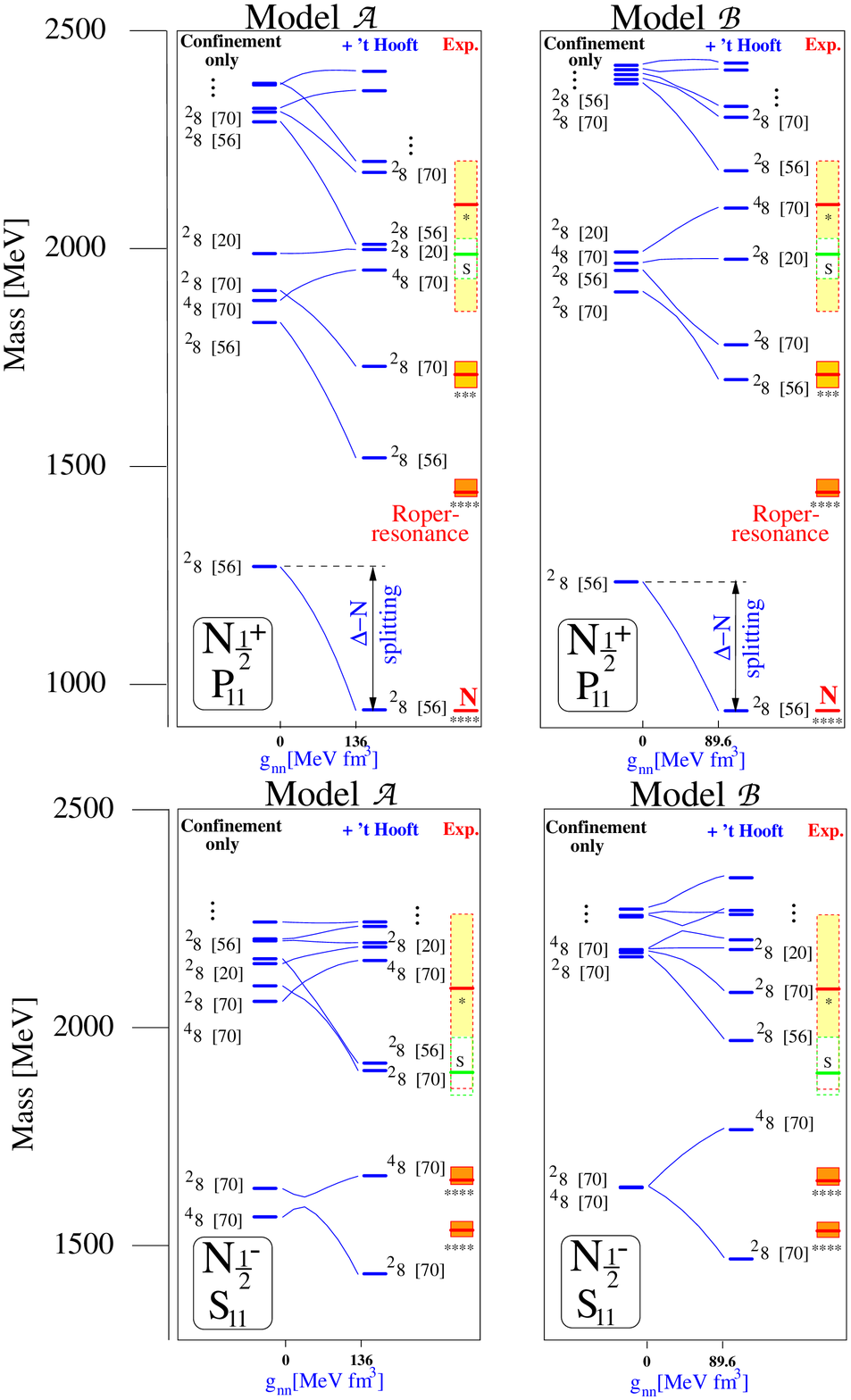},width=130mm}
  \end{center}
\vspace*{-5mm}
\caption{Comparison of the different instanton-induced effects in model
  ${\cal A}$ (left) and model ${\cal B}$ (right) in the $J=\frac{1}{2}$
  sector due to different spin-orbit effects of the two confinement versions.
  In both cases, each state is labeled by its dominant
  spin-flavor $SU(6)$ contribution, see tables \ref{tab:ConfMixA2hw}, \ref{tab:ConfMixB2hw}, \ref{tab:ConfMixA1hw} and
  \ref{tab:ConfMixB1hw}. See the text for a detailed discussion and explanation.}
\label{fig:RoperM1M2_S11M1M2}
\end{figure}

To investigate the most striking differences in the influence of 't~Hooft's
force compared to model ${\cal A}$ let us now have a closer look at the
negative-parity $1\hbar\omega$ shell and the positive-parity $2\hbar\omega$
shell, where, as mentioned before, the biggest deviations show up in the
$N\frac{1}{2}^\pm$ sectors. We start with the states of the $1\hbar\omega$
shell. For this discussion we also show the explicit masses
and the configuration mixing of the $1\hbar\omega$ states for 
the cases without and with instanton force in table
\ref{tab:ConfMixB1hw}.
\begin{table}[!h]
\begin{center}
\begin{tabular}{ccc}
&\textbf{without 't~Hooft's force} & \textbf{with 't~Hooft's force}\\
\begin{tabular}{|c|}
\hline
$J$\\[1mm]
\\
\hline
\hline
$\frac{1}{2}$\\
\\
\\
\\
\hline
\hline
$\frac{3}{2}$\\
\\
\\
\\
\hline
\hline
$\frac{5}{2}$\\
\\
\hline
\end{tabular}
&\hspace*{-4mm}
\begin{tabular}{|c|c|cccc|}
\hline
Mass & 
pos. & 
$\!\!{}^2 8 [56]\!\!$&
$\!\!{}^2 8 [70]\!\!$&
$\!\!{}^4 8 [70]\!\!$&
$\!\!{}^2 8 [20]\!\!$\\[1mm]
$[$MeV$]$& 
neg. & 
$\!\!{}^2 8 [56]\!\!$&
$\!\!{}^2 8 [70]\!\!$&
$\!\!{}^4 8 [70]\!\!$&
$\!\!{}^2 8 [20]\!\!$\\
\hline
\hline
1634 &   99.9 & 0.0 & 12.2 &  {\bf \underline{87.7    }} &      0.0\\
     &    0.1 & 0.0 & 0.0  &      0.1                    &      0.0\\
\hline
1636 &   99.9 & 0.0 & {\bf \underline{87.6    }} &  12.2 &      0.0\\
     &    0.1 & 0.1 & 0.0  &      0.0                    &      0.0\\
\hline           
\hline
1623 &   99.9 & 0.0 & {\bf \underline{64.8    }} & 35.1  &      0.0\\
     &    0.1 & 0.0 & 0.0  &      0.0                    &      0.0\\
\hline
1634 &   99.9 & 0.0 & 35.1  &{\bf \underline{64.8    }}  &      0.0\\
     &    0.1 & 0.0 & 0.0  &      0.0                    &      0.0\\
\hline
\hline
1622 &   99.9 & 0.0 & 0.0 &  {\bf \underline{99.9    }} &      0.0\\
     &    0.1 & 0.0 & 0.0  &      0.1                   &     0.0\\
\hline
\end{tabular}
&\hspace*{-4mm}
\begin{tabular}{|c|c|cccc|}
\hline
Mass & 
pos. & 
$\!\!{}^2 8 [56]\!\!$&
$\!\!{}^2 8 [70]\!\!$&
$\!\!{}^4 8 [70]\!\!$&
$\!\!{}^2 8 [20]\!\!$\\[1mm]
$[$MeV$]$& 
neg. & 
$\!\!{}^2 8 [56]\!\!$&
$\!\!{}^2 8 [70]\!\!$&
$\!\!{}^4 8 [70]\!\!$&
$\!\!{}^2 8 [20]\!\!$\\
\hline
\hline
1470 &   99.7 & 6.1 & {\bf \underline{87.1}}& 6.5  &      0.0\\
     &    0.3 & 0.2 & 0.0  &      0.0                    &0.0\\
\hline
1767 &   99.9 & 1.7 & 6.2 & {\bf \underline{91.2}} &      0.9\\
     &    0.1 & 0.0 & 0.0  &      0.1                    &0.0\\
\hline           
\hline
1485 &   99.9 & 6.1 & {\bf \underline{90.7}} & 3.2 &      0.0\\
     &    0.1 & 0.0 & 0.0                    & 0.0 &      0.0\\
\hline
1631 &   99.9 & 0.0 & 3.1 & {\bf \underline{96.7}} &     0.0\\
     &    0.1 & 0.0 & 0.0  &      0.0             &      0.0\\
\hline
\hline
1622 &   99.9 & 0.0 & 0.0 &  {\bf \underline{99.9    }} &     0.0\\
     &    0.1 & 0.0 & 0.0  &      0.1                   &     0.0\\
\hline
\end{tabular}
\end{tabular}
\end{center}
\caption{Masses and configuration mixing for nucleon states of the
negative-parity $1\hbar\omega$ shell without (left) and with (right)
't~Hooft's force in model ${\cal B}$. For each contribution to the
Salpeter amplitude the corresponding Salpeter norm is
given in $\%$. In each row the upper line shows the positive and the
lower line the negative energy contributions. Dominant contributions are bold printed and
underlined.}
\label{tab:ConfMixB1hw}
\end{table}
Again, the $N\frac{5}{2}^-$ state, which fits the
$N\frac{5}{2}^-(1675,\mbox{****})$ reasonably well, remains unaffected as in
model ${\cal A}$.  In the pure confinement case, {\it i.e.} $g_{nn}=0$, we
find in model ${\cal B}$ that all $1\hbar\omega$ states are almost
degenerate at roughly 1630 MeV; for explicit values see table
\ref{tab:ConfMixB1hw}. This is quite in contrast to model ${\cal A}$,
where, due to moderate spin-orbit effects of the confinement kernel, the
degeneracy of the $N\frac{1}{2}^-$ and $N\frac{3}{2}^-$ states was already
lifted right from start and mass splittings showed up already in the pure
confinement spectrum. This difference is due to the different combinations of
the scalar and time-like vector Dirac structures in the linearly rising part
of the three-body confinement kernels of model ${\cal A}$ and
${\cal B}$: While that of model ${\cal B}$ is chosen to really minimize
relativistic spin-orbit effects owing to a cancellation of effects from the
scalar and time-like vector part, the Dirac structure of model ${\cal A}$
exhibits moderate spin-orbit effects. This we illustrated already in the
earlier discussion of the $\Delta$-spectrum.  Let us analyze what these
different relativistic effects imply for the effect of 't~Hooft's force. Turning
on 't~Hooft's interaction ($g_{nn}>0$) the degeneracy is lifted: In the
$N\frac{3}{2}^-$ sector one of the two states, {\it i.e.}  the state with
dominant $^2 8[70]$ configuration, shows a downward shift while the position
of the other one, with the dominant $^4 8[70]$ configurations, remains almost
unchanged. Thus, the experimentally observed splitting between
$N\frac{3}{2}^-(1700,\mbox{****})$ and $N\frac{3}{2}^-(1520,\mbox{***})$ is
fairly well reproduced and of similar quality as in model ${\cal A}$.  But
now let us focus on the quite different effects in both models for the
$N\frac{1}{2}^-$ sector. For the sake of a better comparison the results of
both models in this sector are depicted together in the lower part of fig.
\ref{fig:RoperM1M2_S11M1M2}.  In model ${\cal B}$, the dominantly $^2
8[70]$ state again shows a downward mass shift as in the $N\frac{3}{2}^-$
sector due to the attractive scalar-diquark correlation, but at the same time
the repulsive part of the relativistic version of 't~Hooft's force, that acts
in the pseudo-scalar diquark sector (and is absent in the non-relativistic
limit), very strongly shifts the state with the dominant $^4 8[70]$
configuration upwards.  Hence, the experimental position of
$N\frac{1}{2}^-(1650,\mbox{****})$ and thus the hyperfine splitting between
$N\frac{1}{2}^-(1635,\mbox{****})$ and $N\frac{1}{2}^-(1650,\mbox{****})$ is
largely overestimated, in contrast to model ${\cal A}$.  There the
situation is improved just because of the moderate relativistic
effects of the confinement force. Recall that the mass
splitting in the pure confinement case exhibits the reversed level ordering
in model ${\cal A}$: the dominantly $^4 8[70]$ state lies below the state
with the dominant $^2 8[70]$ configuration. Thus, in contrast to model
${\cal B}$, the increasing 't~Hooft coupling $g_{nn}$ leads
in model ${\cal A}$ first to a level crossing cancelling the first
confinement-induced splitting before a reversed level ordering is achieved.
Hence the net effect is a weaker and thus improved hyperfine splitting in
model ${\cal A}$ that agrees
better with the experimental findings than in model ${\cal B}$.

We now turn to the instanton effects in the $2\hbar\omega$ shell that show up
in model ${\cal B}$. For this discussion see also table
\ref{tab:ConfMixB2hw}, which shows the explicit mass values and configuration
mixings of the states without and with 't~Hooft interaction.
\begin{table}[!h]
\begin{center}
\begin{tabular}{ccc}
&\textbf{without 't~Hooft's force} & \textbf{with 't~Hooft's force}\\
\begin{tabular}{|c|}
\hline
$J$\\[1mm]
\\
\hline
\hline
$\frac{1}{2}$\\[1.2mm]\\
\\\\
\\\\
\\\\
\\\\
\hline
\hline
$\frac{3}{2}$\\[0.7mm]\\
\\\\
\\\\
\\\\
\\\\
\hline
\hline
$\frac{5}{2}$\\[0.4mm]\\
\\\\
\\\\
\hline
\hline
$\frac{7}{2}$\\[-0.1mm]\\
\hline
\end{tabular}
&\hspace*{-4mm}
\begin{tabular}{|c|c|cccc|}
\hline
Mass & 
pos. & 
$\!\!{}^2 8 [56]\!\!$&
$\!\!{}^2 8 [70]\!\!$&
$\!\!{}^4 8 [70]\!\!$&
$\!\!{}^2 8 [20]\!\!$\\[1mm]
$[$MeV$]$& 
neg. & 
$\!\!{}^2 8 [56]\!\!$&
$\!\!{}^2 8 [70]\!\!$&
$\!\!{}^4 8 [70]\!\!$&
$\!\!{}^2 8 [20]\!\!$\\
\hline
\hline
1234 & 99.9  & {\bf \underline{99.9}} & 0.0  & 0.0  & 0.0\\
     & 0.1   & 0.0 & 0.1 &   0.0                 &0.0\\
\hline
\hline
1900 & 99.9   & 0.2   & {\bf \underline{99.0}}  & 0.0 &0.7\\
     & 0.1    &  0.0                    & 0.0 & 0.0 & 0.0\\
\hline
1949 & 99.9   &{\bf \underline{94.6}}  & 0.1  & 3.9  &1.2\\
     & 0.1    &0.0  & 0.1 & 0.1                    &0.0\\
\hline
1966 & 99.9   &4.6  & 0.1 & {\bf \underline{91.6}} &3.6\\
     & 0.1   &0.0  & 0.0 & 0.1                    &0.0\\
\hline
1991 & 99.9   &0.4  & 0.7  & 4.5  &{\bf \underline{94.3}}\\
     &  0.1  & 0.0 & 0.1 & 0.0                    &0.0\\
\hline           
\hline
1898 & 99.9 & 0.4   & 0.0 & {\bf \underline{99.1}} & 0.4\\
     &  0.1  & 0.0 & 0.0 & 0.1                    &0.0\\
\hline
1922 & 99.9  & {\bf\underline{98.0}} & 1.2  & 0.7  &0.1\\
     &  0.1  & 0.0  &0.0  &  0.0                   &0.0\\
\hline
1953 &99.9    &0.7  & {\bf \underline{85.1}}  &  12.7 & 1.4\\
     & 0.1    & 0.0 & 0.0 & 0.1                    &0.0\\
\hline
1959 &99.9    & 0.8  &  8.3 & {\bf \underline{78.5}}  &12.3\\
     & 0.1   & 0.0 & 0.0 & 0.1                    &0.0\\
\hline
1986 & 99.9   &0.0  & 5.4  &  8.8 & {\bf \underline{85.7}} \\
     & 0.1   & 0.0  &0.0  & 0.1                    &0.0\\
\hline           
\hline
1916 & 99.9   &{\bf \underline{97.4}}  &1.3  & 1.2   &0.0\\
     &  0.1   &0.0  & 0.0 & 0.0                    &0.0\\
\hline
1951 & 99.9   &2.4  &  {\bf \underline{47.0}} & 23.6 &0.0\\
     & 0.1   & 0.0 & 0.0 & 0.0                    &0.0\\
\hline
1952 & 99.9   &0.2  &24.6 &{\bf \underline{75.2}}  &0.0\\
     &  0.1  &0.0  &0.0  &0.0                     &0.0\\
\hline           
\hline
1941 & 99.9   &0.0  &0.0  &  {\bf \underline{99.9}} &0.0\\
     &  0.1  &0.0  & 0.0 & 0.1                    &0.0\\
\hline
\end{tabular}
&\hspace*{-4mm}
\begin{tabular}{|c|c|cccc|}
\hline
Mass & 
pos. & 
$\!\!{}^2 8 [56]\!\!$&
$\!\!{}^2 8 [70]\!\!$&
$\!\!{}^4 8 [70]\!\!$&
$\!\!{}^2 8 [20]\!\!$\\[1mm]
$[$MeV$]$& 
neg. & 
$\!\!{}^2 8 [56]\!\!$&
$\!\!{}^2 8 [70]\!\!$&
$\!\!{}^4 8 [70]\!\!$&
$\!\!{}^2 8 [20]\!\!$\\
\hline
\hline
939 & 99.7   &{\bf \underline{96.1}}  &3.5  & 0.0  &0.0\\
     & 0.3   & 0.1  &0.2  & 0.1                    &0.0\\

\hline
\hline
1698 & 99.8   & {\bf \underline{73.5}} &25.4  & 0.1  &0.8\\
     &  0.2  & 0.1  & 0.1  & 0.1                    &0.0\\
\hline
1778 & 99.9   & 30.2 & {\bf \underline{69.1}} & 0.6  &0.0\\
     &  0.1  & 0.0 & 0.1 &  0.0                   &0.0\\
\hline
1974 & 99.9   & 0.6 & 0.0 & 39.1 &  {\bf \underline{60.2}}\\
     & 0.1    & 0.0   & 0.0 & 0.1 & 0.0\\
\hline
2092 & 99.9   & 1.4 & 3.4 & {\bf \underline{59.1}}& 36.0 \\
     & 0.1    &  0.0 &  0.0& 0.1 & 0.0\\
\hline           
\hline
1762 & 99.9   & {\bf \underline{77.7}} & 20.6  &0.7   &0.9\\
     & 0.1   & 0.0  & 0.0 & 0.0                    &0.0\\
\hline
1904 & 99.9   & 1.0  & 1.2 &  {\bf \underline{95.8}} &2.0\\
     & 0.1   & 0.0 & 0.0 & 0.1                    &0.0\\
\hline
1946 & 99.9   & 15.6  &  {\bf \underline{64.5}} &8.9 & 10.9\\
     &  0.1  & 0.0 & 0.0 & 0.1                    &0.0\\
\hline
1983 & 99.9   & 0.4  & 1.3 & 39.8 & {\bf \underline{58.4}}\\
     & 0.1    & 0.0  & 0.0 & 0.1                    &0.0\\
\hline
2033 & 99.9   &3.0  &15.0  &{\bf \underline{54.3}} & 27.6\\
     & 0.1   &0.0  & 0.0 & 0.0                    &0.0\\
\hline           
\hline
1718 & 99.9   &{\bf \underline{65.1}}  &32.9  & 1.8   &0.1\\
     &  0.1   &0.0  & 0.0 & 0.0                    &0.0\\
\hline
1943 & 99.9   &32.1  &  {\bf \underline{67.4}}&0.4  &0.0\\
     &  0.1  & 0.0 &0.0  & 0.0                    &0.0\\
\hline
1952 & 99.9   & 0.5 & 1.6 & {\bf \underline{97.8}}  &0.0\\
     &  0.1  & 0.0 &0.0  & 0.0                    &0.0\\
\hline           
\hline
1941 &  99.9  &0.0  & 0.0 &  {\bf \underline{99.9}} &0.0\\
     &  0.1  & 0.0 & 0.0 & 0.1                    &0.0\\
\hline
\end{tabular}
\end{tabular}
\end{center}
\caption{Masses and configuration mixing for the nucleon ground-state and the
excited states of the positive-parity $2\hbar\omega$ shell without (left) and
 with (right)
't~Hooft's force in model ${\cal B}$. For each contribution to the
Salpeter amplitude the corresponding Salpeter norm is
given in $\%$. In each row the upper line shows the positive and the lower line
the negative energy contributions. Dominant contributions are bold printed and
underlined.}
\label{tab:ConfMixB2hw}
\end{table}
In figure \ref{fig:Mod1aNgvar} again a lowering of exactly four states of this
shell can be observed, when the 't~Hooft coupling $g_{nn}$ is gradually
increased. In the $N\frac{1}{2}^+$ sector, however, the effect of 't~Hooft's
force obviously strongly fails in generating the low position of the Roper 
resonance, whereas in the $N\frac{3}{2}^+$ and $N\frac{5}{2}^+$ sectors
the situation still is quite similar to model ${\cal A}$.  Counting the
$2\hbar\omega$ states in the non-relativistic oscillator model we expect
exactly one $^2 8[56]$ state in both sectors $N\frac{3}{2}^+$ and
$N\frac{5}{2}^+$.  Accordingly, we find in our models one state in both
sectors, whose embedded Pauli spinors exhibit a dominant $^2 8[56]$
contribution. These are lowered relative to the other states by 't~Hooft's
force, quite similarly in both models. Again, let us focus on the $N\frac{1}{2}^+$
sector, where the largest deviations from the experimental findings and from
the results of model ${\cal A}$ become evident.  For a more detailed
investigation of the different instanton-induced effects the results of both
models are depicted together in the upper part of fig.
\ref{fig:RoperM1M2_S11M1M2}. The totally different results of model
${\cal A}$ and model ${\cal B}$ again can be traced back to different
relativistic effects stemming from the confinement forces. As one
expects from counting states in the non-relativistic oscillator model, we find
in the pure confinement spectra of both models exactly four $N\frac{1}{2}^+$
states belonging to the $2\hbar\omega$ shell. Each of the four possible
spin-flavor $SU(6)$ configurations $^2 8[56]$, $^2 8[70]$, $^4 8[70]$ and $^2
8[20]$ occurs and can be assigned to one of these states. Of course, in the
pure confinement spectra of our Salpeter equation-based models
${\cal A}$ and ${\cal B}$ the degeneracy is lifted due to the
anharmonicity of the linear confinement potential and still more importantly
due to relativistic effects depending on the spinorial Dirac structure
of the confining forces.  Consequently, the two confinement versions induce
different intra-band hyperfine structures of these states.  Comparing the
masses and the corresponding contributions of the spin-flavor $SU(6)$
configurations to the embedded Pauli spinors of the Salpeter
amplitudes (see table \ref{tab:ConfMixA2hw} and \ref{tab:ConfMixB2hw}) one
finds, apart from the different pattern of mass splittings, a totally different
level ordering and configuration mixing of the states in both models. In model
${\cal B}$ we find rather pure $SU(6)$ configurations ($> 91\;\%$) and a
level ordering $^2 8[70] <\; ^2 8[56] <\; ^4 8[70] <\; ^2 8[20]$, whereas in
model ${\cal A}$ we observe, apart from an almost pure $^2 8[56]$ state,
moderate admixtures ($\approx 20-30 \%$) to the dominant configurations and,
in particular, a
different level ordering $^2 8[56] <\; ^4 8[70] <\; ^2 8[70] <\; ^2 8[20]$.
From our considerations at the beginning of this section concerning the effect
of 't~Hooft's force on the different spin-flavor configurations one expects a
lowering of the two states with dominant $^2 8[56]$ and $^2 8[70]$
contributions, where the shift is strongest for the $^2 8[56]$ state and
rather moderate for the $^2 8[70]$ state. In this respect, the crucial
difference between the pure confinement spectra of the two models is that
model ${\cal A}$ exhibits an ordering of the levels with the dominantly 
$^2 8[56]$ state already below the dominantly $^2 8[70]$
state, whereas in model ${\cal B}$ this order is
reversed and the $^2 8[56]$ state lies $\sim120$
MeV above the position in model ${\cal A}$. Consequently, we observe in
model ${\cal A}$ this initial mass splitting being increased by a strong
lowering of the $^2 8[56]$ state of 312 MeV (very similar to the ground-state)
and a more moderate downward shift of the $^2 8[70]$ state of 174 MeV thus
very nicely reproducing the striking low position of the Roper resonance
$N\frac{1}{2}^+(1440,\mbox{****})$ and the position of
$N\frac{1}{2}^+(1710,\mbox{***})$. In model ${\cal B}$, however, the
initial reversed ordering of these two states leads to a level crossing, which
mixes the two configurations. Consequently the net effect of the residual
interaction is weakened and thus neither the low position of the Roper
resonance nor the splitting between the Roper resonance and
the $N\frac{1}{2}^+(1710,\mbox{***})$ can be explained.

To summarize, our discussion convincingly demonstrates the importance of
relativistic effects as they are present in our fully covariant framework
based on the Salpeter equation. To generate prominent hyperfine
structures in the $N^*$-spectrum like the Roper resonance by 't~Hooft's
residual force, the interplay of the residual force and the confinement force
due to relativistic effects of the confinement Dirac structure is
very crucial. We have seen that two different
confinement Dirac structures, which exhibit the same non-relativistic
limit and work almost equally well for the description of the
$\Delta$-spectrum and the spectrum of the light ground-states, might cause a
totally different behavior of particular excited $N^*$ states under the
influence of the residual 't~Hooft interaction. The different results of both models
concerning the Roper resonance thus strongly favor the confinement
version of model ${\cal A}$ to be the more realistic confinement force
in combination with 't~Hooft's residual force.\\
Let us conclude this subsection by drawing the following general conclusions
from this discussion:
\begin{itemize}
\item The very different effects of 't~Hooft's residual force on
  the Roper state in two models that have the {\it same non-relativistic
  limit}, convincingly demonstrates the importance for describing baryons in a
  genuine \textit{fully relativistic framework} in order to decide whether a
  possible choice for a residual interaction is a realistic candidate for
  explaining prominent hyperfine structures of the excited baryon spectrum.
\item In particular, the effect of a residual interaction cannot be
  considered independently from a suitable assumption on the spinorial Dirac 
structure of the confinement force. This is often done by
  non-relativistic or 'relativized' potential models which either completely
  neglect the resulting relativistic effects of the confinement Dirac
  structure by just using central confining potentials or which take only a
  part of the terms of the rather poorly converging ($|{\bf p_q}|/m_q\approx
  1$) generalized Fermi-Breit expansion of the confinement force
(and also of the residual interaction itself) into account.
Relativistic effects of both, the residual force {\it and} the confinement
force, must be consistently taken into account.
\item From our point of view this implies that statements concerning the role
  of a special choice of the residual interaction for generating the hyperfine
  structures in the excited light baryon spectrum, which are based on
  'relativized' or even non-relativistic potential models, are highly context
  dependent.
\end{itemize}
 
\subsection{Summary for the Nucleon spectrum}
After this rather lengthy and detailed discussion of the excited nucleon spectrum, which
emphasized the role of 't~Hooft's instanton-induced interaction for
explaining all the various, striking features of the mass
spectrum, it is worth to summarize our main results 
and statements.\\

With the confinement parameters and the non-strange quark mass fixed by the
phenomenology of the global $\Delta$-spectrum and the 't~Hooft coupling
$g_{nn}$ chosen to fit the $\Delta-N$ splitting we subsequently calculated
the complete excited nucleon mass spectrum. All states of the $N^*$-spectrum thus
were {\it real predictions} within the models ${\cal A}$ and ${\cal B}$.

In a first detailed investigation of the predicted structures in the
positive and negative parity shells we indeed found excellent
agreement between the empirical nucleon mass spectrum and the
predictions of our model ${\cal A}$.  Both the lower resonance
regions and the higher (and even highest) mass regions could be
uniformly described:
\begin{itemize} 
\item As in the $\Delta$ sector, we found an excellent description of
the positive parity Regge trajectory $N\frac{1}{2}^+(939)$,
$N\frac{5}{2}^+(1680)$, $N\frac{9}{2}^+(2220)$,
$N\frac{13}{2}^+(2700)$, $\ldots$ up to highest orbital excitations
$J=\frac{13}{2}$. In particular, the model yields the correct
phenomenological characteristic $M^2\propto J$ with the right slope of
the trajectory. We pointed out that similar to the ground-state all
excited states of the trajectory likewise are lowered by 't~Hooft's
force showing the nontrivial property of the instanton-induced force
to be compatible with the observed linear Regge characteristics.
\item All striking hyperfine intra-band structures in the even- and odd-parity
  shells can be nicely reproduced in at least qualitative (but mostly even
  in completely quantitative) agreement with the hitherto established,
  experimentally observed patterns. As a particular nice feature of model
  ${\cal A}$ we found that it is able to explain the conspicuous low
  position of the Roper resonance $N\frac{1}{2}^+(1440)$ in the
  $2\hbar\omega$ band (Roper {\it problem}), {\it i.e.} the first
  scalar/isoscalar excitation of the nucleon. Apart from the Roper state,
  the model can also account for the positions of three  other rather low
  lying states of the positive-parity $2\hbar\omega$ band around 1700 MeV,
  {\it i.e.} the $N\frac{1}{2}^+(1710)$, $N\frac{3}{2}^+(1720)$ and
  $N\frac{5}{2}^+(1680)$. In the negative parity $1\hbar\omega$ shell, the
  hyperfine splittings of the five experimentally observed states
  $N\frac{1}{2}^-(1535)$, $N\frac{3}{2}^-(1520)$, $N\frac{1}{2}^-(1650)$,
  $N\frac{3}{2}^-(1700)$ and $N\frac{5}{2}^-(1675)$ are reasonably well
  reproduced.
\item Concerning the experimental indications for the three new resonances
  showing up around 1900 MeV in photoproduction experiments with the SAPHIR
  detector at ELSA in Bonn, we found that our model indeed predicts possible
  candidates of states in the corresponding sectors. These predicted states
  nicely match the determined resonance position of the $P_{11}(1986)$ in the
  $N\frac{1}{2}^+$ sector and also the comparatively low
  positions of the $S_{11}(1897)$ and $D_{13}(1895)$ in the negative parity
  sectors $N\frac{1}{2}^-$ and $N\frac{3}{2}^-$, respectively.
\item Comparing the relative arrangement of the different even- and odd-parity
  bands we indeed found overlapping substructures of shells with positive and
  negative parity leading to the occurrence of more or less degenerate states
  with the same spin and opposite parity. Hence, our model is able to account
  for the  striking observed pattern of approximate parity doublets in the
  excited nucleon spectrum as {\it e.g.}
  $N\frac{1}{2}^+(1710)-N\frac{1}{2}^-(1650)$,
  $N\frac{3}{2}^+(1720)-N\frac{1}{2}^-(1700)$,
  $N\frac{5}{2}^+(1680)-N\frac{1}{2}^-(1675)$ and
  $N\frac{9}{2}^+(2220)-N\frac{9}{2}^-(2250)$.
\end{itemize}
The corresponding discussion of model ${\cal B}$ has shown that it works
less well and in particular strongly fails in describing one of the most
prominent features, namely the low position of the Roper resonance.  Hence we
favor model ${\cal A}$ as the more realistic model.

In the subsequent, second part of our discussion we then
analyzed in some detail the instanton-induced effects in the excited
nucleon spectrum of model ${\cal A}$ in order to clarify to what
extent 't~Hooft's force actually is responsible for generating all
these striking features of the $N^*$-spectrum simultaneously.
Starting from the case with confinement only we investigated how 't~Hooft's 
force affects the energy levels when its strength is gradually
increased from zero until the correct $\Delta-N$ splitting is
achieved. Due to the selection rules of 't~Hooft's force we observed
in each shell a systematic lowering of those states which exhibit a
dominant $^2 8[56]$ or $^2 8[70]$ spin-flavor $SU(6)$ configuration.
In fact, we found that really all the prominent features of the
$N^*$-spectrum like {\it e.g} the Roper resonance or the patterns of
approximate parity doublets are simultaneously generated along with
the $\Delta-N$ splitting in this way. Accordingly, we thus could
convincingly demonstrate that the instanton-induced interaction indeed
provides a consistent, systematic and uniform explanation not only for
the ground-state splitting between the $\Delta$ and the nucleon, but
really for all the observed phenomena of the complete $N^*$-spectrum
listed above.  All prominent features of the nucleon spectrum can be
remarkably well described in our fully relativistic model
${\cal A}$. 

To understand the shortcomings of model ${\cal B}$
in the $J=\frac{1}{2}^\pm$ sectors, we finally analyzed the effects of
't~Hooft's force also in model ${\cal B}$. Comparing the
instanton-induced effects in both models in the $N\frac{1}{2}^+$
sector, we demonstrated that the failure of model ${\cal B}$ in
describing the low position of the Roper resonance indeed is caused by
a quite different influence of the instanton induced interaction on
the excited $N\frac{1}{2}^+$ states of the $2\hbar\omega$ shell.  We
found that the different instanton-induced effects have their origin
in a different initial level ordering and a different initial
configuration mixing of the embedded Pauli spinors in the pure
confinement case. This is caused by the different relativistic effects
that are induced by the two distinct confinement Dirac structures of
both models in combination with the embedding map of the Salpeter
amplitudes. Hence we found the interplay of the residual force with
the relativistic effects of the confinement Dirac structure to be very
crucial. This clearly shows the importance to describe the light
baryons in a fully relativistic framework.\\

The discussion of the whole non-strange baryon mass spectrum is thus
completed and we have shown that with our model version ${\cal A}$ a
simultaneous, consistent description of the complete $\Delta$ and nucleon mass
spectra can indeed be achieved with only five parameters.

\section{Summary and conclusion}
\label{sec:concl}
In this paper we have computed the non-strange baryon spectrum on the basis of
the three-particle Bethe-Salpeter equation with instantaneous interactions
kernels.  More precisely we used a string-like three-body confinement force
and a regularized version of 't~Hooft's instanton-induced interaction.  We
could construct a more or less unique model concerning the best fit of the
model parameters. The following points (concerning the favored model ${\cal
  A}$) are particularly remarkable in comparison with the nonrelativistic or
relativized calculations. Regge trajectories up to the highest observed
angular momenta could be reproduced.  The hyperfine structure of the mass
spectrum and in particular the Roper resonance and the parity doublets found a
natural explanation.  The lowering of the Roper resonance was due to a
specific interplay of relativistic effects, the confinement potential (in
model ${\cal A}$) and 't~Hooft's force. This lowering was not isolated to
the Roper resonance alone but could be identified also for higher resonance
states. The question arises whether these achievements are purely accidental.
Of course a clear answer can not be given for any possible phenomenological
interaction, but it can be answered at least for another QCD-inspired
candidate, namely one-gluon-exchange. From phenomenological grounds this
alternative can be discarded for light flavors: We have demonstrated
this by a Bethe-Salpeter calculation in the same spirit as described in
this paper, see appendix \ref{sec:OGE} for a short description.

The way in which we have introduced 't~Hooft's interaction in our
spectroscopic calculation may look a bit too {\it ad hoc} for experts. Indeed
more elaborate treatments of the characteristic instanton effects exist in
the literature \cite{SS98,DP84,DP85,DP86,DP92,NVZ89,NRZ96}.  The result is however identical in form with the
final two-body force used in this paper. The difference shows up mainly in the
form factor of the force which we determined in a purely phenomenological way anyhow.

It remains to compute the spectrum of strange baryons in order to complete our
unified description of the light baryon masses.  This will be done in a
separate paper \cite{Loe01c}.

\begin{acknowledgement}
\textbf{Acknowledgments:}
We have profited very much from scientific discussions with
V.~V.~Anisovich, G.~E.~ Brown, E.~Klempt, K.~Kretzschmar,  
A.~V.~Sarantsev and E.~V.~Shuryak
to whom we want to express our gratitude. We also thank the
Deutsche Forschungsgemeinschaft (DFG) for financial support. 
\end{acknowledgement}

\appendix
\section{Appendix: One-gluon-exchange forces}
\label{sec:OGE}
De Rujula, Georgy and Glashow \cite{DGG75} initially suggested to explain the
hyperfine structure of hadrons by using the OGE as a residual interaction.
Later it was applied in several non-relativistic potential models. To adopt
this interaction in non-relativistic quark models, the fully relativistic
expression for the perturbative OGE interaction is expanded in powers of
$|{\bf p_q}|/m_q$ up to order ${\cal O}(|{\bf p_q}|^2/m_q^2)$, leading to
the so-called Breit-Fermi interaction.  Although various parts of this
interaction have been applied in non-relativistic quark model calculations,
{\it e.g.}  perturbatively for each shell in a naive oscillator model by Isgur
{\it et al.}  \cite{IsKa78,IsKa79}, the full expression, however, has never
been used.  Moreover, due to the fact that the quarks are not really slow,
{\it i.e} $|{\bf p_q}|/m_q\approx 1$, the convergence of this expansion should
be quite bad at least for light quarks such that even higher order
relativistic corrections might contribute significantly. In fact, such a
treatment of relativistic effects of the OGE is therefore not justified. It
should be emphasized that particular terms of the Breit-Fermi expansion are
{\it deliberately neglected}. Employing the full Breit-Fermi interaction, a
severe difficulty appears, which is referred to as the so-called ''{\it
  spin-orbit-problem}''. While the hyperfine structure of ground-state baryons
can be nicely explained by the short-range spin-spin part (Fermi contact
term), the inclusion of spin-orbit forces arising from OGE leads to large
splittings for excited states spoiling the agreement with phenomenology, since
the experimental mass spectrum of excited baryon resonances indicates that
such a strong spin-orbit force should not exist between quarks. In order to
obtain a reasonable description of the spectra these approaches are therefore
forced to remove the spin-orbit components of the OGE by hand and to include
as spin-dependent part the color magnetic hyperfine interaction only. However,
leaving these interactions out is rather unsatisfactory and inconsistent.
Isgur argued heuristically \cite{IsKa78,CaRo00} that spin-orbit forces might
cancel with the Thomas term from a confinement force with scalar Dirac
structure (see also ref. \cite{HK83} and references therein). But in view of
several terms of the Breit-Fermi expansion that are left out anyway, this
explanation is not really convincing and thus still remains rather {\it ad
  hoc}.

In order to correct the flaws in the non-relativistic model, other potential
model calculations, which retain the one-gluon-exchange picture of the
quark interaction, have gone beyond the original model of Isgur {\it et
  al.}.  The so-called 'relativized' extension has been investigated by
Godfrey, Capstic and Isgur for mesons \cite{GoIs85} and subsequently also for
baryons \cite{CaIs86}. Also in this attempt, which still is based on the
ordinary Schr\"odinger equation (with the kinetic energy replaced by its
relativistic expression), the usual terms of the Fermi-Breit reduction of the
OGE (including spin-orbit forces) as in the non-relativistic approach have
been used. But compared to the non-relativistic version the expressions have
been modified in order to {\it qualitatively} parameterize the momentum
dependence of the relativistic corrections away from the $|{\bf
  p_q}|/m_q\rightarrow 0$ limit. However, this effective parameterization of
relativistic effects is at the cost of introducing several new,
non-fundamental parameters which were not derived from first principles.
Using the new freedom to fit the additional parameters, the spin-orbit
interactions could indeed be suppressed relative to the contact interaction.
These effects, along with a partial cancellation of OGE-induced spin-orbit
effects with those of a scalar three-body confinement (calculated within a
two-body approximation) could finally reduce the size of spin-orbit effects to
an acceptable level. Nonetheless, in view of the crude method of
''relativizing'' the non-relativistic quark model, the spin-orbit puzzle
connected with the employment of OGE as residual interaction is in fact
unsolved. 

More reliable investigations ultimately require the description of
baryons in a fully relativistic framework, where relativistic effects of the
residual force and the confinement force are {\it consistently} and {\it
  fully} taken into account.  In this respect, a suitable approach to
investigate this problem in fact is given by our covariant Salpeter framework.
We therefore used this framework to verify the statements of the relativized
quark model \cite{CaIs86} concerning the spin-orbit problem in an analogous,
but fully covariant model which we shall discuss here briefly. Recall that our
approach does not introduce any additional parameters; all relativistic effects
are {\it uniquely} and {\it fully} determined by means of the embedding map
in the Salpeter amplitudes.

According to the assumptions of the relativized quark model \cite{CaIs86} we
parameterize confinement by a $\Delta$-type three-body string potential which
is assumed to have a {\it scalar} Dirac structure:
\begin{equation}
V^{(3)}_{\rm conf}({\bf x_1}, {\bf x_2}, {\bf x_3})
=
\left[3a + b\sum_{i<j} |{\bf x_i}-{\bf x_j}| \right]\;\;
\Id\tens\Id\tens\Id. 
\end{equation}
In view of the instantaneous treatment of the OGE the natural gauge for the
gluon propagator is the Coulomb gauge \cite{Mur83}, which will be applied in the following.
This specific gauge has the advantage that the gluon propagator given by 
\begin{equation}
\gamma^\mu D_{\mu\nu}\gamma^\nu
=
4\pi
\left(\frac{\gamma^0\tens\gamma^0}{|{\bf q|^2 }}
+
\frac{ 
{\bfgrk\gamma}\cdot\!\tens\;{\bfgrk\gamma} 
- 
({\bfgrk\gamma}\cdot {\bf \hat q})\tens({\bfgrk\gamma}\cdot {\bf \hat q})}
{q^2+{\rm i}\epsilon}
\right)
\end{equation}
with ${\bf \hat q}:= {\bf q}/|{\bf q}|$ is already instantaneous in its
component $D_{00}(q)$ which describes the ordinary coulomb potential. In the 
instantaneous approximation we substitute $q^2$ in the second term by $-|{\bf q}|^2$. The
two-quark OGE kernel in coordinate space then reads \cite{Mur83}
\begin{equation}
V^{(2)}(x_1,x_2;\;x_1',x_2')
=
V^{(2)}_{\rm OGE}({\bf x})\;
\delta^{(1)}(x^0)\; \delta^{(4)}(x_1-x_1')\;\delta^{(4)}(x_2-x_2'),
\end{equation}
with $x=x_1-x_2$ and
\begin{equation}
\label{OGEPot}
V^{(2)}_{\rm OGE}({\bf x})
=
-\frac{2}{3}\frac{\alpha_s}{|{\bf x}|}
\left(
\gamma^0\tens\gamma^0 
- 
\frac{1}{2}{\bfgrk\gamma}\cdot\!\tens\;{\bfgrk\gamma} 
- 
\frac{1}{2}({\bfgrk\gamma}\cdot {\bf \hat x})\tens({\bfgrk\gamma}\cdot {\bf \hat x})
\right),
\end{equation} 
where ${\bf \hat x}:= {\bf x}/|{\bf x}|$. Here $\alpha_s$ is the running
strong coupling constant, which in momentum space is assumed to saturate at a
maximal value for $q^2\rightarrow 0$. For the sake of simplicity we shall treat the
coupling $\alpha_s$ as a constant. This coupling together with the other
parameters listed in table \ref{tab:ModelParamOGE} has been fixed by a common
fit to the spin-$3/2$ decuplet and spin-$1/2$ octet ground-states baryons as
well as to the states $\Delta\frac{7}{2}^+(1950, \mbox{****})$ and
$\Delta\frac{11}{2}^+(2420,\mbox{****})$ of the $\Delta$-Regge trajectory in
order to obtain also the correct positions of the higher band structures. The
Salpeter equation has been solved by expanding the states in a rather large
harmonic oscillator basis up to $N_{\rm max}=14$ for the spin-$\frac{1}{2}$
baryons and $N_{\rm max}=12$ for all other spins.
\begin{table}[!h]
\center
\begin{tabular}{cccrl}
\hline
Constituent  & non-strange     & $m_n$     & 450& MeV\\
quark masses & strange         & $m_s$     & 675& MeV\\
\hline
Confinement  & offset          & $a$       & -395& MeV\\
parameters   & slope           & $b$       & 926&MeV fm$^{-1}$\\
\hline
OGE          & strong coupling & $\alpha_s$& 1.10&\\                  
\hline
\end{tabular}
\caption{Model parameters of the scalar confinement, the quark masses and the strong coupling constant}
\label{tab:ModelParamOGE}
\end{table}

\begin{figure}[!h]
  \begin{center}
    \epsfig{file={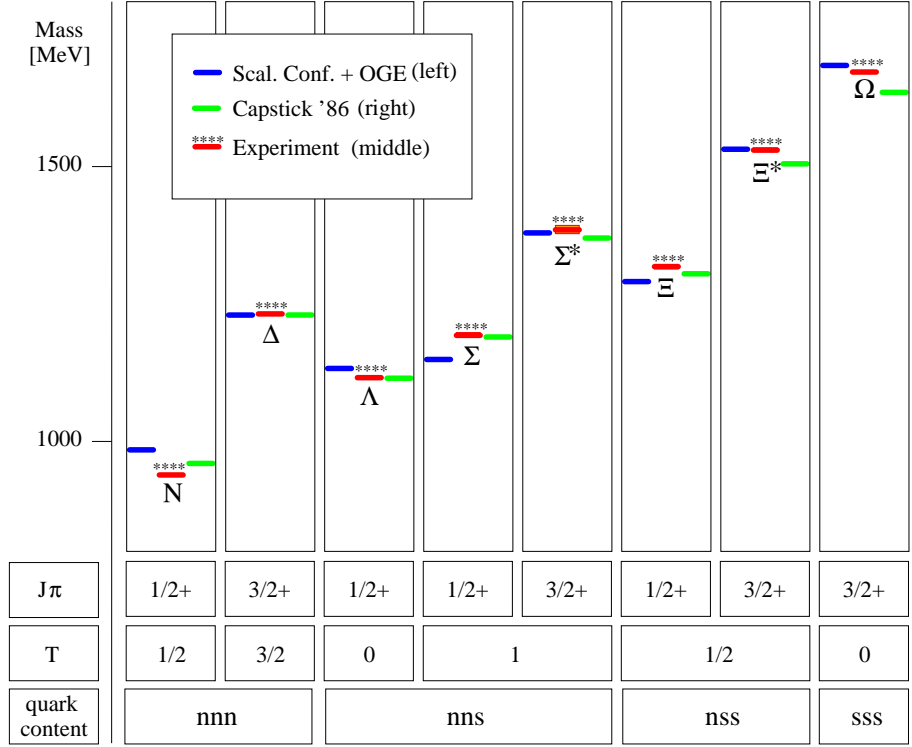},width=120mm}
  \end{center}
\caption{The resulting spin $1/2$ octet and spin $3/2$ decuplet ground-state baryons calculated with
  a scalar three-body confinement and residual OGE interaction (on the left in each column). For
  comparison the results of the 'relativized' model \cite{CaIs86} are shown on the
  right in each column. In the middle of each column the experimental masses
  \cite{PDG00} are displayed.}
\label{fig:grounstatesOGE}
\end{figure}
Figure \ref{fig:grounstatesOGE} shows the resulting masses of the
octet and decuplet ground-state baryons in comparison with the
experimentally observed positions \cite{PDG00}.  In addition, also the
results of the relativized model \cite{CaIs86} are displayed. Since
the ground-states are dominantly S-wave states the spin-orbit forces
are irrelevant.  As can be seen also a fully relativistic treatment of
the one-gluon-exchange interaction allows a reasonable description of
the hyperfine splittings $\Delta-N$, $\Sigma^*-\Sigma$ and $\Xi-\Xi^*$
which is comparable to the mass splittings generated by the
spin-dependent Fermi contact interaction in the non-relativistic and
relativized approach. Note however, that the $\Sigma-\Lambda$
splitting turns out much too small in comparison to the observed
splitting of roughly 75 MeV. It is worth to mention that the value of
the strong coupling constant $\alpha_s=1.1$ in our fully relativistic
approach is {\it not} reduced relative to the usual values
$\alpha_s\simeq 1.0$ determined in nonrelativistic quark
models. Treating $\alpha_s$ really as running coupling constant would
require a saturated value even bigger than 1.1.  This is quite in
contrast to the result of the relativized model \cite{CaIs86}, where
this value became significantly smaller, namely $\alpha_s =0.6$.

\begin{figure}[!h]
  \begin{center}
    \epsfig{file={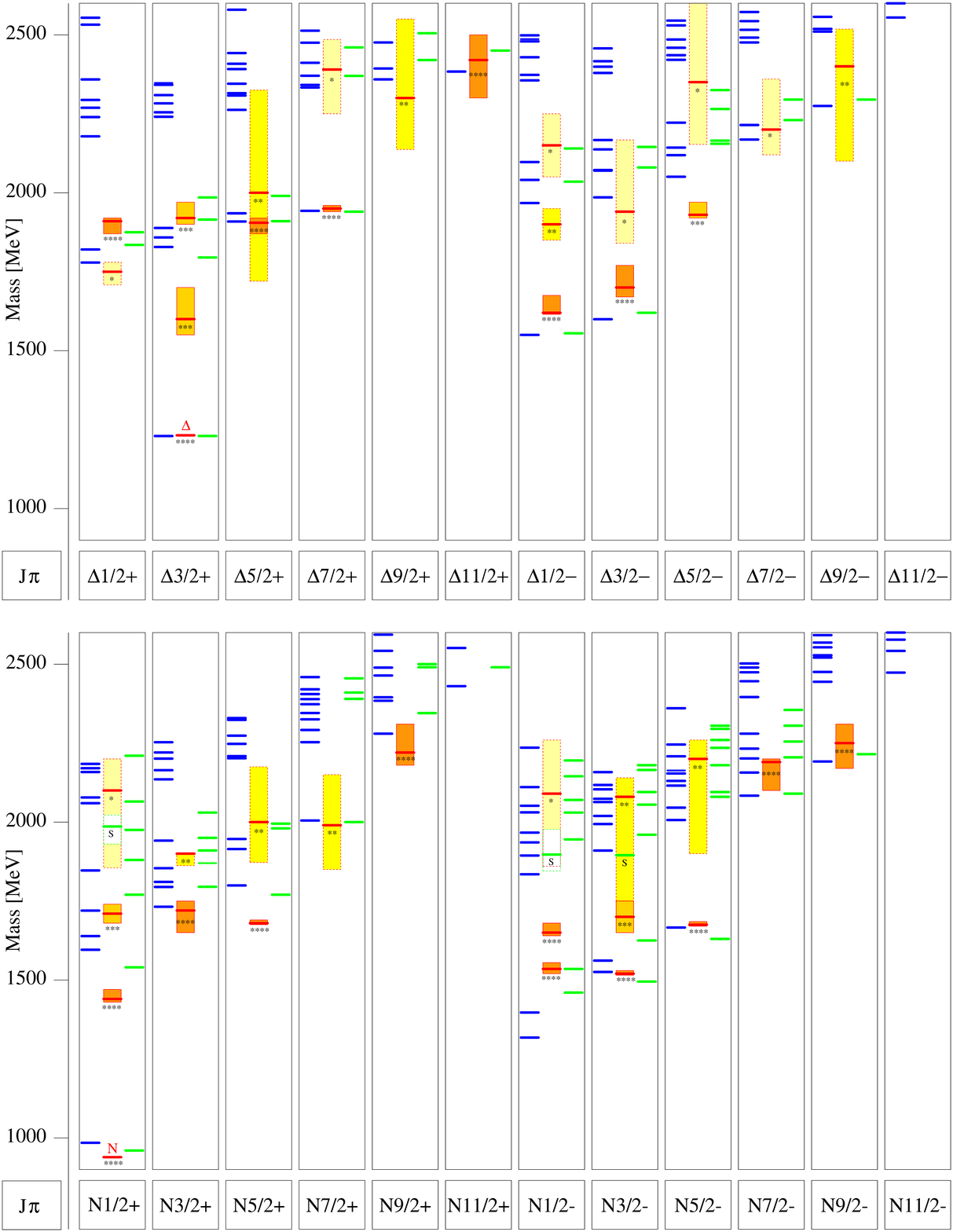},width=159mm}
  \end{center}
\caption{The positive- and negative-parity
  \textbf{$\Delta$- and $N$-resonance spectrum} calculated in a scalar
  three-body confinement model with residual OGE interaction (left in each
  column) in comparison to the experimental spectrum \cite{PDG00} (middle of
  each column) and the results of the 'relativized' model \cite{CaIs86,CaRo93} (right
  in each column). Experimental positions are indicated by a bar, the
  uncertainties by the shaded box which is darker for better established
  resonances; the rating is additionally indicated by stars.}
\label{fig:DelNucOGE}
\end{figure}

Concerning the spin-orbit problem, which in fact is our major
interest here, let us as an example consider the spectrum of the
excited non-strange baryons calculated with the parameter set of table
\ref{tab:ModelParamOGE}.  Our results for the $\Delta$ and $N$ sector
up to 2600 MeV are depicted in fig. \ref{fig:DelNucOGE}, once again
in comparison with the currently experimentally known resonance
spectrum reported by the Particle Data Group \cite{PDG00} and the
corresponding results of the relativized quark model \cite{CaIs86,CaRo93}. 
In each column the resonances are classified by the total
spin and parity $J^\pi$ and in each sector from $\frac{1}{2}^\pm$ to
$\frac{11}{2}^\pm$ at most the first ten predicted excitations are
displayed.  Altogether, the structure of the $\Delta$ spectrum can
still be reasonably well accounted for. The spin-orbit effects are
moderate enough to be compatible with the uncertainties of the
experimentally observed $\Delta$ resonances in each band. Our
calculation roughly agrees with that of the relativized quark
model. For instance the predictions for the two lowest excited
$\Delta\frac{1}{2}^-$ and $\Delta\frac{3}{2}^-$ negative parity states
are quite the same. Notice however that both models predict their
centroid too low with respect to the observed states
$\Delta\frac{1}{2}^-(1620,\mbox{****})$ and
$\Delta\frac{3}{2}^-(1700,\mbox{****})$. This is due to the fact that
the OGE interaction shifts the $\Delta$ ground-state upwards relative
to these two states.

Much bigger discrepancies between the model
predictions and the phenomenological spectrum emerge in the nucleon
sector. First of all the predicted intra-band structures do not well
agree with prominent empirical structures such as {\it e.g.} the four
comparatively low lying states of the positive-parity $2\hbar\omega$
band (including the Roper resonance) or the hyperfine splittings in
the negative-parity $1\hbar\omega$ band. In fact, the most striking
effect of the spin-orbit interaction is seen for the two lowest
excitations of the negative parity $N\frac{1}{2}^-$ sector. The center
of gravity of these states is lowered by roughly 300 MeV with respect
to the $N\frac{5}{2}^-$ state. Hence the predictions are far below the
empirical positions of $N\frac{1}{2}^-(1535,\mbox{****})$ and
$N\frac{1}{2}^-(1650,\mbox{****})$.  We should note that this effect,
although less pronounced, is likewise indicated in the relativized
model \cite{CaIs86}, and also Isgur \cite{IsKa78} predicted within a
perturbative calculation of OGE-induced spin-orbit effects a quite
large downward mass shift of roughly 500 MeV relative to the
$N\frac{5}{2}^-$. However, we obviously cannot confirm the conjecture
that these effects might cancel against corresponding equally large
spin-orbit effects stemming from the scalar confinement force. We
should remark that similar results are obtained also for the
corresponding strange sectors.  We tried to cure this problem by using
other spin structures than scalar.  In fact it turned out that
hyperfine structures of excited states strongly depend on the Dirac
structure chosen. But so far no appropriate choice has been found that
could suppress the large OGE-induced spin-orbit effects in the
$J^\pi=\frac{1}{2}^-$ sectors. In our opinion a solution of the
spin-orbit puzzle connected with the residual OGE interaction thus
seems highly questionable even in fully relativistic quark model.
Our results concerning OGE in the Bethe-Salpeter framework indicate that it
can be discarded for light flavors from phenomenological grounds.  

In fact, apart from the spin-orbit problem of the residual OGE force, there
are even more basic objections against the perturbative OGE interaction.
Actually, it is only valid in the asymptotic-free domain of QCD, where the
strong coupling constant is small and perturbation theory is expected to work.
However, baryon spectroscopy obviously belongs to the domain, where the strong
coupling is large such that perturbation theory fails and therefore
complicated higher order multi-gluon-exchange contributions should be of
roughly the same order of magnitude as the lowest order OGE contribution
itself. The strength of OGE determined in an empirical way by a fit to the
hyperfine splittings turned out to be roughly $\alpha_s\approx 1$ which indeed
makes it hard to treat it as a perturbative effect.  Furthermore, the OGE is
explicitly flavor independent and one thus obtains in the mesonic spectrum
degenerate $\pi$ and $\eta$ mesons in clear contradiction to experiment. In
order to cure this discrepancy one in fact would have to take into account
higher order QCD diagrams \cite{GoIs85}. All these arguments thus call into
question the justification for applying the perturbative OGE in a 
nonperturbative sector of QCD. It can at best be added with a small coupling
to the stronger force which we favor in this paper and would only modify the 
results slightly.

%
%
%
%
%

\end{document}